# Aiming for AI Interoperability: Challenges and Opportunities

November 2025

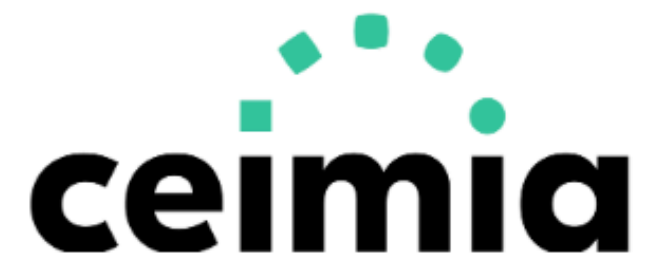

with support from Google.org

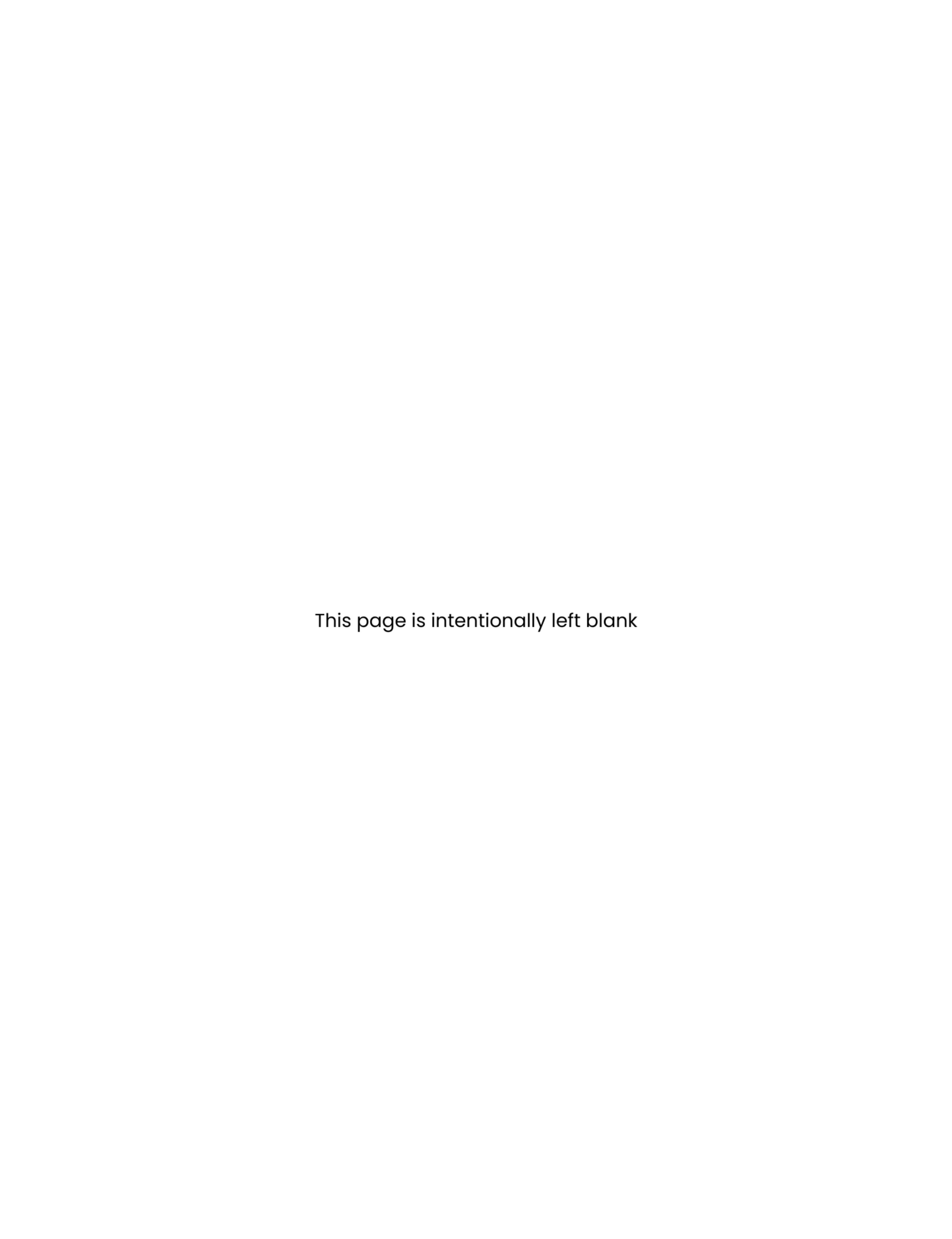

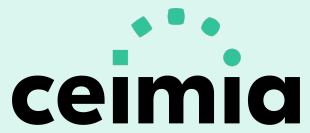

# About CEIMIA

In an era of rapid development in artificial intelligence (AI), including the arrival of generative AI, governments are faced with the crucial task of effectively navigating the complexities surrounding the deployment of AI and its impact on society. It is in this context that the International Centre of Expertise in Montreal on Artificial Intelligence (CEIMIA) supports the work of the Global Partnership on Artificial Intelligence (GPAI), a multi-stakeholder initiative aiming to bridge the gap between theory and practice on delivering responsible AI. GPAI does this by supporting cutting-edge research and applied activities on AI-related priorities. Built around a shared commitment to the OECD Recommendation on AI, GPAI brings together engaged minds and expertise from science, industry, civil society, governments, international organizations and academia to foster international cooperation.

With its unique position supporting GPAI, CEIMIA mobilizes international experts and resources (from the academic, private, and civil society sectors) to promote the responsible development and use of AI for the benefit of humanity. CEIMIA is therefore acting as a key player in the responsible development of AI based on human rights, inclusion, diversity, innovation, economic growth and the well-being of society, while seeking to achieve the United Nations' sustainable development goals.

# Acknowledgements

**Report Authors:**

The report was written by:

**Benjamin Faveri**, Senior Project Manager, CEIMIA;

**Craig Shank**, Affiliate Researcher, CEIMIA;

**Richard Whitt**, President, GliaNet Alliance; and Affiliate Researcher, CEIMIA;

**Phillip Dawson**, Head of AI Policy, Armilla AI; and Affiliate Researcher, CEIMIA.

**Project Advisory Group:**

CEIMIA recognizes the meaningful contribution of the Aiming for AI Interoperability Project Advisory Group members who reviewed the report, presented at webinars, and offered support when needed throughout the project (in alphabetical order):

**Alyssa Lefaivre Škopac**, Director of AI Trust & Safety, Amii;

**Ashley Casovan**, Managing Director, IAPP's AI Governance Center;

**Cornelia Kutterer**, Managing Director, Considerati;

**Gary Marchant**, Regents' Professor and Faculty Director, Arizona State University;

**Graeme Auld**, Professor, Carleton University;

**Kim Lucy**, Director of GRC Standards, Microsoft;

**Laura DeNardis**, Professor, Georgetown University;

**Marta Janczarski**, National Standards Officer, Microsoft;

**Mary Saunders**, Senior VP for Government Relations and Public Policy, ANSI;

**Monique Crichlow**, Executive Director, Schwartz Reisman Institute for Technology and Society.

Administrative and Communications Support:

**Gwenaëlle Le Peuch**, Communications Manager, CEIMIA;

**Myriam Séguin**, Social Media Manager, CEIMIA;

**Caroline Renaud**, Executive Assistant, CEIMIA;

**Nathalie Noel**, Senior Manager, Projects and Ecosystems, CEIMIA;

**Janick Houde**, Partnerships and Ecosystems Manager, CEIMIA.

Supervised by:

**Stephanie King**, Director of AI Initiatives, CEIMIA.

# Disclaimer

This report was researched and developed from November 2024 to November 2025. Given the rapid developments in the AI regulatory and technical landscape, this paper stopped updating sections as of October 2025 to finalize the report and prepare it for publication. Google.Org provided the financial support for this report and its corresponding deliverables, such as three online webinars, the three op-eds, and an in-person workshop and networking event. The authors hold no conflicts of interest. This report reflects the research findings of the author team and does not necessarily reflect the views of CEIMIA, Google.Org, or Project Advisory Group members or their organizations.

# Citation

# Part 1: The State of AI Governance and Need for Regulatory and Technical Interoperability

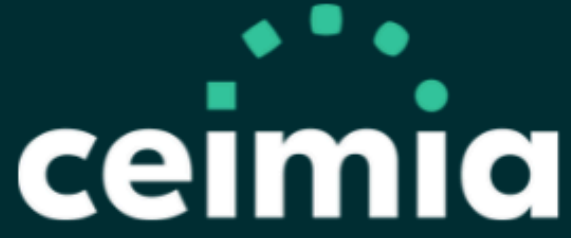

# Part 1: The State of AI Governance and Need for Regulatory and Technical Interoperability

## 1. Introduction

AI technical and regulatory governance efforts are proliferating at a staggering pace through hard and soft laws. Countless national and subnational governments are proposing, passing, and implementing AI regulation and legislation; several international and national standard-setting bodies are drafting and publishing various technical and non-technical AI standards; non-profits and international organizations are releasing AI principles, frameworks, codes of conduct, and certifications; and industry firms are developing and offering various AI governance services and products.[1] While these efforts are being made to address AI's sector-specific, general, and contextual challenges, there are now emerging challenges around the sheer number of AI technical and regulatory governance efforts being released. The current AI governance landscape is becoming increasingly fragmented and convoluted. As a result, firms are unsure which principles, frameworks, standards they should comply with. For those that operate transnationally, it can be unclear whether and how subnational rules apply to their activities. Moreover, firms lack clarity on which rule checking firms they should hire to check their compliance against these various rules and standards (Cihon, 2019; Cihon et al., 2020; Dennis et al., 2024). Similarly, entities making these rules are stuck between trying to decide if they should create rules that are distinct from existing rules or replicate, in whole or in part, existing rules. The former further complicates the governance landscape but can improve or iterate existing rules while the latter simplifies but further ingrains existing rules.

Previous works have compared AI principles, policies, regulations, and standards, finding a mix of similarities and differences (CAIDP, 2024; CEIMIA 2024; Jobin et al., 2019; Marchant & Gutierrez, 2023). However, this work did not look at specific mechanisms or paths forward to lead ongoing AI technical and regulatory governance efforts towards overlap, similarity, harmonization, or interoperability.[2] For the purpose of this report, interoperability is divided into two components: technical and regulatory. Technical interoperability is the connection between more than one system, service, or platforms' infrastructures to promote data management, productivity and scalability, and user experiences (Internet Governance Forum, 2024). Regulatory interoperability is the intentional overlap and consistency between rules – such as those between countries, the same standards being incorporated into various countries' regulation, or national policy efforts aligning with international organization's principles or codes of conduct, among others – to facilitate international trade, reduce barriers to entry for new market entrants, and avoid regulatory venue shopping by transnationally operating firms (Brinsmead, 2021; Internet Governance Forum, 2024). The combination of technical and regulatory interoperability could address the AI governance landscape's emerging fragmentation, convolution, and challenges associated with sector-specific vs. overarching approaches – assuming some well-informed paths forward can be identified and mapped onto the existing AI governance landscape (Google, 2024; Onikepe, 2024; World Bank, 2024).

[1] See Faveri & Marchant (2024a: 2024b) datasets on 200+ US State and Federal laws and regulations around AI and 1400+ technical and non-technical international AI and AI-related standards; OECD's AI Policy Observatory for global AI policy efforts; and the EAIDB's 2024 Responsible AI Ecosystem Market Map for 300+ startup firms offering AI products and services.

[2] Each of these terms effectively gets at the same concept - simplifying the governance landscape. This report uses and focuses on "interoperability."

Identifying these interoperability pathways and mapping them onto the current AI governance landscape is the primary goal of this report. To accomplish this goal, this report develops in three stages. First, we provide a cursory overview of the current AI governance landscape, focusing on hard laws, soft laws, and ongoing interoperability and harmonization efforts. Second, we detail and justify the need for AI technical and regulatory interoperability, drawing attention to the implications of diverging governance regimes, the role of hard and soft law, and challenges to implement technical and regulatory interoperability. Third, we provide our preliminary potential pathways to encourage and implement AI technical and regulatory interoperability, each offering an objective, rationale, and implementation plan.

## 2. AI Governance Landscape

The AI governance landscape is vast, with efforts being advanced through hard laws, such as government legislation and regulation at the subnational, national, and, to a lesser extent, international levels, and soft laws, such as standards, certifications, codes of conduct, and principles, among others, which tend to be developed by industry firms, non-profit and civil society organizations, standard-setting organizations, and industry associations, among others. We do not provide a comprehensive landscape in this report. Instead, we focus on some of the major hard and soft law AI governance efforts across this landscape, and ongoing coordination, harmonization, and interoperability efforts.

### (a) Hard Laws: Legislative and Regulatory Efforts

Hard law typically refers to two types of legal instruments: *laws* passed by legislatures and signed into law by executive bodies and *regulations* or *policies* adopted by the executive branch. In Canada and the United States, the three primary jurisdictions for such implements are the federal, province/territory/state, and counties/municipality governments. These implements could also be considered essential tools in the government's "power of the pen," which can be further divided into two targets: the private and public sector. A related tool, the "power of the purse" relates to the funding mechanisms utilized by government bodies, such as procurement requirements, tax policy, and fiscal grants for research and development projects. These latter mechanisms can be thought of as creating incentives to sway corporate behaviour or bolster national industries, as opposed to the mandatory elements typically contained in "power of the pen" legislation and regulations - a mix of carrots and sticks.

The essential substantive frameworks for AI governance that have evolved over the past several years are largely guardrails, or accountability approaches. This approach entails defining certain risky behaviors related to AI and constraining that behaviour or adoption by two categories of providers: developers and deployers. While developers are the software engineers and others who devise AI models and offerings, deployers encompass all individuals and entities who bring those inventions to the market. For example, in the data protection area, the EU has taken the global lead on regulating AI systems with the EU *AI Act*. This *Act* uses accountability framing by divvying up AI systems into three risk levels: high, limited, and minimal risk. High-risk systems are subject to extensive requirements, including mandatory documentation and monitoring. Conversely, at the US federal level, no laws or regulations currently apply to AI systems like LLMs and agents. Instead, the previous Biden Administration signed an Executive Order (EO) establishing certain policies applicable to the government's use of AI (Executive Office of the President, 2023). Various duties were assigned to NIST in this EO, including establishing guidelines and best practices to promote standardization for developing and deploying "safe, secure, and trustworthy" AI systems. On January 23, 2025, the new Trump Administration revoked the Biden AI EO, requested all departments and agencies to rescind all actions taken under that EO, and adopted a new EO that focused on a commitment to sustain and enhance US AI dominance through human flourishing, economic competitiveness, and national security (White House, 2025). Shortly thereafter, the White House's Office of Science and Technology Policy issued a call for public comments[3] that received over 8700 comments,[4] and has been developed into what is now their *AI Action Plan*.[5]

At the US State level, various legislatures have introduced, are debating, passed, or have signed AI bills into force, which were primarily focused on public sector uses of AI (IAPP, 2025). Most of these created task forces and research studies required publication of inventories of AI use cases and added new procurement requirements. Colorado has adopted the first comprehensive AI laws applicable to the private sector. SB 24-205[6] put in place a "reasonable care" standard to avoid discrimination by AI developers and deployers. A task force is finalizing work to modify the original law, with input from affected businesses and consumer groups. California adopted some 17 separate laws, but in a narrower and sector-specific way (IAPP, 2025). State governments have also made use of the (2025) EO path. As of January 2025, some 13 States have issued EOs that primarily address whether or how AI should be used by State governments (Anex-Ries, 2025). The Center for Democracy and Technology observes that these State EOs do not use consistent definitions of AI and rely on task forces or pilot projects to help realize the benefits, and avoid the pitfalls, of using such technologies (Dwyer, 2025). Given the Trump Administration's recent EO and shifting AI priorities, state-level regulation and legislative efforts may become more fragmented in the short-term as these efforts may or may not shift to align with the new EO. These shifts will likely create additional fragmentation but, in the long-run, lead to less fragmentation (Basu, 2025). At the local level, New York City passed *Local Law 144* in 2021 which regulates AI use in employment decisions. Employers are prohibited from using automated employment decision tools unless it has been subject to an independent bias audit within one year of its use.

In Canada, the *Artificial Intelligence and Data Act* (AIDA) was announced in 2021 and went through part of legislative process but then died on the order papers because of the Canadian Parliament's recent prorogation. While it is unlikely that *AIDA* will restart the legislative process in its current state, there are signals around a potential new AI Bill focused on trust and data privacy from the Canadian *Minister of Artificial Intelligence and Digital Innovation*.[7] In November 2024, shortly before the recent prorogation, the Canadian government launched the *Canadian Artificial*

---

[3] https://www.whitehouse.gov/briefings-statements/2025/02/public-comment-invited-on-artificial-intelligence-action-plan/

[4] https://www.federalregister.gov/documents/2025/02/06/2025-02305/request-for-information-on-the-Development-of-an-artificial-intelligence-ai-action-plan

[5] https://www.whitehouse.gov/wp-content/uploads/2025/07/Americas-AI-Action-Plan.pdf

[6] https://leg.colorado.gov/bills/sb24-205

[7] https://www.cbc.ca/news/politics/ai-canada-regulations-innovators-researchers-9.6935017

*Intelligence Safety Institute*[8] "to ensure that governments are well-positioned to understand and act on the risks of advanced AI systems." In 2025, the Canadian Government has put together an *AI Strategy Task Force*[9] and put out a *Call for Comments*[10] to renew their *Pan-Canadian AI Strategy*.[11]

Most of these hard law efforts have focused on constraining "risky" corporate behaviors outside the protective "guardrails." What happens "within the guardrails?" At most, it is left to transparency requirements. How a certain technology is applied, in the guise of AI agents, and whether and how ordinary citizens can be empowered by these technologies remains largely unregulated. In terms of technical AI interoperability, the analysis is straightforward: no hard laws currently exist around the technical specifications of AI. Some discussions of corporate Application Programming Interfaces (API) and standards development have taken place within industry bodies, but have not led to any legislative, executive, or regulatory actions. Oftentimes, technical interoperability is taken up through soft laws, such as standards, certifications, codes of conduct, and principle efforts, among other mechanisms (Auld et al., 2022).

### (b) Soft Laws: Standards, Certifications, Codes of Conduct, and Principle Efforts

The AI soft law landscape consists of standards (like ISO/IEC 42001), certification programs (like *Eticas Certified AI*[12]), principle documents (like OECD's AI Principles), and other non-enforceable governance efforts. These soft law efforts have developed far quicker and are more plentiful than hard laws. For instance, Marchant & Gutierrez (2023) identified and examined 600+ AI soft law programs; Faveri & Marchant (2024b) created a dataset of 1400+ AI standards across the IEEE, ISO, ETSI, and ITU-T (and a similar dataset came out in 2025[13]); Jobin et al. (2019) provided an early analysis of global AI ethics guidelines. The sheer number of these soft law efforts has led to several challenges, such as overlapping, fragmented, and competing efforts. These challenges are now contributing to increased compliance burdens (e.g., audits, conformity assessments, and other assurance mechanisms to demonstrate compliance against various standards, principles, and other soft laws) and locked-in effects as soft law efforts are all vying for market share. This vying for market share can reduce regulatory and technical innovation and protectionism – where industries or countries will favour their own soft law efforts over others to ensure their own survival and market share. For example, with the *AIDA* pause, the Government of Canada seems to be coalescing around their *Voluntary Code of Conduct*,[14] *Algorithmic Impact Assessment*,[15] and *AI Strategy for the Federal Public Service*.[16] While some AI soft law efforts have seemingly 'won the war,' like ISO/IEC 42001[17] and OECD's *AI Principles*, there is demand for technical and regulatory interoperability to simplify this landscape (Internet Governance Forum, 2024; Onikepe, 2024).

---

[8] https://ised-isde.canada.ca/site/ised/en/canadian-artificial-intelligence-safety-institute

[9] https://www.canada.ca/en/innovation-science-economic-evelopment/news/2025/09/government-of-canada-launches-ai-strategy-task-force-and-public-engagement-on-the-development-of-the-next-ai-strategy.html

[10] https://ised-isde.canada.ca/site/ised/en/public-consultations/help-define-next-chapter-canadas-ai-leadership

[11] https://ised-isde.canada.ca/site/ai-strategy/en

[12] https://eticas.ai/

[13] AI Standards Exchange Database: https://aiforgood.itu.int/ai-standards-exchange/

Technical and regulatory interoperability can reduce compliance cost burdens as AI models or systems developed according to interoperable efforts would be technically consistent (Faveri & Auld, 2024; 2025). Compliance efforts that are uniform across those models and systems also reduce the cost of compliance efforts. Alongside this compliance cost reduction, regulatory interoperability would promote safety and accountability through shared standards for identifying and addressing AI risks. Interoperability efforts would also support the development of globally applicable soft law approaches. Representatives from firms, academia, and government can collaborate to establish and endorse existing standards surrounding specific uses of AI to tackle its risks and challenges. Interoperability supports this approach by ensuring consistency across different soft law initiatives and promoting innovation by allowing new AI models and systems to be adopted and adapted in various contexts, allowing developers worldwide to address both local and global challenges effectively (Soler et al., 2024). Leveraging international technical standards in regulatory approaches facilitates interoperability by aligning commonly accepted protocols and best practices with AI governance across jurisdictions, industries, and sectors, such as through incorporation by reference – where one or more standards are referenced within a law or regulation (Bremer, 2013; Faveri, 2025). This alignment provides a common framework within which developers, providers, deployers, and users can operate effectively while managing AI risks (Dennis et al., 2024). By utilizing

---

[14] https://ised-isde.canada.ca/site/ised/en/voluntary-code-conduct-responsible-development-and-management-advanced-generative-ai-systems

[15] https://www.canada.ca/en/government/system/digital-government/digital-government-innovations/responsible-use-ai/algorithmic-impact-assessment.html

[16] https://www.canada.ca/en/government/system/digital-government/digital-government-innovations/responsible-use-ai/gc-ai-strategy-overview.html

[17] While it is unlikely that there will be a single universal standard for any technology, ISO/IEC 42001 comes fairly close to a widely adopted standard as anything could this early in the innovation process

technical and regulatory interoperability efforts, soft law initiatives in AI governance can achieve greater coherence, effectiveness, and global applicability.

### (c) Coordination, Harmonization, and Interoperability Efforts

Establishing alignment and coherence across borders demands focused policy and standardization efforts. In a rapidly expanding landscape of regulatory models, technical standards, and best practices, efforts often diverge rather than align. National interests and independent operation of policy-making bodies mean that rules can have a natural tendency toward fragmentation, particularly as discussions around AI sovereignty are ramping up - which could undermine interoperability efforts depending on how those discussion's recommendations are implemented. At the same time, aiming to fight fragmentation with prescriptive transnational consistency from the top-down seems naïve at best, and potentially highly counterproductive. Governments will tailor their AI rules and frameworks to their own national security, economic, and ethical priorities, just as they do with financial regulation, health care guidance, and environmental policy. The challenge, then, is to ensure that those national level rules serve an additional interest that is likely to be similar between nations – the ability for AI systems and services to function across borders without unnecessary regulatory friction. To achieve that end, transnational efforts should focus on achieving regulatory and digital trade coherence – recognizing that the rules across borders may be different, but they can still permit cross-border AI interactions and commerce.

> As an illustrative example of 'coherence,' consider the Twenty-foot Equivalent Unit (TEU), a group of standards in the global shipping of physical goods market. The TEU collection of standards enables shipping and trade to transact across a global network, enabling trade, contracting, safety, and security – regardless of whether the trucks that drive the containers on the road within economies drive on the left or the right side of the road, whether they are allowed to go 80 KPH or 150 KPH, how often they must stop at night, or even what the containers can contain. The regulations within a given economy do not need to be the same as others, they just cannot prescribe, for example, a container size that is incompatible with roads, rails, shipping, and freight systems of other nations.

This example, while simpler than AI, illustrates what can be gained by advancing regulations and policies that rely on coordinated mechanisms like international standards and other soft law systems in ways that keep citizens safe, encourage building trusted technologies, and enable effective economic opportunity within national borders. Regulatory interoperability – or coherence – for AI does not mean a nation's laws need to be identical to those in other nations, but rather that laws are structured in ways that enable AI models, services, and compliance mechanisms to be portable and interoperable across jurisdictions. This coherence helps foster competition and innovation by allowing smaller firms to scale across multiple markets. Fragmentation disproportionately benefits large, well-resourced firms – large enterprises are far better positioned to build different AI systems for different legal environments. So which players are most proactive in helping establish coherence? Several governance frameworks and international initiatives aim to bridge regulatory differences and promote coherence in AI regulation through technical and policy standards. The following three subsections describe some of these key participants and mechanisms - intended to be representative, not exhaustive.

### (i) **International Coordination**

The OECD's (2019) *AI Principles* established a well-developed, coherent reference point for AI policy, specifically outlining broad norms for regulatory consideration and for further work by industry. The OECD has since built on the Principles to develop an *AI Policy Observatory*[18] and its *OECD Network of Experts on AI*[19] to build understanding of practices to support these principles, and explicitly to address "interoperability and policy coherence among leading risk management frameworks."[20] The OECD's *AI Principles* have played a significant role in developing regulation and policy in the EU, Japan, Canada, and Latin America, and have helped move regulatory interoperability discussions forward in the G7 and G20.

In 2023, the G7 established the *Hiroshima AI Process*,[21] aiming to discuss and develop "guiding principles and code of conduct aimed at promoting the safe, secure and trustworthy advanced AI systems" with interoperability among AI governance frameworks as a key objective. At the 2024 G7 meetings in Italy, G7 leaders committed to "step up our efforts to enhance interoperability amongst our AI governance approaches to promote greater certainty, transparency and accountability while recognizing that approaches and policy instruments may vary across G7 members," also referring to deepening cooperation among the AI Safety Institutes across the G7 and advancing international standards for AI (Kumar, 2024). At the G20 level, discussions led by India over the past year have demonstrated that even in a broader forum where consensus may be harder to achieve, broadening the network of transnational discussion about regulatory interoperability can advance recognition of fundamental principles that contribute to regulatory coherence (Kerry et al., 2025). In 2025, the G7 published their *Statement on AI for Prosperity*[22] which outlined their efforts to economically support SMEs to improve their competitiveness, AI adoption ability, and attract talent.

[18] https://oecd.ai/en/

[19] https://oecd.ai/en/network-of-experts

[20] https://oecd.ai/en/site/risk-accountability

[21] See here for *Hiroshima AI Process* documents: https://www.soumu.go.jp/hiroshimaaiprocess/en/documents.html

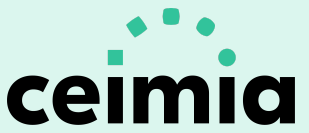

Building on the UK's 2023 *Bletchley Park AI Safety Summit*, several nations established dedicated AI Safety Institutes (Onikepe, 2024). While these institutes vary in structure, staffing, and regulatory authority, they are all state-sponsored and generally aim to address AI governance, safety, and economic issues, while being positioned to understand new AI risks. In 2024, many of these institutes formed the *International Network of AI Safety Institutes* as a forum for communication and collaboration amongst these institutes and to exchange information about testing, safety, and risk, and to promote mutual recognition of national level AI risk assessments across borders (International Network of AI Safety Institutes, 2024).

The UN has also positioned itself as an actor in shaping global AI governance. In 2023, the UN's Secretary-General António Guterres established the *High-Level Advisory Body on AI* (HLAB), which was explicitly tasked to offer diverse perspectives and options on "how AI can be governed for the common good, aligning internationally interoperable governance with human rights and the Sustainable Development Goals" (United Nations, 2023). The HLAB's (2024) final report emphasized the importance of interoperability across AI governance efforts by proposing several specific initiatives that directly aimed at promoting regulatory interoperability in AI governance, including broadening global cooperation to include nations not currently engaged on multilateral initiatives on AI governance, establishing a standing (twice yearly) global AI policy dialogue to enhance international interoperability of AI governance, creating an AI standards exchange, and developing a global data framework for data flow and data interoperability (both the panel and dialogue are now being created within the UN).[23]

The 2025 *Paris AI Action Summit* provided a recent political expression of interest in multilateral regulatory interoperability for AI. The final statement of the participants committed to "international cooperation to promote coordination in international governance," though lacked specific follow up mechanisms or measurable benchmarks. Nonetheless the convening was significant as the largest to date on the topic, with 62 signatories to the final statement (Élysée, 2025) - the United

[22] https://g7.canada.ca/assets/ea689367/Attachments/NewItems/pdf/g7-summit-statements/ai-en.pdf

[23] https://www.un.org/global-digital-compact/en/ai

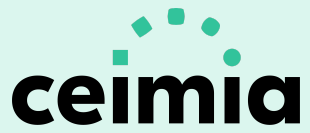

States and the United Kingdom did not join in signing. Notably, the summit marked a shift in emphasis: innovation and adoption of AI technologies took precedence over strong anchoring in AI safety as a regulatory foundation. Though that shift to innovation-aligned governance may signal differences in nations' regulatory emphasis (including less emphasis on multilateral work on "AI safety"), the corresponding emphasis on standards-based regulation underscores the possibility of advancing foundational regulatory and technical interoperability through the international standards system, driven by the emphasis on innovation and adoption. Additional signals reinforcing this trend included the UK's decision to rename its "AI Safety Institute" to the targeted "AI Security Institute," and the European Union's scaling back of its previously ambitious *AI Liability Directive*. Moreover, the absence of key AI ecosystem actors - the US and UK - from the summit's signatories further underscores ongoing challenges to reaching comprehensive international alignment. Addressing these divergences will be critical if symbolic commitments are to transition into meaningful regulatory coherence.

**(ii) <u>National and Regional Efforts</u>**

The European Commission advanced its implementation and enforcement capabilities for the EU's *AI Act* by relying on Europe's "new approach" to standards as a tool to harmonize technical and legal requirements for AI systems (European Trade Union Institute, 2011). Through a European Commission standardization request, CEN and CENELEC - the key European standardization organizations for this work - were tasked with developing European harmonized standards that align with the EU *AI Act's* requirements (European Commission, 2024). AI systems that meet the established European standards will be presumed to meet the EU AI Act's thresholds – aligning compliance across the 27 member states. The potential for collaboration and cross-adoption of standards between CEN/CENELEC and other international standard-setting bodies like ISO/IEC, IEEE, and ITU, offers the possibility that specific aspects of compliance and standards conformance may be consistent across the EU and interoperable with other jurisdictions beyond the EU, like the US or China. However, at the time of writing this report, the current geopolitical climate is contributing to less collaboration between CEN/CENELEC and ISO/IEC, leading to less interoperability potential, especially with the current direction of the standardization requests by CEN/CENELEC.

The US has largely aimed at regulatory governance at industry verticals and agency-led efforts specific to a given area. More recently, federal efforts have sought to minimize fragmentation across state laws but continue to rely on existing laws (such as anti-discrimination, the Federal Trade Commission's rules on privacy and consumer protection, and product liability), as well as overall guidance frameworks, such as NIST's (2023) *AI Risk Management Framework*. While the new US administration is still in its early stages, there is some expectation that the US will continue and lean further into its historical and cultural alignment with a light touch, risk-based regulatory approach, aimed at fostering the US AI industry, using several forms of leverage to move toward increasingly open digital trade (with low regulatory requirements) across its primary trading partners, such as California's SB 53.[24] In his speech at the 2025 *Paris AI Action Summit*, Vice President JD Vance emphasized a pro-innovation, deregulatory approach to AI governance (which is mirrored in the US's *AI Action Plan*). He stated, "[W]e believe that excessive regulation of the AI sector could kill a transformative industry just as it's taking off, and we'll make every effort to encourage pro-growth AI policies." Overall, the return of the Trump administration's assertive, transactional approach to diplomacy, trade, defense, and tariffs may make cooperative multilateral interoperability among nations more challenging, at least initially. The administration's preference for bilateral negotiations over multilateral engagement, combined with increased tariffs and stringent export controls as instruments of geopolitical leverage, may push rival nations, like China, into deeper technological isolation, accelerating parallel and competing AI ecosystems.

China took an early step toward AI competitiveness in its (2017) *New Generation AI Development Plan*, which aimed for Chinese AI leadership by 2030. Though the approach has evolved over the last eight years - particularly in their (2025) *Action Plan for Global Governance of AI*[25] - it is still characterized by proactive government oversight, a top-down regulatory strategy, and the aim of balancing global AI ambitions with concerns about national social stability, economy, security, and values. Crucial to China's domestic regulation is a complex regime of national Chinese standards for AI that form specific requirements within the nation. Essentially, while China has focused its regulatory capabilities on sectoral and subject matter regulation, its domestic standards are mandatory and therefore create a broad horizontal layer of compliance requirements and eventual capacity building for enforcement. In parallel, a key part of China's effort to play a leadership role in the international AI arena is its extensive engagement with the international AI standards work at ISO/IEC and elsewhere.

The heightened security lens through which AI is viewed across world capitals could lead to an even sharper dual-use regulatory divide, prioritizing national defense capabilities over global integration or standards harmonization. At the same time, these pressures may, over time, create incentives, greater clarity, and urgency around interoperability, especially among allied democracies and middle powers concerned about fragmentation's negative economic and security implications. Countries like Canada, Japan, Australia, EU Member States, and actors from the Global South could become more motivated to foster alternative or complementary multilateral frameworks – building on work at the OECD, Global Partnership on AI (GPAI), or emergent AI-specific trade coalition – to counterbalance US-China tensions and protect the role these nations play in global innovation ecosystems. These nations also play a key role in the international standards ecosystem. Their motivation to ensure stable trade flows, data transfer relationships, and integrated supply chains provides a strong impetus to bridge divergent regulatory systems.

[24] https://www.gov.ca.gov/2025/09/29/governor-newsom-signs-sb-53-advancing-californias-world-leading-artificial-intelligence-industry/
[25] https://www.gov.cn/yaowen/liebiao/202507/content_7033929.htm

Countries adept at fostering mutual recognition and use of international frameworks could gain disproportionate influence over global AI governance and innovation ecosystems, effectively turning regulatory alignment into a strategic form of soft power. Pragmatic economic and security incentives may ultimately drive a deeper, targeted international consensus around both technical and regulatory interoperability.

#### (iii) Networked and Distributed Governance

The various examples above identify some of the aims and incentives across disparate stakeholders interested in advancing regulatory and technical interoperability. But these examples also provide us with a window into the complexity of the coordination problem. National interests may diverge. Coalition partners may come and go or may evolve in their views in ways that are incompatible with other partners. Initiatives are many, but resources to staff are limited. No central body exists with either capacity or authority to coordinate these efforts. And yet this complex, multidimensional environment may well be better suited to the difficulties of interoperable AI governance than alternatives that rely on hierarchy or single point of control. Ostrom's (1990) foundational studies of water and fisheries management demonstrated real world governance operating through complex "polycentric" interactions that include governmental institutions, private actors, civil society and intergovernmental organizations. These polycentric interactions are needed to address collective action problems, particularly of the commons. AI governance may well track domains from finance to climate to food safety and forest sustainability, where no single central authority, or even a group of governments, controls policy, regulatory standards, and regulation (Faveri & Auld, 2024). Indeed, a transatlantic team writing for Brookings recently urged that this "iterative and networked approach to AI governance will be key" (Kerry et al., 2025). With AI technologies' rapid pace of change, their application, and regulatory responses, its complex, networked, and distributed governance will likely remain the key coordination mechanism that can address AI's governance challenges.

## 3. Need for AI Technical and Regulatory Interoperability

### (a) Interoperability - State of Play

AI technical interoperability today has solid grounding in existing structures in the IT ecosystem – open datasets, standardized APIs, widely adopted protocols, and an emerging form of semantic interoperability accelerated by generative AI's evolving capabilities. Enterprises have a workable but uneven starting point as they benefit from growing adoption of containerization, cloud-native architectures, and robust APIs, yet still grapple significantly with legacy infrastructure, inconsistent vendor standards, and practical difficulties around integrating diverse AI systems. At the consumer and small-business level, interoperability gaps are even more pronounced. Data and application portability remain constrained by technological barriers - like inconsistent privacy implementations and demand for robust consent frameworks - and economic incentives that discourage platforms from offering easy portability. Addressing these gaps would unlock significant potential: seamless platform transitions for users, modular assembly of AI applications by developers, and streamlined regulatory oversight. However, achieving these scenarios requires concerted efforts beyond technological solutions, including thoughtful regulation, collaborative standard-setting, incentives alignment, and realistic governance approaches to overcome economic resistance and legacy inertia.

From a user standpoint, regulatory and technical interoperability fundamentally shapes individual agency, freedom from lock-in, and the ability to seamlessly exercise data rights across jurisdictions. From a regulatory and implementation perspective, users – whether consumers, businesses, or governments – will reasonably expect AI systems to deliver consistent experiences and outputs regardless of geographical location or the underlying platform. From a technical interoperability perspective, data portability and application portability are key tools, empowering users to migrate their data and services across interoperable systems, thus avoiding vendor lock-in and enabling meaningful choice. Open APIs and other connectivity methods also foster innovation by enabling connections between otherwise disparate platforms, ensuring that users continuously benefit from competition and improvements without sacrificing convenience or experiencing service degradation as they move from one jurisdiction to another.

From a regulatory perspective, interoperability at the user level must address consistency in rights and protections. Users increasingly interact with AI systems that make significant decisions affecting their lives, from financial transactions to healthcare outcomes. Interoperability standards should therefore ensure that users can clearly understand, contest, and seek redress for adverse AI-driven outcomes, regardless of jurisdictional boundaries. Different societal values may dictate regulations that create different outcomes between jurisdictions, but those differences should be transparent and visible to users moving between different nations so individuals are able to understand and appropriately address AI decisions that may affect them across jurisdictions – and can gain redress when harms occur.

### (b) Diverging and Conflicting Governance Regimes

AI governance has emerged as a fragmented and contentious global landscape, with Canada, the EU, UK, US, and other jurisdictions adopting starkly different regulatory regimes. These divergences reflect varying priorities, legal traditions, and risk appetites, creating a complex patchwork of compliance requirements for multinational organizations. The EU has positioned itself as a global standard-setter through its (2024) *AI Act*, which introduces a binding, risk-based classification system that bans certain AI applications outright and imposes stringent obligations on high-risk systems, like biometric identification and critical infrastructure (CEIMIA, 2024). This horizontal framework contrasts sharply with the UK's sector-specific approach, which avoids new legislation in favour of empowering existing regulators to apply five cross-cutting principles – safety, transparency, fairness, accountability, and redress – tailored to industries like healthcare and finance (Casovan et al., 2023). While the EU has centralized oversight, the UK relies on decentralized enforcement by bodies like the *Information Commissioner's Office*, prioritizing innovation over prescriptive rules (Information Commissioner's Office, 2025).

Canada's *AIDA*, while terminated, blended elements of both models but faced criticism for delayed implementation and overreliance on future regulations. For example, the EU's *AI Act* categorizes AI systems based on risk while *AIDA* categorized AI systems by impact levels, with a focus being placed on high-impact systems (AI & Data Act, 2025; IAPP, 2025; Zajko, 2023). Under *AIDA*, those AI systems that were assessed to be high-risk AI could be audited by (Section 15) and made available to (Section 17) to the Minister. This model contrasts with the US, where federal inertia has led to a chaotic mix of state-level laws and non-binding federal guidelines. States like California and Illinois have enacted sector-specific AI laws targeting hiring algorithms and biometric data, while the Biden administration's (2022) *Blueprint for an AI Bill of Rights*, remained aspirational. China's hybrid model further complicates the global landscape, combining state-driven ethical guidelines for general AI research with hard laws targeting specific technologies like generative AI. Unlike Western frameworks emphasizing individual rights, China's regulations focus on aligning AI development with national security and their government's values, requiring algorithmic transparency to curb "disorderly expansion" of private firms (CEIMIA, 2024). This state-centric approach diverges sharply from the EU's human rights focus and the US's market-driven ethos. Meanwhile, emerging economies like Brazil struggle to balance corporate lobbying with public interest, as seen in debates over Bill 2338/23, where Big Tech firms successfully diluted provisions on AI accountability and information integrity (Zanatta & Rielli, 2024).

These conflicting regimes create friction for global AI deployment. A healthcare algorithm compliant with the EU's strict data governance rules may violate US state laws permitting broader biometric data collection, while China's mandatory security reviews for AI exports clash with Canada's emphasis on ethical innovation. The lack of interoperability is particularly apparent in risk assessment methodologies. The EU mandates third-party conformity assessments for high-risk systems, whereas the UK and US favor self-audits and post-market monitoring. Even voluntary standards diverge, with the EU promoting regional standards and frameworks for AI systems and the US relying on the NIST's (2023) *AI Risk Management Framework*. Such fragmentation risks reducing cross-border collaboration and innovation. For instance, Canadian AI firms face regulatory limbo as *AIDA's* delayed implementation[26] contrasts with the EU's enforceable rules, while UK startups must navigate overlapping guidance from multiple sectoral regulators. These disparities underscore the urgent need for regulatory and technical interoperability, particularly as AI products are quickly going to market and transcending national boundaries. Without coordinated efforts, this diverging and conflicting regulatory and technical regime will exacerbate compliance costs, hinder accountability, and deepen geopolitical divides in AI governance.

### (c) National and International Trade Benefits

AI interoperability, both technical and regulatory, can be a fundamental enabler of digital trade, removing friction that would otherwise hinder cross-border, state, or provincial commerce. When nations develop interoperable AI norms and governance structures, businesses avoid costly customization for each market, reducing compliance burdens and increasing efficiency. The *United States-Mexico-Canada Agreement* (USMCA) sets an important precedent in its *Digital Trade Chapter* by requiring its members to base domestic regulations on international standards, ensuring AI-driven businesses can operate with fewer barriers to entry (Office of the United States Trade Representative, 2020). Regulatory fragmentation, on the other hand, disrupts this progress. As AI systems become integral to industries from finance to healthcare and beyond, diverging standards risk creating economic silos where valuable AI services (or their underlying models) are unavailable. This fragmentation isolates economies and favours large players who can bear the costs of separate development for each nation. As history has shown, fragmentation favours the powerful: incumbent firms, with ample resources and entrenched market positions, can readily navigate the complexity, costs, and inefficiencies of technical and regulatory incompatibility. For example, the global fragmentation between *Global System for Mobile Communications* and *Code Division Multiple Access* cellular standards during the 1990s and early 2000s allowed large, resource-rich players like Qualcomm, Nokia, and Verizon to thrive, while effectively blocking smaller, innovative handset manufacturers and startups from competing, as they lacked the resources needed to support multiple incompatible technologies. Thus, fragmentation does not merely complicate markets - it can entrench existing power structures, suppress innovation, and limit consumer choice. Without a commitment to interoperability, both technical and regulatory, society risks entrenching unchallenged, dominant structures rather than fostering innovation.

---

[26] Granted, there are other relevant laws in place that these Canadian AI firms must comply with, such as privacy and consumer protection laws, there is an uncertainty within the market around AI-specific laws.

Interoperability can ease regulatory burdens and be a mechanism for ensuring fair market access. For example, the Asia-Pacific Economic Cooperation (2023) found that effective digital trade rules "increase the flows of digitally ordered and digitally deliverable trade by between 11% and 44% in successive years." One additional but less noticeable benefit of regulatory interoperability is its ability to build on the strongest technologies available anywhere in the world. When economies align their AI regulations, they create an ecosystem in which companies, governments, and researchers can adopt best-in-class AI innovations, regardless of where they originate.

Beyond cost savings, AI interoperability fosters trust in AI-driven products and services across markets. Consumers and businesses rely on AI systems for decision-making in critical sectors like healthcare, law, and finance. Yet, without regulatory interoperability, the safety, fairness, and privacy standards of these systems can vary dramatically from one jurisdiction to another. Establishing a shared baseline of AI governance principles ensures that users in different markets can trust AI applications regardless of where they were developed. Trade agreements increasingly recognize this need. Building on long standing commitments like the World Trade Organization's (1995) *Agreement on Technical Barriers to Trade*, USMCA and similar frameworks; emerging commitments should emphasize prohibiting forced data localization, facilitating cross-border data flows, and promoting mutual recognition of privacy protections, all of which can help ensure that AI adoption is unhindered by artificial regulatory barriers. When businesses can trade under global norms for digital trust, consumers gain confidence in AI innovations across markets, fostering more widespread adoption and economic growth. Regional frameworks like the Asia-Pacific Economic Cooperation[27] (APEC) and Association of Southeast Asian Nations[28] (ASEAN) have also recognized the importance of digital trade coherence. APEC has urged economies to pursue "regulatory interoperability" to prevent market fragmentation, while ASEAN's ongoing development of a *Digital Economy Framework Agreement* aims to dismantle digital trade barriers by harmonizing national AI policies (Tech for Good Institute, 2024). These frameworks provide a roadmap for governments seeking to align AI regulations in a way that promotes both trade efficiency and consumer trust.

## 4. Current Challenges to AI Interoperability

The challenges to regulatory and technical interoperability are related: (a) a lack of substantive and jurisdictional uniformity; (b) obsolescence; (c) limited focus; and (d) specific technical interoperability challenges.

### (a) Lack of Substantive and Jurisdictional Uniformity

The biggest challenge for regulatory interoperability is a clash of jurisdictions. The EU *AI Act* and its extra-jurisdictional reach can be problematic for non-EU-based companies that face increased compliance costs. Within the US, the initial jurisdictional clashes are between the States, and between specific States and the federal government. Industry representatives are lobbying the US Congress to pre-empt State AI laws. However, in the absence of an actual federal or national standard that is inconsistent with one or more State treatments, pre-emption is not an option. So, in the near term, the jurisdictional clashes will occur between different States.

The State of Colorado's *Senate Bill 24-205* (Colorado's AI Act), originally passed in 2024 and now delayed until June 2026, serves as an important example as it quickly attracted attention from several actors, including industry representatives seeking to weaken or even repeal it.[29] Most of this *Statute's* provisions are focused on the outcomes of "consequential decisions" by AI systems. These decisions have a "material legal or similarly significant effect" on a consumer or worker's major economic and life opportunities – specifically, employment, education, finances, health care, housing, insurance, legal services, or essential government services. This Colorado law requires that developers and deployers provide transparency – basic proactive disclosures about the AI systems – while deployers must provide an annual impact assessment that analyzes whether an AI decision system creates a risk of algorithmic discrimination. A "duty of care" also applies to those who operate these systems to take reasonable steps to prevent algorithmic discrimination. Importantly, the law does not hinge on a company's status as being based in Colorado; rather, it applies to consumers and workers who live in Colorado.

---

[27] https://www.apec.org/

[28] https://asean.org/

[29] https://www.clarkhill.com/news-events/news/colorados-ai-law-delayed-until-june-2026-what-the-latest-setback-means-for-businesses/

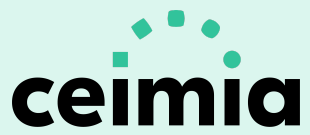

### (b) Obsolescence: Technology is Moving Rapidly

AI is not a new technology, it is rather a set of ever-evolving technologies based on algorithms, data, and compute power. From "traditional" AI, such as algorithms in search and social media platforms, we are now moving to generative/LLM AI platforms. Now the focus is on "agential AI," with software-based agents undertaking tasks on behalf of an individual or entity. Future versions of AI could move away from LLMs to cognitive or inferential AI. In each instance, new forms of AI present new types of challenges. For example, in a world of AI agents, how can one discern that a particular bot or assistant has the requisite authorization to act on someone's behalf as there are little efforts in-place to do so.

### (c) Limited Focus of Regulatory Measures

The current and proposed laws addressing AI focus on one particular aspect, the accountability of the underlying developers and deployers for the systems they create and release. The duty of care, as established by the Colorado AI law, makes this focus even more tangible, as these firms are charged with taking reasonable steps to avoid harm for certain decisions they impose on individuals. Under the EU *AI Act*, the crucial determiner of regulation is how "risky" a particular system can be deemed. This focus on consequential decisions and risky behaviors by the deployers and developers is missing another key aspect - empowering end users. Rather than treating individuals as passive recipients of the AI-based actions imposed by others, policymakers can also recognize and incentivize ways for individuals to utilize AI to promote their own best interests. The focus on the "existential" risks of AI systems may be a tactic of the AI platforms to get policymakers to focus on the company's behaviours outside the "guardrails," rather than the policy implications of what goes on inside the guardrails.

As a potential approach, policymakers could shift their attention to granting individuals concrete rights vis-a-vis AI systems. These concrete rights could include the right to: (1) interoperability of their AI agents with other AI agents and platforms; (2) portability of their personal data to other AI agents and platforms; (3) query and contest consequential decisions rendered by such systems; (4) recourse for adverse impacts from such consequential decisions; and (5) delegate their rights to a trusted third party entity. Policymakers also could assign a general, fiduciary law-based duty of care where AI systems make consequential decisions or utilize personal data. Finally, as a stronger measure, no decision by an AI system on behalf of an individual could be considered legally binding unless the decision was made via an entity with a fiduciary law-like duty of loyalty towards that individual.

### (d) Specific Technical Interoperability Challenges

A specific challenge to technical interoperability is that, to date, it has failed to receive priority as a governance issue. No hard laws or regulations exist, nor do incentivizing measures for adoption of technical interoperability. The specific form is also unresolved. As we discuss below, uniform technical standards would appear to be superior to unilateral and inconsistent corporate APIs, even if the latter would be necessary in the shorter term. Both horizontal and vertical versions of technical interoperability would open the commercial developer and deployment area "between the guardrails." More competition, innovation, and user choice and empowerment should follow as a result.

## 5. Potential Pathways Toward AI Regulatory and Technical Interoperability

### (a) Encourage Public Sector Incorporation of Global AI Policy, Standards, and Frameworks

Governments could use globally recognized AI principles and standards as the foundation for shaping domestic AI policy. In developing or updating AI regulations, they should draw upon, reference, or directly incorporate international standards. Utilizing established standards as compliance mechanisms will enhance regulatory interoperability, minimize fragmentation, and align domestic frameworks with global norms. This pathway fosters safer, fairer, and more trustworthy AI systems while promoting innovation and facilitating cross-border scalability. Policymakers could enhance global AI interoperability by adopting a maturity-based framework to guide the selection of appropriate standards (or other soft law mechanisms) and assure connection to the formal standardization system. This approach would include using

‘softer’ international standards products such as ‘technical reports’ to document the state of the art for nascent systems and guide future formal standardization efforts. Mature AI systems, in respect of which there is a common understanding and maturing techniques for assessment, could follow international standards, and, where there is public policy importance, incorporate conformity assessments and certifications.

### (b) Technical Interoperability through Open Software Standards

Today there is no meaningful way for one AI system to connect directly with another. Instead, most such systems are being developed within corporate silos that preclude interoperability, whether between agents (horizontal) or between agents and platforms (vertical). This situation precludes end user choice, robust marketplace competition, and edge-based innovation. While one potential option would be the widespread utilization of APIs promulgated by large corporate-owned AI platforms, these interfaces tend to remain closely controlled and restrictive. An ideal path forward is through the development of open software standards by recognized expert bodies, such as IEEE, ISO/IEC, IETF, or ITU, and organizations like the W3C, Linux Foundation, and Open Specification Foundation. Granted, ISO/IEC and ITU are not set up or intended to develop open software standards. The challenge for these national body/government-based standard-setting bodies is to recognize the value and contributions of open-source software projects and write standards that can serve as a basis for open-source applications.

While open AI-to-AI interoperability standards may well develop in the marketplace, free from any government involvement, such an outcome seems unlikely. The incumbent players can be assumed to seek to protect their new AI silos from encroachment by others. In this situation, governmental bodies can step in to create meaningful incentives for open AI interop. The classic governmental intervention in the market is the “power of the pen” – laws and regulations mandating certain behaviors. While potentially effective, these power of the pend measures can be difficult to design, implement, and enforce. Instead, less restrictive implements can be considered, such as the “power of the purse” (procurement, tax benefits, grants) or the “power of the pulpit” (akin to preaching by influential figures to favour particular actions or behaviours).

### (c) Frameworks, Sandboxes, and Measurement/Mapping Tools and Roadmaps

Creating an AI interoperability framework that offers a more structured understanding of AI system layers and identifies critical points for interoperability and regulatory oversight would provide a baseline for public and private actors to use when making interoperability efforts. Like NIST's (2010) *Framework and Roadmap for Smart Grid Interoperability Standards* or (2011) *Cloud Computing Reference Architecture* and ISO/IEC's (2023) *22123-3 Information technology — Cloud computing Part 3: Reference architecture*, this framework would clarify technical and governance needs at each layer of AI systems. This framework would define layers (e.g., data, model, inference, deployment) and map interoperability and oversight requirements and could be piloted in specific use cases to validate its use. Similarly, developing regional governance frameworks as scalable models for global alignment would allow economically mid-sized countries to collaborate on shared priorities, serving as testbeds for broader initiatives.

To test interoperability frameworks, and other possible solutions and governance mechanisms in controlled environments, regional or multilateral sandboxes would provide a safe regulatory or technical environment to carry these tests out. The US, UK, UAE, and Singapore have similar sandbox efforts. For example, Senator Ted Cruz unveiled a new bill to loosen the rules for AI innovators - the first step in a new policy framework he also rolled out on September 10, 2025. According to Cruz's office, the bill is part of the wider Republican effort to push President Donald Trump's *AI Action Plan* forward. The *Strengthening Artificial Intelligence Normalization and Diffusion By Oversight and eXperimentation*[30] (the *SANDBOX Act*) was introduced at a Senate Commerce subcommittee hearing featuring Michael Kratsios, the architect of the *AI Action Plan*. Kratsios' office would be responsible for setting up the sandbox for AI developers and deciding who gets what waivers. Setting up these sandboxes could be done by establishing cross-border partnerships with GPAI, OECD, and national governments. Similarly, partnering with regional bodies, like the African Union or ASEAN, could allow for a focus on sectors where cross-border AI use is prevalent, like finance, transport, healthcare or defense.

Alongside these framework and sandbox efforts, there is a need to develop tools and methodologies that enable systematic measurement/mapping between technical frameworks and regulatory frameworks. With diverse AI governance frameworks and sandboxes emerging globally, tools for cross-referencing and aligning frameworks are essential to identify overlaps, gaps, and opportunities for harmonization. This approach would also help with the creation of unified, or at least coherent, controls frameworks, defining shared control points. To implement this approach, there must be design mapping methodologies and tools to illustrate relationships between frameworks, leveraging existing methodologies from NIST

[30] https://www.commerce.senate.gov/2025/9/sen-cruz-unveils-ai-policy-framework-to-strengthen-american-ai-leadership

(2023), OECD (2019), etc. Additionally, testing to refine these tools into crosswalk exercises between major frameworks, like the EU *AI Act*, recent US AI Executive Orders, OECD's *AI Principles*, and G7's (2023a) *Hiroshima Principles* would be needed (like NIST's crosswalk exercise[31] between its RMF to the EU *AI Act* (when it was proposed), OECD *Recommendation on AI*, and the US' *AI Bill of Rights* and *EO 13960*).

Using these measurements and mapping efforts' outcomes, sector-specific interoperability roadmaps should be created to provide a structured, phased, or element-based approach to implementing interoperability. A roadmap ensures alignment, prioritization, and scalability of interoperability efforts, guiding sectors toward shared goals. For example, the rapid advancements in AI agents would demand standards work on interoperable "zero-trust" models for authentication, authorization, and account tracking. Perhaps standardized tracking of what these AI agents have done would provide an accountability model. When identifying these priority areas, one must also identify relevant standards in that area, following the previous example, standards around what data travels along the AI agent's journeys (and with whom it can be shared) and what is not shared (like privacy and security).

### (d) Build the Economic and Trade Case for Interoperability

There are significant economic and trade benefits to achieving technical and regulatory interoperability for nations and industry. Incentivizing development and adoption of key interoperability standards and interoperable regulatory frameworks as described above demonstrate how interoperability can enhance economic growth, facilitate trade, and strengthen national defense. Motivating policymakers and stakeholders to prioritize these efforts is critical and will face some resistance. Moving toward this path would require economic impact studies on the trade benefits of interoperability and developing case studies showing how aligned AI governance strengthens various sectors, like defense and security cooperation (e.g., shared AI systems for cybersecurity or defense alliances; APEC, 2023). Similarly, addressing interoperability challenges within specific industries can demonstrate practical progress. Certain sectors, such as finance, transportation, healthcare, and defense, are more ready for regulatory alignment due to their existing norms and shared priorities, such as NATO's ongoing interoperability efforts.[32] This sector-specific approach can be worked towards by collaborating with industry leaders and sectoral organizations to develop tailored agreements and use those agreements to align technical and ethical standards for cross-border AI use. Absent concerted efforts to advance interoperability, fragmentation emerges naturally as governments, industries, and organizations often default toward short-term, localized optimizations that promise immediate competitive or security advantages.

[31] https://www.nist.gov/system/files/documents/2023/01/26/crosswalk_AI_RMF_1_0_OECD_EO_AIA_BoR.pdf

[32] https://nhqc3s.hq.nato.int/apps/architecture/nisp/volume1/index.html

Interoperability, on the other hand, requires intentional and sustained investment to establish effective frameworks and implement and maintain them as technology evolves. Yet this investment typically aligns with stakeholders' overall long-term self-interest. Interoperability delivers strategic value through expanded economic opportunities, broader market access, enhanced innovation driven by competition, and strengthened security from more predictable, trusted interactions. Given the accelerating global reliance on interconnected systems, understanding these benefits is crucial to overcoming short-term biases and motivating stakeholders toward cooperative interoperability.

# Part 2: Lessons Learned from Other Industries' Successful and Failed Regulatory and Technical Interoperability Efforts

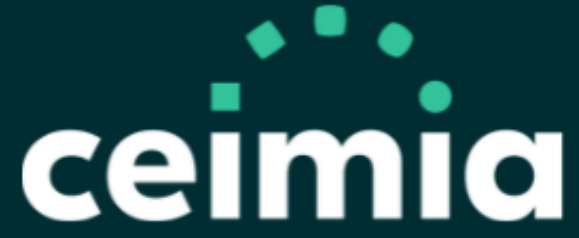

# Part 2: Lessons Learned from Other Industries' Successful and Failed Regulatory and Technical Interoperability Efforts

## 1. Introduction

Building off Part 1's AI governance landscape and need for and pathways toward AI technical and regulatory interoperability, Part 2 explores how regulatory and technical interoperability has developed in four established sectors. As AI rapidly evolves and proliferates across sectors and jurisdictions, the challenge of achieving effective technical and regulatory interoperability has become increasingly urgent. The current AI governance landscape is characterized by fragmentation and convolution, with proliferating standards, principles, and regulations creating compliance burdens and potential lock-in effects as different initiatives compete for market dominance. Rather than adding to this already complex ecosystem, there is considerable value in examining how other established sectors have navigated similar interoperability challenges, extracting actionable insights that can inform more effective approaches to AI technical and regulatory interoperability efforts.

The examination of regulatory and technical interoperability efforts across four distinct domains – nanotechnology through the *NanoDefine* project; environmental sustainability via the EU's *INSPIRE Directive*; telecommunications spanning 19th-century telegraphs to post-Snowden architectures; and internet/web architecture development – reveals patterns that transcend individual technological contexts. These case studies collectively demonstrate that interoperability is not merely a technical endpoint but an ongoing negotiation between competing interests, involving complex dynamics of technological evolution, regulatory frameworks, and institutional coordination. The diversity of these cases in temporal scope, sectoral focus, and interoperability outcomes provides a comprehensive foundation for understanding both successful strategies and critical failure modes.

The lessons emerging from these historical experiences prove particularly relevant to AI's challenges. Unlike established technologies that achieved interoperability after lengthy maturation periods, AI systems are being deployed at scale while fundamental questions about definitions, measurement standards, safety protocols, and governance frameworks remain unresolved. The case studies reveal that once incompatible systems become entrenched, achieving interoperability requires exponentially more effort and resources than proactive standardization undertaken during a technology's nascent stages. This temporal reality creates urgency around establishing common frameworks for AI governance before divergent approaches become too institutionalized to reconcile effectively.

Four critical lessons emerge from these case studies. First, the need for adaptive governance frameworks that can evolve alongside rapid technological advancement, avoiding the rigid specifications that hindered long-term effectiveness in cases like the *INSPIRE Directive*. Second, the importance of establishing early definition and measurement standards during technology development lifecycles, as demonstrated by both successful early interventions in telecommunications and the costly retrofitting required in nanomaterials governance. Third, the necessity of building trust through robust verification mechanisms rather than relying solely on cooperative goodwill, a lesson emphasized by telecommunications' evolution from trust-based systems to zero-trust architectures following security crises. Fourth, the value of bridging technical and regulatory divides through structured collaboration that aligns diverse stakeholder interests while maintaining implementable standards. These lessons collectively suggest that successful AI interoperability can be achieved by moving beyond the current proliferation of disconnected regulatory efforts toward coordinated approaches that balance necessary standardization with the flexibility required for rapidly evolving technologies. The window for implementing such coordinated frameworks may be rapidly closing as AI technologies mature and regulatory approaches become institutionalized, making the insights from these historical cases both timely and essential for avoiding the costly fragmentation patterns that have characterized other technological domains.

Part 2 of this Report develops in three stages. First, we present the methodology which describes the rationale for case studies, how the cases were selected, against which criteria each case was examined, and the limitations of the methodology. Second, the four cases are provided - nanotechnology, environmental sustainability, telecommunications, and internet/web architecture - with each divided into four case criteria: history around the need for regulatory or technical interoperability; regulatory, industry, and geopolitical climate; intervention points used for regulatory or technical interoperability efforts; and how those interventions played out. Third, and lastly, four lessons are extracted across these case studies to inform emerging AI regulatory and technical interoperability efforts.

## 2. Methodology

This report adopts a comparative case study approach to analyze historical and contemporary regulatory and technical interoperability efforts across four sectors: nanotechnology (*NanoDefine*), environmental sustainability (EU's *INSPIRE*

*Directive*), telecommunications (telegraphs to post-Snowden), and internet/web architecture (HTTP/IP). Case studies were chosen to identify patterns, challenges, and success factors across these interoperability efforts; lessons to be drawn from these efforts, and their implications emerging AI interoperability efforts. This methodology section is divided into four parts: (a) case study methodology selection and rationale; (b) case selection justification; (c) case study criteria and analytical framework; and (d) limitations.

### (a) Case Study Methodology Selection and Rationale

Case studies were chosen as the methodological approach given their ability to provide granular insights into complex systems – technical and regulatory systems in this report – where controlled experiments are impractical. This approach allows for systematic examinations of *how* and *why* specific interoperability efforts succeeded or failed across diverse contexts. By focusing on historical and sectoral examples, this analysis avoids over-reliance on speculative AI governance frameworks and instead grounds recommendations in empirically observable outcomes. The case study method also aligns with this report's goal of identifying transferable lessons for AI interoperability. As interoperability challenges often involve path dependencies, institutional inertia, and geopolitical tensions, case studies allow for a nuanced exploration of these dynamics. Case studies also accommodate the heterogeneous nature of interoperability efforts-ranging from 19th-century telegraph standardization to 21st-century data sovereignty conflicts-while maintaining a structured comparative framework.

### (b) Case Selection Justification

Four cases were selected to represent critical junctures in interoperability governance. First, *Nanotechnology: Nanodefine* analyzes early-stage standardization efforts in a nascent field, highlighting challenges in anticipatory governance, the risks of premature regulatory fragmentation, and a focus on safety standards. This case provides insights into managing interoperability for technologies still in foundational development phases. Second, *Environmental Sustainability: The EU INSPIRE Directive* examines a mature regulatory interoperability framework for spatial data infrastructure. The *INSPIRE Directive's* mixed outcomes but persistent coordination gaps illustrate the tensions between top-down harmonization and bottom-up implementation in multi-jurisdictional regulatory and technical systems. Third, *Telecommunications: Telegraphs to Post-Snowden* traces a 150-year arc from the (1865) International Telegraph Union's creation to post-Snowden zero-trust architectures. This longitudinal case reveals how security crises reshape interoperability paradigms, demonstrating the interactions between technical standards, geopolitical trust, and economic incentives. And fourth, *Internet/Web Architecture*: *HTTP/IP* assesses the unplanned evolution of internet interoperability from ARPANET to modern splinternets. The HTTP/IP case highlights the consequences of divorcing technical interoperability from governance frameworks, particularly around content moderation and digital sovereignty. These cases were selected for their diversity in temporal scope, sectoral focus, and interoperability outcomes. Collectively, they address the three pillars of interoperability identified in the AI governance literature: technical compatibility, regulatory alignment, and institutional coordination while also providing other lessons for emerging AI interoperability efforts.

### (c) Case Study Criteria and Analytical Framework

Each case is evaluated against four criteria that are derived from Part 1 of this report: (i) History around the Need for Regulatory or Technical Interoperability; (ii) Regulatory, Industry, and Geopolitical Climate; (iii) Intervention Points Used for Regulatory or Technical Interoperability Efforts; and (iv) How those Interventions Played Out. These criteria enable systematic comparison across the chosen case studies while accommodating context-specific variations within them. These criteria are an effort to balance the *processes* and *outputs* within each case's interoperability efforts, recognizing that interoperability efforts evolve with changing technologies, regulations, and power dynamics, among other factors. By combining historical depth with structured comparison, this methodology allows for analyzing AI interoperability challenges while remaining cognizant of its contextual boundaries. The cases collectively highlight that interoperability is not a technical endpoint but an ongoing negotiation between competing interests.

### (d) Limitations

Three key limitations are found within this methodology: selection bias, temporal compression; and reproducibility. The focus on European and North American cases risks overlooking interoperability models from Asia, Africa, and South America. For instance, China's AI standardization initiatives and ASEAN's digital economy framework agreement could yield contrasting lessons about state-led interoperability. Longitudinal cases like telecommunications condense complex histories into narrative arcs, potentially oversimplifying possible causal relationships. For example, the 1914 cable-cutting analysis, while illustrative, cannot fully capture the multi-decadal erosion of trust in pre-WWI diplomatic networks. Case study findings are inherently sensitive to interpretation. The internet architecture case's emphasis on technical design choices (e.g., IP's "hourglass model") might lead different researchers to prioritize economic factors (e.g., US government funding) or cultural drivers (e.g., hacker ethos). Additionally, the nanotechnology case uses nanotechnology as a proxy for AI's early-stage challenges, but this analogy may not fully capture generative AI's unique regulatory environment. These limitations are partially mitigated through triangulation with primary sources – including treaty texts, technical standards, and policy evaluations – and explicit acknowledgment of the report's Western and Eurocentric focus.

## 3. Case Studies

### (a) Case 1: Nanotechnology: Standardizing Nanomaterial through NanoDefine

#### (i) History around the Need for Regulatory or Technical Interoperability

The nanotechnology sector represents a convergence of physics, chemistry, biology, and engineering, focused on manipulating matter on the atomic and molecular scale (1–100 nanometers (nm)) to create materials and devices with novel properties (Bayda et al., 2019). This sector has catalyzed breakthroughs across industries, like targeted drug delivery systems in medicine to lightweight high-strength composites in aerospace (Malik et al., 2023). Nanomaterials – particles, tubes, or structures within the nanoscale – are the foundational building blocks of this sector, enabling innovations such as quantum dots for imaging, nanoencapsulation for controlled release of agrochemicals, and nanofilters for environmental remediation (Elsaid et al., 2023; Gil et al., 2021; Kumar & Raj, 2025). A standardized definition of what a nanomaterial is, is critical for navigating this sector's regulatory and technical challenges. Without consensus on what constitutes a nanomaterial, inconsistencies arise in safety assessments, product labeling, and international trade compliance (Allan et al., 2021; Sylvester et al., 2009). For example, the European Commission's (2011) *Recommendation on the Definition of Nanomaterial* established a threshold of ≥50% particles by number in the 1–100 nm range, yet disparities in measurement techniques led to conflicting classifications of identical materials – leading to technical and regulatory interoperability challenges within the sector. The (2019) *NanoDefine* project emerged as a response to these interoperability challenges.

The demand for regulatory and technical interoperability in nanomaterial classification – especially during the 2010s in the EU – emerged from the absence of harmonized methods to determine whether materials qualified as nanomaterials under existing regulatory frameworks. In 2011, the European Commission published its *Recommendation on the Definition of Nanomaterial*, establishing that a nanomaterial consists of natural, incidental, or manufactured material containing particles where 50% or more have one or more external dimensions in the size range of 1-100 nm. However, this definition created an immediate regulatory implementation problem, as there were no standardized and validated methods available to reliably measure and classify materials according to this recommended criterion. The regulatory landscape was becoming increasingly complex, with nanomaterials falling under various EU regulations including the (2006) *Registration, Evaluation, Authorisation and Restriction of Chemicals*, (2012) *Biocidal Products Regulation*, and (2017) *Medical Devices Regulation*. Each regulatory framework required accurate identification of nanomaterials, but manufacturers, enforcement laboratories, and regulatory authorities lacked the technical tools and standardized procedures necessary to make consistent determinations about what constituted a nanomaterial, creating considerable challenges for industry compliance, regulatory enforcement, and market harmonization across EU Member States.

The technical interoperability challenge was multifaceted, involving the development of measurement methods and standardization of sample preparation, instrument calibration, data analysis procedures, and decision-making frameworks (Allan et al., 2021). Different analytical techniques produced varying results, and there was no consensus on which methods were most appropriate for specific types of materials or regulatory contexts (van Rijn et al., 2022). The European Commission recognized that addressing this challenge required a coordinated and evidence-based approach that would bring together expertise from researchers, industry, and regulatory bodies (European Commission, 2022). The complexity of nanomaterial characterization, combined with the need for regulatory acceptance and international harmonization, meant that no single organization or country could develop adequate solutions independently. This realization led to the conceptualization of a large-scale collaborative project – *NanoDefine* – that would systematically address the technical and regulatory interoperability challenges inherent in implementing the EU nanomaterial definition (Brüngel et al., 2019).

#### (ii) **Regulatory, Industry, and Geopolitical Climate**

The *NanoDefine* project was conceived and implemented from 2011 to 2018 and was characterized by rapidly evolving regulatory frameworks for nanotechnology and increasing international cooperation on nanomaterial safety and standardization (Gaillard et al., 2019). The regulatory climate was marked by growing precautionary approaches to nanomaterial governance, often driven by public concerns about potential health and environmental risks, combined with recognition of the significant economic potential of nanotechnology applications (Gottardo et al., 2021). The EU was positioning itself as a global leader in responsible nanotechnology development, emphasizing the integration of safety considerations throughout the innovation process. During this period, the ISO's (2005) *Technical Committee 229 on Nanotechnologies*[33] formed and was working to establish international standards for terminology, measurement methods, and safety practices. Similarly, the OECD's (2006) *Working Party on Manufactured Nanomaterials*[34] was actively developing test guidelines and guidance documents for nanomaterial safety assessment (Rasmussen et al., 2016). However, these international efforts were still in their early stages, and there was limited coordination between different standardization bodies and regulatory frameworks.

The geopolitical landscape and industry climate were fraught with competition between major economic blocs to establish leadership in nanotechnology regulation and standardization (Allan et al., 2021). Industry was experiencing rapid growth in nanotechnology applications across sectors from electronics and materials to cosmetics and food packaging, but also by significant uncertainty around regulatory requirements. Companies were investing heavily in nanotechnology research and development while simultaneously struggling to navigate evolving regulatory landscapes. There was particular concern among European manufacturers about potential competitive disadvantages if EU regulations were more stringent or difficult to comply with than those in other regions. Industry stakeholders were calling for clearer guidance, standardized methods, and harmonized approaches that would enable efficient compliance while supporting innovation. Alongside industry, the scientific and technical communities were also grappling with fundamental challenges in nanomaterial characterization and risk assessment. Measurement techniques were rapidly evolving, but with limited understanding of their performance characteristics, applicability to different material types, and suitability for regulatory purposes. Research institutions across Europe were developing new analytical methods and approaches, but these efforts were largely fragmented and lacked coordination. The need for validated standardized methods that could support regulatory decision-making was becoming increasingly urgent as the number of nanomaterial-containing products entering the market continued to grow (Allan et al., 2021; Brüngel et al., 2019; Gottardo et al., 2021).

### (iii) Intervention Points Used for Regulatory or Technical Interoperability Efforts

The *NanoDefine* project, funded under the EU's *7th Framework Programme*[35] with 29 partners from 11 European countries, implemented a comprehensive multi-tiered intervention strategy designed to address technical and regulatory interoperability challenges (European Commission, 2022). This project sought to develop an approach that combined method validation, performance evaluation, and decision support frameworks for consistent implementation of the European Commission's (2011) *Recommendation on the Definition of Nanomaterial* across different regulatory contexts (Mech et al., 2020).

[33] https://www.iso.org/committee/381983.html
[34] https://www.oecd.org/en/topics/chemical-safety-and-biosafety.html
[35] https://cordis.europa.eu/project/id/604347/reporting

There were four intervention points used throughout *NanoDefine* to address the underlying interoperability challenge. The first involved establishing a comprehensive knowledge base through systematic evaluation of existing measurement techniques and development of new analytical approaches (European Commission, 2017; Mech et al., 2020). This knowledge base evaluated 40+ different characterization methods, including electron microscopy, dynamic light scattering, single particle inductively coupled plasma mass spectrometry, and various separation techniques, assessing their performance against standardized quality criteria (Mech et al., 2020). This evaluation process involved extensive inter-laboratory validation studies conducted under the auspices of the *Versailles Project on Advanced Materials and Standards*,[36] ensuring that method performance data would be internationally recognized and accepted. The second was the development of Standard Operating Procedures (SOPs) and harmonized validation protocols that could be implemented consistently across different laboratories and regulatory contexts (European Commission, 2017; Mech et al., 2020). Within the *NanoDefine* project, 23 detailed SOPs were created covering sample preparation, measurement protocols, data analysis procedures, and quality control measures. These SOPs were designed to be adaptable to different analytical techniques while maintaining consistency in critical performance parameters and decision criteria. The third intervention involved creating a tiered decision support framework to guide users through the complex process of selecting appropriate measurement methods and interpreting results according to the European Commission's (2011) *Recommendation on the Definition of Nanomaterial*. This framework, implemented as the *NanoDefiner e-tool*, incorporated expert knowledge about method performance, material properties, and regulatory requirements into a user-friendly software platform that could provide method recommendations and automated material classification capabilities (Brüngel et al., 2019). This e-tool was designed to be expandable and adaptable, allowing for incorporation of new methods and updates to regulatory definitions as they became available. Lastly, *NanoDefine* implemented targeted outreach and training to ensure widespread adoption and proper implementation of the developed tools and methods. This outreach and training included collaboration with standards and international bodies, like *CEN/TC 352 on Nanotechnologies*[37] and contributions to the OECD's test guideline development processes,[38] ensuring that project outcomes would be integrated into formal standardization and regulatory frameworks.

[36] https://www.vamas.org/

**(iv) How those Interventions Played Out**

The interventions implemented through the *NanoDefine* project achieved significant success in addressing technical and regulatory interoperability challenges, with outcomes that continue to influence nanomaterial governance frameworks globally. The integrated approach developed by this project provided a practical solution to the complex challenge of implementing the EU nanomaterial definition, resulting in widely adopted tools and methods that have been incorporated into regulatory guidance and industry practice. The systematic evaluation of measurement techniques and development of SOPs successfully established a robust technical foundation for nanomaterial classification. The validation studies demonstrated that multiple analytical techniques could provide reliable and consistent results when applied according to standardized protocols, addressing previous concerns about method reliability and reproducibility. The 23 SOPs developed by the project have been adopted by laboratories across Europe and beyond, providing a harmonized technical framework that supports consistent implementation of nanomaterial identification procedures. The *NanoDefiner e-tool* proved to be particularly successful, providing an accessible platform that democratized access to expert knowledge about nanomaterial characterization. This tool has been downloaded and used by hundreds of organizations worldwide, including regulatory authorities, industry laboratories, and research institutions (Mech et al., 2020). Independent case studies have confirmed the consistency and reliability of this tool's recommendations, validating the underlying decision support framework (Brüngel et al., 2019). The e-tool's success demonstrated that complex technical knowledge could be effectively packaged and disseminated through user-friendly digital platforms.

Perhaps most significantly, *NanoDefine's* outcomes have been explicitly recognized and incorporated into updated EU regulatory frameworks. When the European Commission published its (2022) revised *Recommendation for the Definition of Nanomaterial*, it specifically referenced the *NanoDefine* project and its methodological contributions. The framework and tools developed by the project were successfully adapted to accommodate the updated definition, demonstrating their flexibility and sustainability. This regulatory recognition validates the project's approach and ensures the continued relevance of its outputs. The international impact of the *NanoDefine* has been substantial, with its methodological approaches and decision support frameworks influencing nanomaterial governance initiatives worldwide. The project's emphasis on harmonized validation protocols and standardized decision criteria has been adopted by international organizations including the OECD and ISO, contributing to global coherence in nanomaterial characterization approaches (Allan et al., 2021). The success of the *NanoDefine* project has also served as a model for subsequent European initiatives addressing regulatory science challenges in nanotechnology, demonstrating the value of large-scale collaborative approaches to technical and regulatory interoperability challenges.

[37] https://standards.cencenelec.eu/dyn/www/f?p=205:7:0::::FSP_ORG_ID:508478&cs=105D77E18D80442539DAD7D6A6B7EC5FA

[38] https://www.oecd.org/en/topics/sub-issues/testing-of-chemicals/test-guidelines.html

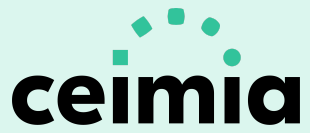

### (b) Case 2: Environmental Sustainability - The EU's INSPIRE Directive

#### (i) History around the Need for Regulatory or Technical Interoperability

The EU's (2007) *INSPIRE Directive* was enacted as European environmental policy makers grappled with the limitations of incompatible spatial data systems across EU-Member States. Cross-border challenges, like managing the Danube River Basin – which spans 10 EU-Member States with disparate water quality monitoring formats – exposed critical gaps in data interoperability (Massimo & Alessandro, 2007; Vanderhaegen & Muro, 2005). Air quality assessments faced similar hurdles, with cross-border smog modeling requiring laborious manual reconciliation of datasets using conflicting coordination systems and metadata standards (Cho & Crompvoets, 2018). These operational inefficiencies coincided with growing pressures to meet the EU's environmental targets under frameworks like the (2000) *Water Framework Directive*, which demanded harmonized data for transnational reporting.

The (1985) *CORINE Land Cover Project* laid early groundwork by standardizing land use mapping across Europe. However, its 1990 findings revealed that nearly 30% of environmental datasets remained unusable for cross-border analysis due to technical discrepancies (Land Monitoring Service, 2021). National mapping agencies operated in silos. For example, Germany's topographic databases employed different classification schemes than France's, while newer EU members lacked digitized spatial data entirely (Keil, 2017). This classification scheme fragmentation undermined critical initiatives, such as assessing biodiversity loss under the (1992) *Habitats Directive* or tracking deforestation impacts under the (1997) *Kyoto Protocol*. The (2001) *Gothenburg Summit* marked a turning point, explicitly linking data interoperability to climate change mitigation strategies. European Environment Agency reports highlighted that 40% of environmental policymaking resources were wasted reconciling incompatible datasets (Cho & Crompvoets, 2018). Meanwhile, the 2004 EU enlargement intensified these challenges, integrating 10 member states with varying technical capacities—Lithuania's partially paper-based systems contrasted sharply with the UK's advanced Ordnance Survey digitization efforts (Cetl et al., 2017).

Negotiations for a unified framework began in 2001 between environmental agencies seeking strict harmonization efforts to enable transnational analyses and national institutions who resisted harmonization efforts based on perceived threats to data sovereignty. The (2007) *INSPIRE Directive* emerged as a compromise, targeting 34 environmental themes while allowing phased implementation. High-priority datasets, like hydrography and protected ecosystems, received earlier deadlines, acknowledging the complexity of overhauling legacy systems (INSPIRE Knowledge Base, 2025). This approach recognized the diversity of Europe's technical landscapes - from Denmark's centralized geospatial infrastructure to Spain's regionally devolved models (Cetl et al., 2017). This Directive's development also responded to evolving technological demands. Early 2000s advancements in GIS and remote sensing outpaced existing data-sharing frameworks, leaving initiatives like the (2008) *Marine Strategy Framework Directive* reliant on patchwork solutions. *INSPIRE* aimed to future-proof interoperability by mandating machine-readable formats like *Geography Markup Language*, while the *CORINE Land Rover Project's* success in standardizing land cover classifications informed its metadata protocols (Copernicus, 2019). By addressing both technical and governance gaps, the *INSPIRE Directive* shifted Europe's environmental data infrastructure from fragmented to a cohesive analytical resource.

### (ii) Regulatory, Industry, and Geopolitical Climate

The *INSPIRE Directive's* implementation (2007–2013) unfolded against a backdrop of fragmented regulatory frameworks, competing industry priorities, and geopolitical tensions stemming from the EU's rapid expansion. Regulatory disparities were stark. While this *Directive* mandated harmonized spatial data standards, Member States retained distinct legal frameworks for environmental governance. For instance, Poland's land use classification systems conflicted with INSPIRE's thematic requirements, necessitating costly adjustments to align cadastral data with EU specifications (Dawidowicz et al., 2020; Ogryzek et al., 2020). Only Belgium transposed the *Directive* by the 2009 deadline, while 26 Member States faced infringement proceedings for non-compliance, reflecting institutional resistance to centralized EU mandates European Commission, 2010. This regulatory friction was compounded by the 2008 financial crisis, which diverted resources from technical upgrades in Southern European Member States like Greece and Spain, where austerity measures prioritized economic stability over data infrastructure modernization (Van, 2011).

Industry stakeholders exhibited divergent responses. Private geospatial firms in the Netherlands championed *INSPIRE's* open standards as a market opportunity, whereas French and Italian mapping agencies fiercely protected proprietary systems, fearing loss of competitive advantage (Cho & Crompvoets, 2018; Gabele, 2008; Ogryzek et al., 2020). National mapping bodies, such as Germany's *Länder*-controlled agencies, struggled to reconcile decentralized data governance with INSPIRE's interoperability requirements, leading to fragmented metadata catalogs (Keil, 2017). Meanwhile, emerging technologies like cloud computing and IoT sensors exposed gaps in the directive's original technical specifications, which were designed for static datasets rather than real-time analytics (Woolf, 2010). This misalignment sparked debates about the feasibility of enforcing rigid standards in a rapidly evolving tech landscape.

Geopolitically, the EU's enlargement to 27 Member States amplified implementation challenges. Newer entrants, like Croatia, lacked the institutional capacity of older members like Denmark, which leveraged its centralized *Basic Data Program* to integrate *INSPIRE* with broader e-government reforms (European Commission, 2025a). Federalized states faced unique hurdles: Spain's autonomous communities developed conflicting interpretations of Annex III[39] themes like soil erosion risk, while Germany's regional governments resisted ceding control over environmental datasets (Keil, 2017). This *Directive* also intersected with global climate commitments, as *INSPIRE's* infrastructure became critical for tracking progress under the (1997) *Kyoto Protocol* and (1998) *Aarhus Convention*. However, geopolitical priorities shifted post-2008, with many member states adopting minimalist compliance strategies to conserve resources – a trend the European Commission later attributed to domestic fiscal pressures (European Commission, 2020). These dynamics created a paradox. While *INSPIRE* aimed to unify environmental data governance, its implementation exposed latent tensions between EU integration and national sovereignty. This *Directive's* reliance on non-binding technical guidelines allowed flexibility but bred inconsistency, as seen in Sweden's full adoption vs. Greece's parallel systems for domestic and EU data (Kjellson, 2017; Tsiavos, 2010). This period underscored the complex interplay between regulatory ambition, technological change, and geopolitical realities in shaping transnational regulatory and technical interoperability efforts.

[39] https://eur-lex.europa.eu/legal-content/EN/TXT/PDF/?uri=CELEX:32007L0002

### (iii) Intervention Points for Regulatory and Technical Interoperability Efforts

The *INSPIRE Directive's* implementation relied on a multifaceted strategy combining legal mandates, technical standardization, and adaptive governance structures to bridge Europe's environmental data divides. At the regulatory level, this *Directive* compelled Member States to transpose its requirements into national law by May 2009, establishing legally binding obligations to inventory, harmonize, and publish spatial data. This top-down approach ensured political accountability, with non-compliant nations facing infringement proceedings. For instance, 26 Member States initially missed the transposition deadline, prompting the European Commission to launch legal actions that accelerated adoption. However, this *Directive* permitted flexibility in implementation methods, allowing countries like Germany to integrate *INSPIRE's* requirements into existing federal frameworks, while newer EU members, like Croatia, relied on centralized systems funded by EU grants (European Commission, 2025b).

Technically, this *Directive* introduced rigorous specifications to standardize data formats, metadata, and service interfaces. Central to this effort was the phased rollout and adoption of the Geography Markup Language (GML) for data exchange, which replaced incompatible national formats like France's EDIGÉO[40] and Germany's ATKIS.[41] The phased rollout prioritized high-impact environmental themes: Annex I datasets (e.g., hydrography, protected sites) faced a 2013 compliance deadline, while Annex II (e.g., elevation, land cover) and Annex III (e.g., soil erosion) deadlines extended to 2020.[42] This staggered timeline acknowledged the complexity of retrofitting legacy systems, particularly in nations with paper-based archives. To address technical capacity gaps, the EU funded transnational projects such as the *Baltic Marine Environment Protection Commission* (HELCOM),[43] which pooled resources from nine countries to harmonize marine spatial data using INSPIRE specifications.

A critical intervention was the establishment of the *INSPIRE Maintenance and Implementation Group* (MIG),[44] which acted as a collaborative hub for translating regulatory requirements into technical action. The MIG developed non-binding guidelines, such as the 2012 Technical Guidance on Coordinate Reference Systems,[45] which resolved conflicts between Italy's regional geological surveys. Yet, the voluntary nature of these guidelines created variability. Sweden adopted full harmonization, migrating its entire *National Land Survey* to INSPIRE-compliant services, while Greece maintained dual systems – one for EU compliance and another for domestic needs – due to technical debt in its cadastral databases (Rannestig & Nilsson, 2008). Funding mechanisms played a pivotal role, with the EU allocating over €300 million through programs like ISA² (Interoperability Solutions for European Public Administrations).[46] These funds supported technical overhauls, such as Spain's €48 million investment to convert cadastral data to GML (Femenia-Ribera et al., 2022). However, the 2008 financial crisis diverted resources in Southern Europe, delaying implementation in Greece and Portugal by nearly five years. Private-sector engagement was limited but influential: Dutch geospatial firms developed open-source Extract, Transform, Load tools like *hale»studio*,[47] reducing data conversion costs for smaller nations. The *INSPIRE Directive's* most innovative intervention was its hybrid governance model, which balanced centralized standards with localized adaptations. For example, the UK aligned *INSPIRE's* requirements with its Ordnance Survey's "OS OpenData" initiative, leveraging existing infrastructure to minimize costs (Carpenter & Watts, 2013). Conversely, Germany's decentralized approach required coordinating 16 federal states, each with distinct environmental data policies, through a national geoportal. This flexibility allowed *INSPIRE* to accommodate diverse administrative systems but also led to fragmented outcomes, as seen in inconsistent metadata quality across borders. Ultimately, these interventions underscored the tension between regulatory enforcement and technical pragmatism. While *INSPIRE* succeeded in establishing a baseline for interoperability, its reliance on non-binding technical guidelines and variable funding created an uneven landscape.

---

[40] https://docs.safe.com/fme/html/FME-Form-Documentation/FME-ReadersWriters/edigeo/edigeo.htm
[41] https://www.adv-online.de/Products/Geotopography/ATKIS/
[42] https://eur-lex.europa.eu/legal-content/EN/TXT/PDF/?uri=CELEX:32007L0002
[43] https://helcom.fi/helcom-revamps-its-metadata-catalogue-improving-the-search-of-over-1000-baltic-sea-maps/
[44] https://wikis.ec.europa.eu/spaces/InspireMIG/overview

#### (iv) How those Interventions Played Out

The *INSPIRE Directive's* implementation yielded mixed outcomes, revealing both the transformative potential and inherent limitations of large-scale regulatory and technical interoperability frameworks. By 2023, over 95% of mandated spatial datasets were accessible through the *EU INSPIRE Geoportal*,[48] enabling cross-border analyses previously deemed impractical. For instance, harmonized river basin data facilitated transnational flood risk modeling in the Rhine region, improving emergency response coordination between Germany, France, and the Netherlands. Denmark's integration of *INSPIRE* with its national digitalization strategy reduced environmental reporting costs by 40%, while Estonia leveraged interoperable wildfire risk datasets to automate real-time alerts during the 2018 peatland fires. These successes underscored interoperability's value in enhancing environmental governance efficiency.

[45] https://inspire-mif.github.io/technical-guidelines/data/rs/dataspecification_rs.pdf
[46] https://ec.europa.eu/isa2/home_en/
[47] https://wetransform.to/halestudio/
[48] https://inspire-geoportal.ec.europa.eu/srv/eng/catalog.search#/home

However, operational challenges persisted. The European Commission's (2022) INSPIRE evaluation found that 35% of hydrography datasets remained outdated, and coordinate reference system mismatches disrupted 20% of cross-border analyses. Technical debt accumulated as Member States grappled with legacy system overhauls. For example, Spain spent €48 million converting cadastral data to *INSPIRE's* GML only to encounter compatibility issues with modern cloud platforms. Smaller nations like Latvia struggled with capacity gaps, relying on EU-funded projects to maintain basic metadata catalogs. Regulatory compliance also proved uneven. While Sweden achieved full dataset harmonization, Greece and Italy maintained parallel systems for domestic and INSPIRE-compliant data, doubling maintenance costs. *INSPIRE's* most significant lesson lies in the tension between regulatory precision and technical adaptability. *INSPIRE's* rigid 2007-era specifications, designed for static datasets, clashed with emerging technologies like real-time sensor networks and AI-driven analytics. For example, the Netherlands developed costly hybrid APIs to bridge INSPIRE's batch-processing architecture with dynamic air quality monitoring systems. Similarly, this *Directive's* structured GML schemas proved incompatible with machine learning models requiring unstructured inputs, limiting its utility in predictive analytics. The European Parliament's (2016) *Regulatory Fitness and Performance* evaluation highlighted these mismatches, noting that "overly prescriptive metadata requirements" discouraged private-sector participation. Subsequent reforms aimed to reduce compliance burdens but faced criticism for lagging behind technological advancements.

Geopolitical and governance disparities further shaped outcomes. Member States with centralized data infrastructures, like Denmark, achieved faster implementation by aligning *INSPIRE* with broader e-government reforms. In contrast, federalized states like Germany faced protracted negotiations between federal and regional authorities, delaying dataset harmonization by up to seven years. The EU's 2008 financial crisis exacerbated these divides, as austerity measures diverted funds from technical upgrades in Southern Europe. These disparities underscored the importance of tailoring interoperability frameworks to diverse administrative contexts – a lesson now informing the EU's 2025 *Data Spaces Initiative*, which adopts a more flexible "interoperability-by-design" approach (Data Spaces Support Centre, 2025). Crucially, *INSPIRE* demonstrated that regulatory mandates alone cannot ensure interoperability. While binding legislation secured minimum compliance, the lack of enforceable technical standards led to fragmented data quality. This *Directive's* reliance on non-binding guidelines allowed nations to prioritize politically expedient datasets over high-impact ones, resulting in gaps in critical areas like soil erosion monitoring. Conversely, initiatives combining regulatory pressure with capacity-building incentives – like the HELCOM project – achieved more sustainable outcomes by pooling technical resources across borders. The legacy of *INSPIRE* lies in its proof-of-concept for transnational data governance. It advanced Europe's position as a global leader in spatial data infrastructure while exposing the need for adaptive regulatory and technical interoperability frameworks. Current reforms to these frameworks emphasize machine-to-machine interoperability and reduced human mediation, addressing past fragmented adoption. However, this *Directive's* central lesson is clear: successful interoperability requires continuous dialogue between policymakers, technologists, and end-users, ensuring systems evolve alongside environmental and technological realities.

### (c) Case 3: Telecommunications - Telegraphs to Snowden

#### (i) History around the Need for Regulatory or Technical Interoperability

The mid-19th-century expansion of international rail networks significantly increased the economic necessity of rapid cross-border communication. Telegraph systems became indispensable for coordinating train schedules, managing logistics, and enhancing safety, thereby creating powerful economic incentives to standardize telegraph interoperability (Science Museum, 2018). The related surge in intra-European trade and financial markets amplified demands for reliable cross-border communication. Telegraph interoperability became critical for timely market information, reducing uncertainty, and enhancing economic coordination among Europe's major financial hubs (ITU, 2024). During the spring of 1865, delegates from twenty nations gathered in Paris to sort out how to efficiently transmit telegraph messages across national borders - to which the (1865) *International Telegraph Convention* was agreed upon and signed, forming the International Telegraph Union (ITU), the precursor to today's International Telecommunication Union (ITU, 2025). Until then, individual nations independently adopted incompatible telegraph systems. Each telegraph system differed in coding methods and how signals were electrically transmitted and received. For example, the French Breguet system used needle telegraphs with distinctive signaling patterns while Germany's Siemens system utilized different electrical standards and coding conventions. Such differences required tedious, error-prone manual transcription, and re-transmission at national borders (Marsh, 2018; Siemens, 2025). This fragmented approach resulted in delays, errors, and significant costs that undermined developing potential for cross-border trade and diplomatic communication. Banks, rail lines, trade companies, and diplomats needed reliable cross-border communications to ensure accurate and timely financial transactions, contract negotiations, transportation, and diplomatic exchanges. Delays or errors directly translated into financial losses and diplomatic tensions, amplifying economic and political pressures for immediate solutions (Standage, 2014).

Around 50 years later, by 1914's WWI outbreak, telegraph networks had evolved into critical infrastructure supporting national security and global diplomacy. Initially designed for peaceful cooperation, commerce, and diplomacy, these evolved interconnected networks rapidly became strategic assets and vulnerabilities during wartime. The geopolitical climate shifted abruptly as communication channels once symbolizing international collaboration became targets of military strategy. Although interoperability initially promoted international cooperation and economic efficiency, connectivity itself became a critical vulnerability as alliances fractured (Boghardt, 2012). The strategic vulnerability of interoperable communications was starkly illustrated by the Zimmermann Telegram incident in 1917, where British intelligence intercepted a secret German diplomatic communication routed through British-controlled cables. This incident revealed how dependent nations had become on infrastructure they did not control, making technical interoperability a double-edged sword (Boghardt, 2012).

Nearly a century after WWI, Edward Snowden's 2013 disclosures dramatically highlighted vulnerabilities inside the global network, threatening interoperability by undercutting the fundamental trust that interoperability has been built on. Snowden, a US National Security Agency contractor, exposed extensive surveillance programs exploiting digital communications infrastructure's inherent openness and interconnectedness (Lyon, 2014). His revelations severely disrupted international trust, exposing profound vulnerabilities in the internet's technical and regulatory interoperability efforts. This case revealed how the very features that enabled global digital connectivity - like open protocols, shared standards, and integrated networks - could be leveraged for surveillance efforts at unprecedented scales. This revelation created a fundamental tension between the technical requirements for interoperability, and the security needs of nations and individuals, echoing earlier historical patterns, but with greater stakes in a digitally dependent world.

### (ii) **Regulatory, Industry, and Geopolitical Climate**

The mid-19th century telegraph moment represented a period of expanding international trade and diplomatic exchange, creating strong economic incentives for cross-border communication standardization. Despite these shared economic interests, nations remained protective of their sovereignty and wary of binding international agreements that might limit their autonomy. The challenge was to create a framework that balanced these competing priorities. During the 1865 Paris conference, delegates were mindful of preserving national sovereignty while acknowledging the substantial mutual benefits of standardized communication (ITU, 2025). This diplomatic balancing act - voluntary rather than compulsory standardization - was pivotal in ensuring widespread acceptance of the conference outcomes and creation of the International Telegraph Union. Economic imperatives for efficient cross-border communication proved sufficient to motivate cooperation despite sovereignty concerns.

The pre-WW1 regulatory environment had prioritized expanding connectivity without adequately addressing security vulnerabilities in increasingly interoperable systems. When geopolitical tensions escalated into open conflict, the existing governance frameworks - designed primarily for peacetime cooperation - proved inadequate for protecting strategic communication assets. The WWI outbreak dramatically transformed the telecommunications landscape from one of expanding peaceful commerce to one characterized by military necessity and broken trust. International frameworks that had facilitated cooperation suddenly became vectors of vulnerability. On August 5, 1914, Britain severed Germany's submarine telegraph cables, weaponizing communication infrastructure as part of its war strategy (Ashdown, 2021). Trust in the network evaporated overnight, and nations rapidly sought ways to isolate and protect their communications infrastructure. In WWI's aftermath, nations recognized the need to rebuild cooperative frameworks while integrating security considerations. During the 1920s, the ITU expanded its mandate to include telephone and radio services, acknowledging the growing complexity of international communications (ITU, 2024).

Snowden's revelations elicited widespread global condemnation and shock at the extent of US surveillance, prompting major reactions from traditional adversaries and close allies. European nations accelerated regulatory efforts like the *General Data Protection Regulation* (GDPR), while China intensified its push towards sovereign digital infrastructure and standards (Fazlioglu, 2023; Harrell, 2025; Rossi, 2018; Thumfart, 2025). The geopolitical landscape had evolved significantly from earlier telecommunications moments, with digital networks now forming the backbone of global economic, political, and social systems. Unlike earlier moments, this crisis emerged from peacetime surveillance activities rather than traditional warfare, raising new questions about the appropriate balance between national security, privacy, and international cooperation. The post-Snowden era witnessed increased tension between globalized technical interoperability and growing demands for digital sovereignty. Nations sought to maintain the benefits of connected systems while protecting their citizens and interests from surveillance and security vulnerabilities. This tension continues to shape the regulatory and geopolitical climate for telecommunications interoperability today.

### (iii) **Intervention Points Used for Regulatory or Technical Interoperability Efforts**

Responding to economic pressures for more efficient cross-border communications, delegates reached the (1865) *International Telegraph Convention* agreement and ITU establishment, introducing voluntary international standards for telegraph interoperability. The ITU soon developed guidance standards enabling seamless message transmission across national boundaries, including specific technical requirements for on premises equipment (no longer the latency and guesswork of the dial and pointer systems) and messaging protocols (like Morse Code), and regulatory requirements on cross-border fees. This intervention represented a diplomatic innovation by creating voluntary standards that nations could adopt without threatening sovereignty while establishing ITU's ongoing institutional framework to manage and update these standards as technology developed. The voluntary nature of adoption was crucial to building broad participation.

Britain's 1914 severance of Germany's submarine telegraph cables represented a negative type of intervention, the strategic disruption of interoperability for military advantage. This negative intervention highlighted the security vulnerabilities in internationally connected systems. During WWI, the intervention's focus shifted to rebuilding interoperability frameworks with increased attention to security concerns. The ITU expanded its mandate beyond telegraphs to telephone and radio communications, developing more comprehensive standards that acknowledged the strategic importance of these technologies. However, these efforts still primarily emphasized technical compatibility rather than security by design, leaving vulnerabilities that would be exploited in later conflicts.

In response to the trust crisis triggered by Snowden's revelations, governments and industry accelerated their focus on cybersecurity, particularly zero-trust architectures - systems that continuously verify identities and assume every network component is potentially compromised. Unlike previous models that implicitly trusted internal network components once authenticated, zero-trust architectures fundamentally changed assumptions by continuously verifying identities in each part of the network, encrypting communications end-to-end, and treating all network interactions as potentially compromised. This transformative shift re-established interoperability models on a foundation of explicit verification rather than assumed trust (Rogers & Eden, 2017; Rose et al., 2020).[49] Establishing mutually agreed upon frameworks, like NIST's (2020) *Zero Trust Architecture*, requires consensus on protocols, identity management, and data handling practices. This technical intervention was complemented by regulatory responses like the EU's GDPR, which established new legal frameworks for data protection and privacy across borders. This dual approach - technical standards for security coupled with regulatory frameworks for compliance - represents a more mature intervention strategy than previous moments, reflecting lessons learned from earlier crises in telecommunications interoperability.

#### (iv) <u>How those Interventions Played Out</u>

The ITU's impact on standardization efforts was swift and substantial. Within a decade, the ITU's standardized protocols enabled direct telegraph connections linking Europe's major economic centers, significantly reducing delays and costs, and directly fueling growth in international trade and diplomatic communications (ITU, 2020; Standage, 2014; Wenzlhuemer, 2007). The ITU's success stemmed from a convergence of mutual economic interests, cooperative diplomatic negotiations, and the non-threatening nature of voluntary standards that respected national sovereignty. Each of these convergences contributed to the necessary trust environment's development for successful technical interoperability. This early experience illustrated the potential of cooperative technical interoperability frameworks to deliver substantial economic and diplomatic benefits - and to set agendas for regulatory responses that respect sovereignty while offering enough coherence for shared communication and benefit.

The 1917 Zimmermann Telegram incident dramatically demonstrated the strategic vulnerabilities created by interoperable communications networks. British intelligence's interception of the German diplomatic message routed through British-controlled cables directly facilitated US entry into WWI by shifting US public opinion against neutrality and exemplifying how dependent nations had become on infrastructure they did not control. These developments demonstrated the potential for interoperable telecommunications networks to become a security liability without robust governance and trust frameworks. Technical connectivity, previously viewed as beneficial, became perilous when geopolitical trust dissolved. Nations subsequently recognized the imperative for secure, independent communication infrastructures capable of operating effectively even during international conflict. The interwar rebuilding of telecommunications interoperability achieved partial success, establishing frameworks for telephone and radio standardization through the ITU's expanded mandate. However, these efforts could not prevent the communication networks from again becoming strategic targets during WWII, highlighting the persistent challenge of balancing connectivity with security in an unstable geopolitical environment.

[49] US/China relations have also been an important part of this shift and the concept has fueled ideas about IT stack elements since no country can make the whole stack any longer.

Zero-trust architectures operate on the principle of "never trust, always verify," emphasizing continuous authentication and strict access controls (Rose et al., 2020). However, implementing zero-trust on a global scale requires a foundational level of trust and collaboration among nations and organizations to develop and adhere to shared frameworks. Economic incentives and broad industry collaboration have facilitated the shift needed to build out the zero-trust framework. This collaborative effort ensures that while individual entities may not inherently trust each other's networks, they can rely on mutually agreed-upon standards to enable secure and efficient communication. The post-Snowden regulatory responses, particularly the EU's *GDPR*, have significantly impacted global data practices, establishing new norms for privacy protection and data sovereignty. These frameworks have created compliance requirements that shape how organizations implement interoperable systems, adding a regulatory dimension to what was previously a primarily technical standards' domain.

### (d) Case 4: Internet/Web Architecture - HTTP/IP

#### (i) History around the Need for Regulatory or Technical Interoperability

The Internet and the World Wide Web (Web) present compelling case studies in the virtues and pitfalls of regulatory and technical interoperability. In both cases, the initial focus on a single uniform set of technical protocols – TCP/IP for the Internet, and HTTP for the Web – led to an array of multilateral and then multistakeholder organizations and processes for ensuring widespread adoption of those protocols. Over time, however, the growing fraying of regulatory interoperability around the world has led to a technical "splinternet" of networks (Whitt, 2013: 2018: 2024).

Before the Internet and Web, US military researchers were aware of a technical challenge: that disparate computing networks could not connect and share resources. In 1968, the US Department of Defense's Advanced Research Project Agency awarded contracting company BBN the first government contract to connect and develop ARPANET, as a way for different computers within the same network to communicate (Whitt, 2013). That same year, Vint Cerf and Bob Kahn took on the bigger task of developing a set of protocols that enabled the interoperability of computers located on entirely different networks. In 1974, their famous paper announced the TCP/IP suite, in which TCP was touted as the means of sharing resources that exist in different data networks (Cerf & Kahn, 1974). IP was later separated out to logically distinguish router addressing - an IP header with routing instructions - from host packet sending (TCP organizing data packets; Whitt, 2013). What happened next was a highly successful model of adoption led by the US Government. In 1980, the US Department of Defense began implementing TCP/IP as a recognized standard, and then formally incorporated it into ARPANET in 1983. Separately, the National Science Foundation recognized the utility of TCP/IP as a means of connecting academic institutions and campuses, and so in 1985 brought the protocols to the research network NSFNET (Whitt, 2013).

Alongside the unilateral US governmental adoption came several multistakeholder governing bodies, each of which carried out specific tasks related to improving and promulgating IP as a standardized protocol. Early bodies included the Internet Configuration Control Board (ICCB), International Control Board, Internet Activities Board, and Internet Engineering Task Force (IETF; Whitt, 2013). Each of these organizations was set up and run by engineers and other stakeholders involved in promulgating the standards and protocols for the nascent Internet. The IETF, launched in 1986, created a multistakeholder process for further developing the Internet's governing protocols. By 1989, when commercial activity was first authorized onto the combined platform of "the Internet," and government control and backing began to dissipate, IP was the official lingua franca of this "network of networks." The Internet Society arrived in 1992, along with the ICCB's replacement, the Internet Architecture Board, to establish various processes and educational campaigns in support of the Internet's adoption.

In addition to the regulatory interoperability environment of a supportive government and numerous multistakeholder bodies, the technical configuration of the Internet itself relied extensively on technical interoperability. Per the IETF's 1958 *Request for Comment*, the Internet itself was set up with four core design principles: the goal of connectivity (why), the structure of layering (what), the tool of IP (the how), and the function of end-to-end intelligence (the where). These design principles led to the famous "hourglass" configuration of the Internet, with IP residing at the narrow waist (the Middle Layers), applications and content at the top (the Upper Layers), and various communications networks at the bottom (the Lower Layers; Whitt, 2013). As Werbach (2007) observes, "the defining characteristic of the Net is … a relentless commitment to interconnectivity." To become part of the Internet, operators of individual networks voluntarily connect to preexisting networks to tap into IP's technical interoperability, as bolstered by the regulatory interoperability of its surrounding institutions.

### (ii) Regulatory, Industry, and Geopolitical Climate

Looking back, the late 1970s and early 1980s were a crucial period to ascertain how IP won out over other potential market outcomes. In 1978, ISO began developing what would become their (1984) *Open System Interconnection* (OSI) reference model (Whitt, 2019). While the OSI reference model subsequently provided a useful seven-layer schematic for protocol development and implementation, it never secured a foothold in the marketplace vis-a-vis IP. In fact, IP quickly achieved universal acceptance and prominence, in large part due to the regulatory, industry, and geopolitical climate of the time.

OSI became synonymous with a more "top down" approach, as championed by France and other European countries. By contrast, the more bottom-up "rough consensus and running code" approach exemplified by IP allowed for a broader base of consensus and quicker evolution of the standard. The US Government's clear preference for IP also nudged aside the European countries' mandates for OSI. To one critic, however, the Internet's technical community of the 1990s was less democratic and inclusive than other industry standards bodies. Russell (2014) believes that "they preferred the rapid dissemination of a pragmatic kludge to a time-consuming pursuit of new technical knowledge," contrasting the development of TCP/IP with the OSI reference model. ISO employs the "open system" moniker to describe its intentions, and the well-specified interfaces between each layer in the seven-layer reference model promote "openness" (Russell, 2014).

Ironically, as Russell (2014) argued, ISO's attempt to utilize more formalized democratic mechanisms of international standardization to establish the OSI reference model may have been its downfall. In his words, "openness was OSI's founding justification, noblest aim, and fatal flaw," cautioning that this little understood history of OSI has some troubling lessons for those who champion inclusivity, openness, and multi-stakeholder governance. These values opened up OSI's process to strategies of delay and disruption and to internal conflicts that were impossible to resolve. Openness does not appear to stand as an explicitly stated engineering design principle of the Internet. Instead, the overt interest is in connecting disparate networks. This overt interest may be because openness is not in itself an explicit design principle. Instead, openness emerges as a phenomenon resulting from the operation of engineering design principles. The other possible outcome would have been neither IP nor OSI winning out, and instead, a reliance purely on the marketplace. Here, it seems clear that interoperability might never have developed on its own. Ferguson (1993) argues that many new technologies like the Internet come about not from the free market or the venture capital industry, but rather from government research agencies and in universities, particularly UCLA, where Stephan Crocker "invented the internet" (Metz, 2012). The Internet's open architecture was a fundamental principle that was a hallmark of the government research effort, one that would not have come about if the Internet had been created by private industry. Without the existence of a ready alternative like the Internet, 'closed' networks may well have become the prevailing marketplace norm. Instead, walled gardens, like CompuServe[50] and Prodigy[51] quickly lost their influence as end users connected to the Web.

The Web's overlay to the Internet developed later, and somewhat differently. There was a need to make the peer-to-peer nature of the Internet easier for ordinary people to utilize. In the late 1980s Tim Berners-Lee invented HTTP as the software protocol to power what became the World Wide Web (Whitt, 2024). The European organization CERN first initiated the Web in 1989, and it was introduced as an "overlay" to the Internet in the early 1990s, meaning that the Web operates as another layer on top of the Internet, much like email does. In 1994, the World Wide Web Consortium (W3C) was formed to evolve the Web's various protocols and standards (Whitt, 2024). As a multistakeholder organization like those attached to the underlying Internet, the W3C produced widely available specifications describing the Web's building blocks.

### (iii) Intervention Points Used for Regulatory or Technical Interoperability Efforts

As we have seen, intervention points run the gamut from the "hard power" of government laws and regulations to the "soft power" of industry protocols and standards, and government "nudge" implements such as procurement actions. The "ISO vs. IP" protocol showdowns of the 1980s ostensibly centered on the soft power considerations of whether and how the relevant standards bodies, and eventually the marketplace of industry adoption and implementation, would favour one of the two approaches. Instead, the US government's explicit adoption of IP within its networks, as well as its employment of the "power of the purse" of procurement, helped IP win the day – merits aside.

Indeed, the 1980s and early 1990s were heady days for technology maximalists, who saw major upside with minimal downside to the widespread adoption of the Internet and Web protocols in a relatively unregulated environment. The US took the lead in this maximalist stance, with the Clinton Administration openly touting the benefits of what Al Gore called the "information superhighway."[52] The political play at the time was to encourage US citizens to get online while also endorsing a "hands off" approach best represented in the slogan "don't regulate the Internet" (Whitt, 2013). In 1998, the *Internet Corporation for Assigned Names and Numbers* was formed pursuant to an exclusive US government contract as the guardian of the domain name system necessary to run the Internet. A nascent industry of local "dial-up" ISPs sprang up over the course of the 1990s, eventually being replaced by broadband providers utilizing telephony-based DSL and cable modem services.

[50] https://www.compuserve.com/
[51] https://en.wikipedia.org/wiki/Prodigy_(online_service)

Another intervention point is the temporary defeat of anti-interoperability forces supporting the *Stop Online Piracy Act* (SOPA) and *Protect Intellectual Property Act* (PIPA; Whitt, 2013). In late 2011, the US Congress was heading towards adopting legislation aimed at stopping websites from hosting content that violates US copyright laws. SOPA and PIPA each sought to impose certain technical requirements on website owners, search engines, Internet Service Providers (ISPs), and other entities, to block the dissemination of unlawful content on the Web. Hundreds of internet engineers raised red flags about these proposed requirements (Whitt, 2013). While most took no issue with the objective of reducing copyright-infringing content, they strongly objected to the proposed technical requirements of filtering the internet's Domain Name System (DNS) of naming and numbering resources. Such an approach was perceived to be grossly over-inclusive as it would affect legitimate users and uses and ineffective because technical workarounds would be available to the purveyors of bad content, overall presenting a lack of fitness to the desired objective. One letter sent to Congress by Internet luminaries observed that the Internet has been designed for reliability, robustness, and minimizing central points of failure or control - universal interoperability (Higgins & Eckersley, 2011). And the proposed SOPA would be a form of anti-regulatory interoperability that would harm technical interoperability. In the end, a robust pushback by internet companies and end users – later termed the "Internet Blackout Day" – convinced US legislators not to move forward with SOPA or PIPA. At the moment, robust regulatory and technical interoperability was saved.

#### (iv) <u>How those Interventions Played Out</u>

Nonetheless, the 2011 defeat of SOPA and PIPA served as the high mark for the heady internet optimism of the 1990s. In the years that have followed, the US and other governments took actions that deliberately blocked Web traffic – whether in the name of protecting against the dissemination of child pornography, extremist terrorist content, or content violating intellectual property laws. These interventions proved similar, attacking perceived ills at the upper layers, content and applications, by instituting network traffic bars at the middle and lower layers. This approach has been seen by some as a "layers violating" intervention that both overreaches (by affecting innocent uses) and under-performs (by allowing technical workarounds; Whitt, 2013).

[52] https://en.wikipedia.org/w/index.php?title=Information_superhighway&oldid=1259689806

In the 2020s, the situation remains murky. Regulatory and technical interoperability are no longer the unalloyed good that several previous generations had assumed. Nations now patrol their borders against unwanted content and applications, constraining and breaking technical interoperability in actions that lead to splinternets. Platform companies exert market power to the extent that countries seek to regulate their cross-border activities. The Web overlay has brought a parade of negatives, like surveilling and tracking end users, proliferating extremist content, and manipulative practices by the leading platform companies. In response, end users have deployed ad blockers and content filters. A truism is that there is no longer a single Web, but instead five billion distinctive Webs – one for each of us who are online and using it.

What went wrong? Part of the problem is that policymakers and others equated the Web and its applications/content with the underlying Internet. As a result, the Internet's open version of interoperability at the middle layers of the hourglass became an easy political target, even though the actual offenders were the companies utilizing the upper layers of the hourglass, including the Web. Various intervention points have been used to affect regulatory and technical interoperability efforts. In some cases, laws, regulations, and treaties have employed a top-down approach, such as the SOPA/PIPA debates. Softer forms of power utilize incentives over mandates but still have governments intervening into existing interoperability protocols and standards. When hard power and soft power collide – typically represented by governments on one side and industry standards bodies on the other – governments tend to prevail. This prevalence is especially the case when government actions are bolstered by incumbent industry players. At some point, however, the politics of solving pressing societal issues supersedes legitimate concerns about regulatory overreach. For too many policymakers, degrading technical interoperability is a relatively easy fix at the middle and lower layers to combat issues arising at the upper layers. Unless policymakers gain a full appreciation of the risks involved with these hard power interventions, however, those anti-interoperability actions will only continue.

**Table 1: Case Study Summary**

| Case / Criteria | History around the Need for Regulatory or Technical Interoperability | Regulatory, Industry, and Geopolitical Climate | Intervention Points Used for Regulatory or Technical Interoperability Efforts | How those Interventions Played Out |
|---|---|---|---|---|
| Nanotech: NanoDefine | Fragmented measurement criteria and no standard nanomaterial definition; EU's (2011) Recommendation highlighted classification and regulatory challenges, industry compliance difficulties; NanoDefine arose to create consensus and tools. | Rapidly evolving regulation, global competition for standards; ISO/OECD efforts nascent; public and industry safety concerns; fragmented research and industry technical approaches; need for harmonization critical for EU leadership. | Multi-tiered EU project (2013-18): analyzed 40+ measurement techniques, validated SOPs and protocols, created NanoDefiner e-tool and training; deep collaboration with labs and regulators; outcomes fed back to ISO/OECD. | NanoDefine tools/SOPs adopted in law and practice; validated methods reduced variance and regulatory gaps; e-tool enabled wider compliance; model for further nanotech regulatory science; recognized/adapted in EU's (2022) definition, spread globally. |
| Environmental Sustainability: INSPIRE Directive | Disparate EU states' environmental data, e.g. for river basins, made cross-border work slow, manual, costly. The EU pushed for harmonized frameworks for reporting on shared environmental goals. INSPIRE was created in 2007 as a compromise. | Fragmented regulations, tech, and industry incentives. Financial crisis, resource disparity, and legal resistance delayed compliance. Private sector split by open/proprietary interests. Geopolitical expansion (new EU members) tested resilience. | Legal mandate for standardization of spatial data (like GML), staggered deadlines, funding incentives; non-binding technical guides and capacity-building for poorer states. Maintenance group created. | By 2023, >95% dataset access achieved, enabling cross-border analyses, but data quality is uneven; some states retained dual systems or outdated data; rigid and dated technical standards mismatched new analytics; governance adaptability improved later through continued stakeholder engagement. |

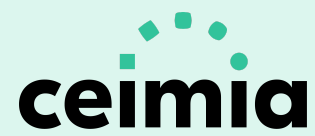

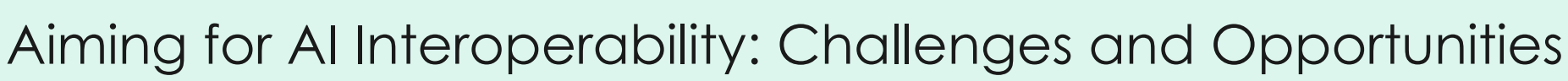

| Telecomm: Telegraphs to Snowden | 19th century telegraph networks incompatible; economic, diplomatic needs forced 1865 convention. Unified protocols under ITU. WWI and II and later Snowden exposed security/trust flaws in assuming interoperability cooperation. The need for zero-trust emerged. | Changing incentives: expansion, then crisis. Voluntary, sovereignty-respecting standardization at first. Security crises shifted priorities networks weaponized. Post-Snowden: global privacy/geopolitical tension, led to GDPR and more. Open networks challenged by demands for sovereignty. | International voluntary standards (ITU); post-crisis: cables cut, trust lost; later, new consensus and technical standards. 21st cent: cybersecurity and zero-trust, continuous verification, legal standards (GDPR), plus industry protocol updates. | Early voluntary standards enabled fast utility/trade; wars revealed vulnerabilities, led to redesign. Zero-trust and GDPR models are now dominant, emphasizing verification/sovereignty interoperability only possible if security/verification in-built, not assumed. |
| --- | --- | --- | --- | --- |
| Internet/Web Architecture: HTTP/IP | US DoD & academia drove TCP/IP; voluntary network joiners; contrasted with top-down failed OSI model and walled-garden networks. Governance via US-led agencies, then multi-stakeholder, rough consensus. The Web evolved from protocols later. | 1980s: US favoured IP, open but pragmatic consensus approach; OSI/Europe more top-down, less adopted. Initial open architecture; private sector/IP dominance. Later, splinternet from diverging regulatory models, content control, digital sovereignty. | US government early adoption, procurement power, plus open standards. Multi-stakeholder bodies (like IETF/W3C) kept protocol adaptation rapid. Attempts to block/interfere (e.g., SOPA) resisted by stakeholders. Societal pivots/Internet Blackout shaped trajectory. | Initial global interoperability: rapid innovation, universal Internet. Subsequent fragmentation (splinternet); new state interventions (content filtering and sovereignty) erode interoperability. Open core remains technically, but top-down vs. bottom-up governance and markets are evolving. |

## 4. Lessons Learned

Across these case studies four lessons emerge that can inform and be applied to ongoing AI interoperability efforts: (a) designing adaptive governance frameworks; (b) establishing early definition and measurement standards; (c) building trust through verification mechanisms; and (d) bridging technical and regulatory divides through collaboration.

### (a) Lesson 1: Designing Adaptive Governance Frameworks

Given AI's rapid technical advancements, there is a need to design and implement adaptive governance frameworks that can evolve alongside these technological advancements. The often-used rigid governance frameworks will likely become cumbersome and outdated as these technical advancements outpace them. This lesson is illustrated across the case studies and directly addresses the fragmentation and convolution challenges identified in Part 1's analysis of the current AI governance landscape. For instance, the EU's *INSPIRE Directive* provides the clearest example of how rigid regulatory frameworks can hinder long-term interoperability goals. This *Directive's* rigid 2007-era specifications that were designed for static datasets clashed with emerging technologies. This inflexibility forced countries like the Netherlands to develop costly hybrid APIs to bridge *INSPIRE's* batch-processing architecture with dynamic air quality monitoring systems, while the *Directive's* structured GML schemas proved incompatible with machine learning models requiring unstructured inputs. The European Parliament's evaluation specifically criticized these overly prescriptive metadata requirements for discouraging private-sector participation, demonstrating how inflexible governance can stifle innovation.

Conversely, the *NanoDefine* project succeeded precisely because it built adaptability into its framework from the outset. The project's *NanoDefiner e-tool* was designed to be expandable and adaptable, allowing for incorporation of new methods and updates to regulatory definitions as they became available. This adaptive design enabled the framework to successfully accommodate the European Commission's (2022) revised *Recommendation for the Definition of Nanomaterial*, demonstrating their flexibility and sustainability. Similarly, the project's emphasis on developing SOPs that were adaptable to different analytical techniques while maintaining consistency in critical performance parameters created a governance model that could evolve with technological progress. The Internet/Web Architecture case further reinforces this lesson by showing how initially successful interoperability can deteriorate when governance frameworks fail to adapt to changing circumstances. While the TCP/IP and HTTP protocols achieved remarkable initial success through their open architecture and multistakeholder governance, the subsequent growing global fraying of regulatory interoperability led to a technical 'splinternet' of networks, demonstrating that even successful interoperability frameworks require ongoing adaptation to address challenges like privacy concerns, content moderation, and AI sovereignty.

This lesson, if taken and implemented within the AI governance landscape, could partially address the proliferation of AI standards, principles, and regulations that create compliance burdens and lock-in effects as different initiatives vie for market share. Rather than adding more rigid frameworks to this fragmented landscape, the case studies suggest that successful AI technical and regulatory interoperability require governance mechanisms that can accommodate technological evolution while maintaining core interoperability principles. This accommodation requires designing AI governance frameworks with built-in mechanisms for incorporating new technical advancements and standards, updating risk assessments as AI capabilities advance, and adapting compliance requirements to emerging use cases. The *NanoDefine* project's success in creating tools that remained relevant across regulatory updates and technological changes provides a concrete model for how AI governance can balance necessary standardization with the flexibility required for rapidly evolving technologies. Without such adaptive frameworks, AI interoperability efforts risk replicating *INSPIRE's* costly rigidity or facing the fragmentation challenges that have emerged in internet governance.

### (b) Lesson 2: Establishing Early Definition and Measurement Standards

Establishing early definitions and measurement standards in a technology's development lifecycle can avoid regulatory and technical lock-in. This lesson proves particularly relevant to AI's emerging governance challenges whereby there is little coordination and harmonization efforts around AI's technical and regulatory requirements across jurisdictions, leading to economic trade disadvantages. The case studies demonstrate that once incompatible systems become established, achieving interoperability requires exponentially more effort and resources than proactive standardization efforts undertaken during a technology's nascent stages, thereby delaying economic trade advantages.

The nanomaterials case study exemplifies this principle most clearly. Despite the European Commission's (2011) *Recommendation for the Definition of Nanomaterial* establishing that nanomaterials consist of particles where 50% or more have external dimensions in the 1-100 nm range, disparities in measurement techniques led to conflicting classifications of identical materials – leading to technical and regulatory interoperability challenges within the sector. The absence of harmonized measurement methods created immediate regulatory implementation problems, demonstrating how definitional precision without accompanying technical standards fails to achieve true interoperability. The subsequent *NanoDefine* project emerged specifically to address these challenges but required substantial resources to retrofit solutions that could have been more efficiently implemented during the technology's early development phase. Similarly, the *INSPIRE Directive* case reveals how delayed standardization efforts compound complexity over time. Environmental data systems across EU Member States had evolved independently, resulting in nearly 30% of environmental datasets remaining unusable for cross-border analysis due to technical discrepancies. National mapping agencies operated in silos, with Germany's topographic databases employing different classification schemes than France's, necessitating laborious manual reconciliation of datasets using conflicting coordination systems and metadata standards. The *INSPIRE Directive's* implementation required extensive retrofitting of existing systems rather than building interoperability from the ground up. The telecommunications case study illustrated where early intervention proved successful. The (1865) *International Telegraph Convention* emerged when telegraph technology was still expanding, allowing delegates from twenty nations to establish common standards before incompatible systems became too entrenched. This early intervention prevented the kind of systemic fragmentation that later plagued other technologies.

For AI governance, this lesson suggests that the current proliferation of disconnected regulatory efforts described in Part 1 risks creating the same fragmentation challenges that required costly retrofitting in nanomaterials and environmental data systems. The window for establishing common definitional frameworks, measurement standards, and interoperability protocols may be rapidly closing as AI technologies mature and regulatory approaches become institutionalized. The case studies collectively demonstrate that while technical interoperability can sometimes be achieved retroactively, the political and economic costs of harmonizing already-established regulatory frameworks increase dramatically over time, making early coordination efforts essential for avoiding future governance challenges.

### (c) Lesson 3: Building Trust Through Verification Mechanisms

Trust and crisis preparedness are inextricably linked in successful interoperability efforts, with considerable implications for AI governance. The case studies consistently demonstrate that interoperability systems designed only for cooperative conditions become severe liabilities during trust crises, while those incorporating robust verification mechanisms from inception prove more resilient and adaptable. For instance, the telecommunications case provides the most striking illustration of this dynamic. The *International Telegraph Union's* framework thrived for decades based on mutual trust and shared economic benefits, enabling global connectivity. However, WWI exposed the fundamental vulnerability of trust-dependent systems when Britain strategically severed Germany's undersea cables and the Zimmermann Telegram incident demonstrated how shared infrastructure could become a weapon. This crisis immediately undermined existing technical interoperability frameworks that had operated on assumptions of continued cooperation. Significantly, the post-Snowden telecommunications environment shows how this lesson was eventually internalized. Rather than abandoning interoperability entirely, the industry developed zero-trust architectures that operate on 'never trust, always verify' principles, enabling secure cooperation even in low-trust environments through continuous authentication and verification mechanisms. Similarly, the *NanoDefine* project demonstrates how incorporating verification and transparency from the outset can build sustained trust across diverse stakeholders. This project's success stemmed from developing technical standards and creating comprehensive validation protocols, transparent inter-laboratory studies, and decision support frameworks that enabled stakeholders to verify compliance independently. This approach built confidence among regulatory authorities, industry laboratories, and research institutions across 11 European countries, leading to widespread adoption and formal integration into EU regulatory frameworks. *NanoDefine's* emphasis on expert knowledge that could be packaged and disseminated through user-friendly digital platforms created accountability mechanisms that sustained trust even as the technology evolved.

Conversely, the *INSPIRE Directive* case illustrates how insufficient attention to trust-building and verification can undermine even well-intentioned interoperability efforts. Despite legal mandates and significant funding, many EU Member States maintained parallel systems rather than fully embracing harmonized approaches. Greece and Italy's decision to operate dual systems for domestic and EU-compliant data reflected underlying trust deficits regarding centralized EU mandates and concerns about data sovereignty. This *Directive's* reliance on non-binding technical guidelines, while providing flexibility, fostered inconsistency that further eroded confidence in the system's reliability.

For AI governance, this lesson is particularly urgent given the current trust crisis surrounding AI systems' transparency, bias, and safety; creating exactly the kind of uncertainty that undermines trust in interoperability frameworks. These case studies suggest that AI interoperability efforts must move beyond aspirational principles toward concrete verification mechanisms that enable stakeholders to independently assess compliance and performance, such as expanding and developing the emerging technical standards for AI explainability, auditing, and risk assessment that function as verification tools rather than mere guidelines. Like the post-Snowden zero-trust architectures, AI interoperability frameworks should assume that trust breakdowns are inevitable and design systems that can maintain functionality through verification rather than faith. The *NanoDefine* model of collaborative validation and transparent decision support tools offers a particularly relevant template for building trust in AI governance across diverse international stakeholders while anticipating future crises that could otherwise fragment emerging interoperability efforts.

### (d) Lesson 4: Bridging Technical and Regulatory Divides through Collaboration

Bridging technical and regulatory divides through collaboration emerged as a lesson learned across the case studies. This bridging can align various stakeholders' interests, leading to greater buy-in and reduced conflict. For example, the internet and web architecture case illustrated this principle through the contrasting fates of TCP/IP and the OSI model. While the OSI reference model relied on formalized, top-down standardization processes, TCP/IP succeeded by prioritizing a rough consensus and running code philosophy. Engineers, academics, and industry stakeholders collaborated iteratively to refine protocols based on practical implementation, creating a bottom-up ecosystem that accommodated these diverse interests. This collaboration mirrored the internet's technical 'hourglass' architecture, where a narrow IP layer enabled expansive innovation at the edges. For AI governance, this underscores the value of interoperable core protocols (e.g., for risk assessment or accountability tracking) that permit jurisdictional adaptations, akin to Part 1's call for regulatory coherence rather than identical rules. Similarly, the *NanoDefine* project exemplifies how structured collaboration can translate regulatory intent into actionable tools. Faced with inconsistent methods for classifying nanomaterials under the European Commission's (2011) *Recommendation for the Definition of Nanomaterial*, this project convened researchers, industry labs, and regulators to develop validated measurement protocols and decision-support software. By integrating technical expertise with regulatory needs, *NanoDefine* created practical compliance pathways. AI governance requires mechanisms to operationalize high-level principles, like the OECD's *AI Principles*, into sector-specific implementation guides, ensuring that technical interoperability aligns with regulatory intent. Part 1 highlights this need through its analysis of soft law fragmentation, where competing certifications and standards create compliance confusion. A *NanoDefine*-like approach – developing shared tools for risk assessment or auditing – could contribute to the harmonization of these efforts.

The *INSPIRE Directive* case reveals both the potential and pitfalls of collaboration. While *INSPIRE* mandated harmonized geospatial data standards, its phased implementation allowed EU Member States to adapt technical specifications to local infrastructures. Denmark's integration with national digitalization strategies reduced costs, whereas federalized states, like Germany, faced delays from conflicting regional policies. This implementation challenge mirrors Part 1's observation that AI governance must balance transnational cooperation with respect for national priorities. AI regulatory and technical interoperability may avoid this, among other, challenges by learning from these case studies. For instance, allowing jurisdictions to tailor transparency requirements for public-sector AI uses while adhering to common risk classification frameworks.

Crucially, these cases demonstrate that interoperability thrives when stakeholders perceive mutual benefit. The telecommunications sector's post-Snowden shift to zero-trust architectures succeeded because industry and governments shared incentives to rebuild user trust through verifiable security protocols. For AI interoperability, analogous incentives exist in cross-border digital trade through interoperable compliance frameworks that could reduce market barriers to entry for startups while giving governments confidence in transnational AI safety.

# Part 3 - Moving Forward: Implementing Lessons Learned through Sectoral Roadmaps

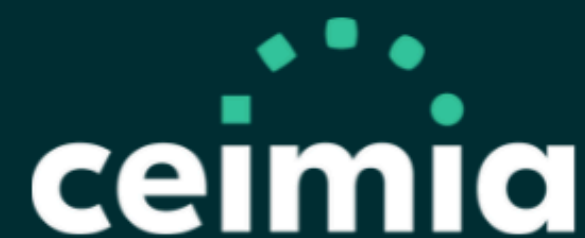

# Part 3: Moving Forward: Implementing Lessons Learned through Sectoral Roadmaps

## 1. Introduction

Parts 1 and 2 of this report have highlighted the urgent necessity for AI regulatory and technical interoperability and the valuable lessons that can be drawn from historical precedents across diverse technological domains when aiming to implement AI interoperability efforts. Part 3 builds directly upon these foundations by transitioning from analysis to action, providing concrete roadmaps for implementing the lessons learned from historical cases while addressing the specific gaps and challenges identified in the current AI governance landscape. This final section of the report recognizes that while Parts 1 and 2 established the problem and extracted valuable lessons, the critical task now lies in translating these insights into practical roadmaps that different sectors can follow to aim toward AI regulatory and technical interoperability. The approach taken in Part 3 acknowledges that achieving meaningful interoperability requires coordinated action across multiple sectors, each with distinct roles, capabilities, and incentives, while recognizing that no single actor or approach can address the complexity of current AI governance challenges.

Part 3 of the Report develops in three stages. The first stage examines the current state of AI interoperability efforts, both regulatory and technical, to establish what foundation exists for future coordination. This examination reveals a landscape where significant initiatives are already underway, like the G7's (2023a) *Hiroshima AI Process* and (2024) *International Network of AI Safety Institutes* to Google's (2025) *Agent-to-Agent Protocol* and Anthropic's (2024) *Model Context Protocol*, yet critical gaps persist. These gaps include geopolitical disruptions affecting international cooperation, regulatory fragmentation creating incompatible standards, governance shortfalls in operational AI oversight and global representation, security and trust deficits in cross-system integration, and fundamental incompatibilities between regulatory frameworks across jurisdictions. Second, four sectoral roadmaps are detailed, one for the standard-setting and international organizations, NGO and civil society, public, and private sectors. These roadmaps are situated within Part 1's AI governance landscape, built in the lessons learned across Part 2's case studies, and address Part 3's interoperability effort gaps where applicable. Each roadmap is summarized in Table 2 below. Each roadmap is structured to provide concrete, actionable guidance while acknowledging that these high-level approaches must be adapted to specific organizational contexts and capabilities. Together, these roadmaps create a comprehensive framework for coordinated action toward AI regulatory and technical interoperability. Third, and lastly, some brief concluding remarks are offered.

## 2. Existing and Emerging AI Regulatory and Technical Interoperability Efforts

### (a) AI Regulatory Interoperability Efforts

The global landscape of AI regulatory interoperability reflects a network of multilateral initiatives, each attempting to harmonize approaches while addressing diverse national interests and developmental stages (Faveri et al., 2025a: 2025b: 2025c). Canada's recent adoption of a national digital identity standard - CAN/DGSI 103.0[53] - based on the (2020) *Pan-Canadian Trust Framework* exemplifies efforts to build interoperable systems that align with international frameworks including Europe's (2014) eIDAS and (2025) *FATF Recommendations*. The G7's (2023a) *Hiroshima AI Process* represents a considerable coordinated effort in AI governance to establish an international framework with guiding principles and a code of conduct for advanced AI systems. This process has generated broader support through the (2024) *Hiroshima AI Process Friends Group*, now comprising ~50 governments beyond the G7, demonstrating the potential for expanding consensus-building mechanisms in AI governance. Similarly, the (2024) *International Network of AI Safety Institutes* brings together technical organizations from Australia, Canada, the European Commission, France, Japan, Kenya, South Korea, Singapore, the United Kingdom, and the United States. This international network prioritizes four crucial areas: research collaboration on AI risks and capabilities, development of common testing practices for advanced AI systems, facilitation of shared approaches to interpreting test results, and active engagement with countries at all development levels to increase capacity for diverse participation in AI safety science. The EU's harmonized standards process under the EU *AI Act* represents another significant interoperability effort, utilizing their 'New Approach' to regulation by delegating technical content development to CEN/CENELEC – the EU-level standard-setting organization (Soler et al., 2024). This new approach attempts to create unified technical standards that can facilitate global compliance and regulatory interoperability.

Regional initiatives also contribute to the interoperability landscape. The *Transatlantic Trade and Technology Council* (TTC) focuses on aligning AI approaches between democratic allies, while APEC economies are advancing AI standards for trust and innovation through collaborative conferences and knowledge exchange (APEC, 2025; Office of the United States Trade Representative, 2021). The ASEAN's (2023) *Digital Economy Framework Agreement* aims to establish progressive rules covering digital trade, cybersecurity, cross-border data flows, and emerging technologies including AI. OpenAI's (2025) letter to Governor Newsom illustrates how private sector entities are advocating for regulatory harmonization, specifically recommending that California recognize international frameworks like the EC's (2025) *General-Purpose Code of Practice* and federal agreements with the *Center for AI Standards* and Innovation as compliance pathways. This approach seeks to avoid creating incompatible regulatory requirements that could fragment the AI development ecosystem.

[53] https://dgc-cgn.org/product/can-dgsi-103-0/

### (b)Technical AI Interoperability Efforts

Google's (2025) *Agent-to-Agent* (A2A) protocol represents the first major industry-standard for multi-agent communication and collaboration. This protocol enables independent AI agents to communicate through standardized message formats, task coordination, and capability discovery mechanisms, allowing agents from different vendors and frameworks to work together seamlessly, like how HTTP enables universal web connectivity. A2A has gained substantial industry support, with contributions from 50+ technology partners, including major firms like Salesforce, MongoDB, and ServiceNow, and consulting firms like Accenture, Deloitte, and McKinsey (Surapaneni et al., 2025). Complementing Google's agent-focused approach, Anthropic's (2024) *Model Context Protocol* (MCP) addresses the vertical integration challenge by standardizing how AI systems connect with external tools, databases, and data sources (agent-to-tool). MCP provides a universal framework that eliminates the need for custom integrations with each data source, transforming what was previously an "N×M" integration problem into a more manageable standardized approach. The protocol has been rapidly adopted by major AI providers including OpenAI and Google DeepMind, demonstrating initial industry momentum toward standardization (Wiggers, 2025a: 2025b). The convergence of these horizontal (agent-to-agent) and vertical (agent-to-tool) integration standards is creating new possibilities for comprehensive AI system ecosystems. Combining A2A and MCP protocols could address complex multi-agent scenarios while maintaining secure and standardized communication pathways. This integration approach has been successfully demonstrated in practical applications such as IT incident response, where multiple specialized agents collaborate using A2A for communication while accessing various tools and systems through MCP. Aside from A2A and MCP, the (2020) *Open Voice Interoperability*

*Initiative* (OVON) represents a comprehensive practical approach to AI interoperability, establishing frameworks for diverse conversational AI agents including chat, voice, and videobots, and human agents to interact effectively. OVON's architecture uses universal APIs based on Natural Language, allowing AI agents developed with different technologies to communicate using standardized natural language interfaces.

There are also novel architectural approaches underway around technical AI interoperability's infrastructure challenges. For example, the (2025) *Networked AI Agents in a Decentralized Architecture* (NANDA) framework represents an ambitious attempt to create comprehensive infrastructure for secure and trustworthy agent ecosystems. NANDA provides global agent discovery, cryptographically verifiable capability attestation, and cross-protocol interoperability across multiple communication standards including A2A, MCP, and other emerging protocols. Security considerations are driving additional innovation in technical AI interoperability frameworks. These security-focused approaches recognize that interoperability without robust security mechanisms could create significant vulnerabilities in distributed AI systems. Similarly, agent discovery and naming services are emerging as critical infrastructure components. The *Agent Name Service* proposes DNS-inspired naming conventions and *Public Key Infrastructure* certificates for verifiable agent identity and trust (Government of Canada, 2023; OWASP, 2025). Such naming services are critical for enabling secure and scalable agent ecosystems where participants can reliably discover and authenticate potential collaboration partners. Beyond protocol and architectural innovations, the technical AI interoperability landscape benefits from robust institutional standardization efforts. For example, Faveri & Marchant (2024b) created a dataset of ~1600 international AI and AI-related standards from the ISO, ETSI, ITU, and IEEE; the *AI Standards Hub*, a UK-based initiative within the Alan Turing Institute, maintains a database of 500+ AI-related standards, covering published and under/in-development AI standards (AI Standards Hub, 2025); and, in 2025, the AI for Good launched an *AI Standards Exchange Database* in collaboration with the IEC, ISO, and ITU which contains 770+ AI standards and technical publications from global standards development organizations (AI for Good, 2025). These databases showcase the extensive ongoing work in AI standardization cataloging.

### (c) Gaps within Regulatory and Technical AI Interoperability Efforts

Within these regulatory and technical AI interoperability efforts, five gaps and challenges emerge. Below, each of these five gaps and challenges are described and will be the foundation for what the following section's roadmaps will be based on.

### (i) Geopolitical Disruptions and Trade Implications

The shifting geopolitical landscape presents additional challenges to AI regulatory interoperability. The 2025 Trump Administration's policy shifts are expected to introduce significant disruptions to existing cooperative frameworks. Proposed changes to AI-chip export controls to China could reduce technical friction while accelerating China's AI self-sufficiency and contribute to global fragmentation of AI regulation and standards (Villaenor, 2025). The implementation of aggressive tariff policies threatens to strain alliances and accelerate efforts by the EU, Canada, Japan, and South Korea to strengthen non-US trade ties. These disruptions could undermine multilateral cooperation mechanisms like the G7 and TTC that have been central to existing interoperability efforts. Similarly, the pivot from 'AI Safety' to 'AI Security' in US AI policy discourse risks reducing focus on safety, fairness, and bias, among other issues, potentially creating friction with other nations that prioritize these aspects of AI governance, putting them at odds with international partners and further driving regulatory fragmentation (Kore & Dhawan, 2025). Export controls following fragmented licensing approaches tied to bilateral bargaining, rather than well-evaluated technical risk assessments or standards alignment, threaten to politicize technical cooperation and undermine efforts to build consensus around shared technical standards. These geopolitical tensions reinforce motivations outside the US for sovereign 'data' and 'technology' approaches, potentially leading to the development of separate, incompatible technological and regulatory ecosystems. The resulting policy volatility means the landscape for AI regulatory interoperability is likely to remain fluid, contested, and unpredictable, with bilateral negotiations potentially fragmenting multilateral consensus-building efforts. The convergence of these factors suggests that achieving AI regulatory interoperability will require sustained commitment to multilateral cooperation, development of flexible frameworks that can accommodate diverse national approaches, and mechanisms for ongoing dialogue and adaptation as both technology and geopolitical circumstances continue to evolve.

### (ii) Regulatory Fragmentation and Incompatible Standards

The most significant challenge to AI regulatory interoperability is the proliferation of incompatible mandatory risk classification models and competing standards, certification, and assurance frameworks across jurisdictions (Faveri et al., 2025; Faveri et al., 2025). While the EU *AI Act* establishes specific categories for 'high-risk' AI applications, jurisdictions like Colorado are developing their own distinct classification systems, creating a patchwork of requirements that companies must navigate (Laszio, 2025). This fragmentation is compounded by the emergence of multiple, potentially incompatible standards, certification, and assurance frameworks without mutual recognition agreements or harmonized assessment criteria. The absence of official or authoritative interoperability tools or regulatory crosswalk frameworks make it difficult for organizations to understand how compliance with one system relates to requirements under another. Many existing frameworks, including the US (2022) *AI Bill of Rights*, establish high-level principles without providing concrete implementation models or operational guidance, creating uncertainty for developers while limiting practical interoperability. This regulatory fragmentation creates significant compliance burdens, increases costs and complexity, and undermines efforts to create coherent global standards for AI development and deployment (Kerry et al., 2025).

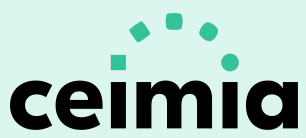

There is a fundamental fragmentation in the standards landscape, exacerbated by competitive dynamics that incentivize proprietary approaches over genuine openness. While initiatives like A2A and MCP promise technical interoperability, concerns have emerged about whether these industry-led standards represent genuine openness or strategic positioning by major technology companies. The proliferation of competing protocols creates a sort of 'AI protocol war'[54] with different frameworks, like A2A, MCP, and ACP, competing for adoption while addressing overlapping but not identical use cases. This fragmentation forces organizations to choose between standards or develop complex integration layers that negate many interoperability benefits. The competitive dimension of this fragmentation reflects broader tensions between collaboration and competition in the AI industry. Major technology companies face conflicting incentives regarding interoperability. While standardization could expand market opportunities, it also risks commoditizing competitive advantages built on proprietary approaches. This fragmentation problem is compounded by the rapid pace of AI development, which often outpaces traditional standards development processes. By the time formal standards emerge, market leaders may have already established de-facto standards through deployment and adoption, creating path dependency that resists later harmonization efforts (Faveri et al., 2025).

Current regulatory approaches reflect strategic fragmentation, where jurisdictions assert regulatory independence in service of national strategic goals, even at the cost of global interoperability. For example, the Trump administration's *Executive Order 14179*, which rescinds previous AI safety measures while mandating development of what is now their (2025) *AI Action Plan*, exemplifies how political considerations can fragment technical standardization efforts (AI Gov, 2025; Federal Register, 2025). Similarly, China's pursuit of sovereign AI strategies and the EU's emphasis on rights-based regulation create incompatible requirements that technical standards alone cannot reconcile (Racicot & Simpson, 2025). This regulatory fragmentation creates practical barriers to technical interoperability implementation. Different jurisdictions impose varying requirements for data governance, algorithmic transparency, human oversight, and risk assessment that make it difficult to develop global technical AI interoperability standards. Organizations are forced to navigate complex and often vague overlapping requirements that change based on deployment geography and use case context. The governance challenge extends beyond regulatory compliance to include fundamental differences in how jurisdictions conceptualize AI risk and appropriate responses. The burden of this incompatibility has an outsized impact on startups and SMEs. It is hard enough for a team of 8-10 people with a robust legal and compliance team to figure this out, let alone a smaller organization without the expertise, personnel, or budget who are simply trying to get their initial customer/client-base.[55] While the EU emphasizes comprehensive risk assessment and algorithmic accountability, the US increasingly focuses on maintaining competitive advantage and innovation speed. These philosophical differences create structural barriers to harmonized approaches that technical interoperability standards cannot address independently. Moreover, the pace of regulatory development often lags technical advancement, creating uncertainty about future compliance requirements that discourages investment in interoperability infrastructure.

[54] Adapted from the "standard wars" literature.

### (iii) Governance Gaps in AI Operations and Global Representation

Current AI governance frameworks suffer from critical gaps in addressing operational aspects of AI systems and ensuring inclusive global participation. There is a general lack of rules and policies governing AI activities, including issues of delegation, consent, authentication, and revocability, as well as insufficient coverage for AI agent activities, particularly regarding limitations on fraudulent actions and accountability mechanisms. Environmental oversight is notably absent, with inadequate policies governing the use of energy and water resources to fuel data centers and AI infrastructure. Similarly, the limited representation of Global Majority countries in the development of rules, principles, standards, certifications, and assurances risks creating governance systems that do not account for diverse developmental needs, cultural contexts, and technological capabilities, potentially exacerbating global digital divides (Menon, 2025). Additionally, there is insufficient coordination of research agendas examining the impacts of AI agents on human decision-making and psychology, limiting the evidence base available for developing effective governance frameworks and understanding broader societal implications.

[55] A "Playbook for SMEs" is being developed for HAIP compliance by Canada as part of their G7 commitment: https://g7.canada.ca/en/news-and-media/news/g7-leaders-statement-on-ai-for-prosperity/

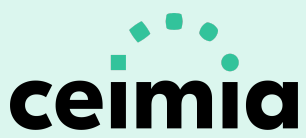

#### (iv) Security, Privacy, and Trust Deficits in Cross-System Integration

The second gap involves challenges related to security, privacy, and trust that emerge when AI systems must share data and capabilities across organizational and technical boundaries. Cross-platform AI interoperability necessarily involves sharing sensitive data, models, and decision-making processes across systems that may have different security models, governance frameworks, and trust boundaries. These challenges are particularly acute in regulated industries where data residency, audit trails, and explainability requirements create additional complexity layers. The question 'where does our data go?' emerges as a barrier to firms' AI adoption, even when technical capabilities meet organizational needs. Privacy challenges in AI interoperability extend beyond traditional data protection concerns to include new risks associated with AI-specific vulnerabilities. AI systems do well with massive data sets for training and improving performance, often involving personal and sensitive information, creating privacy risks that scale with the size of data and system integration. AI models' often opaque nature, particularly deep learning systems, make it difficult to provide the transparency and explainability that firms require for trust and regulatory compliance. The security architecture required for truly interoperable AI systems must address multiple threat vectors simultaneously, such as unauthorized access to shared data, manipulation of inter-agent communications, model poisoning through compromised integrations, and privacy leakage through cross-system inference. Traditional security approaches designed for isolated systems prove inadequate for the distributed trust relationships required by interoperable AI ecosystems. These trust deficits create a defensive posture among firms where procurement processes focus on risk mitigation rather than capability evaluation, fundamentally changing how firms approach AI adoption and creating systematic barriers to interoperability implementation.

## 3. Sectoral Roadmaps to AI Interoperability: Applying Lessons Learned

Drawing on the content from the existing and emerging AI regulatory and technical interoperability efforts and the gaps within these efforts detailed above, the overall challenges in the AI governance landscape outlined in Part 1, and the lessons learned from the case studies in Part 2, we have developed four sectoral roadmaps (listed below) for moving toward AI regulatory and technical interoperability. There are four roadmaps, one for each of the: (a) standard-setting and international organizations; (b) NGOs and civil society; (b) public sector; and (d) private sector. Each roadmap is structured into phases/elements[56] over a five-year timeline, detailed in-text and summarized in Table 2, with an implementation timeline and success metrics section also provided at the end of each roadmap subsection. We acknowledge that these roadmaps are high-level - just as their sectoral scopes are - and recommend reading and thinking through how they may be made more granular in your specific firm, department, agency, or organization.

### (a) Standard-Setting and International Organization Roadmap for AI Regulatory and Technical Interoperability

Standard-setting organizations (SSOs) and international organizations (IOs) occupy a central, often underestimated role in shaping AI regulatory and technical interoperability. They are not passive conveners, but strategic actors that can stabilize cooperation during geopolitical disruption, accelerate coherence when regulatory or political avenues stall, and anchor credibility through principles, frameworks, assurance, and certification. Their independence allows them to provide common reference points that governments, firms, and civil society can adopt, standing somewhat apart from national politics while enabling both technical and regulatory interoperability. As Parts 1 and 2 demonstrated, this ecosystem is already crowded with initiatives and stressed by challenges: the pace of innovation outstrips consensus processes, geopolitical rivalry fragments approaches, proprietary protocols risk capture, and SMEs and the Global South remain underrepresented. The lessons from earlier sections also showed that standards succeed when they balance adaptability, verification, and openness, and fail when they become rigid, fragmented, or captured by narrow interests. Against this backdrop, the roadmap that follows sets out how SSOs and IOs - working in coalition with governments, industry, academia, and funders - can move from convening to active architecture of global AI interoperability. Interoperability is a governance priority and business imperative. Firms that can demonstrate conformity with recognized standards and certifications gain smoother access to international markets, reduce duplicative compliance costs across jurisdictions, and lower liability exposure by showing adherence to accepted benchmarks. For smaller firms in particular, interoperable standards reduce barriers to entry and make participation in cross-border digital trade feasible. Historical experience demonstrates that interoperability efforts succeed when they balance adaptability, verification, and openness, and fail when they are rigid, fragmented, or captured by narrow interests. These lessons directly inform the roadmap that follows, organized around five mutually reinforcing phases. Each represents both a strategic objective and an actionable pathway toward global interoperability.

[56] Phases are used when sequential steps are appropriate for the sector's roadmap and elements are used when a set of non-sequential recommendations are appropriate for the sector's roadmap.

### (i) **Phase 1: Build a Coherence Infrastructure**

The experience of the *INSPIRE Directive* illustrates the cost of failing to provide coherent infrastructure. Rigid frameworks, designed without sufficient regard for evolving technologies, forced costly retrofits and produced uneven adoption. By contrast, developing shared crosswalks and reference architectures at the outset can avoid similar fragmentation and help ensure that new AI standards evolve with the technology. The proliferation of national, regional, and industry-led frameworks for AI has already made it increasingly difficult to determine how compliance in one jurisdiction translates into another. Without a common map, firms face duplicative costs and regulators struggle to monitor outcomes across borders. Standards bodies and IOs are uniquely positioned to address this gap by developing shared crosswalks between regimes like the EU *AI Act*, OECD *AI Principles*, and ISO/IEC 42001, among other standards in the 42000 series. Building on this, the field also requires the establishment of a global reference architecture for AI assurance and governance. Much as NIST's reference models for cloud computing clarified roles and responsibilities within complex technology supply chains, an AI reference architecture, particularly considering advancing agentic AI and evolving training and inference models, could provide clarity about accountability, certification touchpoints, and the chain of trust across developers, deployers, and users (Hadfield et al., 2024).

### (ii) **Phase 2: Accelerate AI Standards Development**

Formal standards processes are notoriously slow, but AI technologies evolve at a pace that risks leaving those processes behind. The TCP/IP protocols and the W3C process showed how fast-moving, open, and iterative standards-setting can outpace more rigid approaches such as OSI. If SSOs cannot shorten the cycle, de facto standards set by corporate platforms will continue to dominate. Yet speed alone is insufficient. The credibility of accelerated outputs depends on their being transparent in design, iterative in their validation, and visibly adopted by regulators and markets. This credibility lens must be applied to new instruments such as Technical Reports, pre-normative work, and hybrid models that combine open-source iteration with consensus-based ratification. In practical terms, this means fast-tracking work on areas like AI agents or cross-agent protocols while ensuring that resulting standards are tested, updated, and trusted by stakeholders across jurisdictions. The window for establishing credible and coherent interoperability frameworks is narrowing. Standards, certification, and assurance ecosystems cannot be retrofitted once fragmented practices and de facto corporate standards are entrenched. Delay risks locking in systems that are opaque, incompatible, or exclusionary.

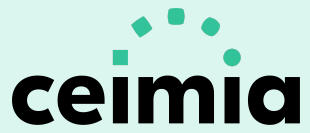

#### (iii) Phase 3: Establish Certification and Assurance as the Unifying Layer

Standards provide frameworks, but assurance and certification translate those frameworks into verified practice. Certification is not only a technical validation mechanism; it is a cornerstone of credibility that enables governments, businesses, and the public to trust that a system has met agreed thresholds of safety, fairness, accountability, or identified requirements. By providing market legitimacy, certification reassures buyers and investors, and by offering political cover, it gives regulators across divergent jurisdictions confidence that oversight can be achieved through accredited third parties. NanoDefine demonstrated that embedding verification protocols directly into standards and building decision-support tools can anchor trust across stakeholders. For AI, certification and assurance must play this same role, translating abstract standards into verifiable practice and enabling trust in cross-border deployments. It is certification that enables governments, businesses, and the public to trust that an AI system has met agreed thresholds of safety, fairness, accountability, and other identified requirements. To achieve this, SSOs and IGOs should promote joint certification models that integrate both management-system evaluations (like ISO/IEC 42001) and product-level assessments, resulting in two certifications (management system and product) that leverage one standard and enabling compliance with regulatory expectations like the EU's CE marking. International accreditation networks, including IAF and ILAC, provide the infrastructure for mutual recognition, but they must be extended to AI. Governments and large procurers also have a crucial role to play by sending clear demand signals, indicating that certified systems will be favored in procurement decisions. These signals are the catalyst needed for certification ecosystems to scale from pilots to widely adopted practice (Hadfield et al., 2024; Shank, 2021). Beyond procurement, market incentives provide a powerful driver for adoption. Firms see conformity assessment not only as a compliance obligation but as a business case: certified systems gain easier access to foreign markets, reduce duplicative compliance costs across jurisdictions, and mitigate liability risk by demonstrating adherence to recognized standards. Framing certification as both a governance tool and a competitiveness strategy makes clearer why private-sector alignment will be central to scaling assurance ecosystems – and scaling interoperability that goes with them.

### (iv) Phase 4: Development of Sovereignty-Compatible Architectures for Standards, Regulation, and AI Technologies

The long life of the ITU shows that cooperative frameworks can survive upheaval if they respect sovereignty. For AI, the equivalent is sovereignty-compatible architecture whereby a narrow set of shared standards ensure interoperability and leave room for local adaptation. Governments are beginning to articulate 'sovereign AI' approaches, seeking greater control over data, models, and operations within their jurisdictions. Interoperability tools, both regulatory and technical, will need to accommodate these objectives while still enabling trusted cross-border use. Divergent national priorities are a reality of the global AI landscape and attempts to impose uniformity will almost certainly fail. The challenge is to identify a baseline set of standards that serve as a 'common backbone,' allowing interoperability across borders while leaving space for local adaptation. This approach mirrors the success of TCP/IP, which became the backbone of global connectivity not by dictating every layer of communication, but by providing just enough commonality for diverse networks to interoperate. Sovereignty-compatible certification schemes can build on this logic, establishing global baselines while allowing countries to embed their own priorities in implementation and oversight (Whitt, 2019).

### (v) Phase 5: Broaden Representation and Participation

The history of Internet governance demonstrates that broad participation sustains legitimacy and adoption. The IETF's principle of 'rough consensus and running code' enabled diverse actors to shape the protocols that underline today's Internet. Legitimacy in standards-setting depends on whose voices are heard. Currently, SMEs, Indigenous communities, and countries in the Global Majority remain underrepresented and face some of the steepest barriers to market access and the highest risks of exclusion (Arai, 2023). Funding mechanisms and commitments (like those from government or industry), liaison roles, and diversity mandates will be needed to broaden representation and participation in standards committees and technical working groups. The aim is not simply equity for its own sake, but the construction of standards that reflect the diversity of contexts in which AI will be deployed. Without this, interoperability frameworks risk privileging the priorities of large economies and major technology firms, further entrenching fragmentation and mistrust.

### (vi) Implementation and Success Metrics

The success of this roadmap should be evaluated across four areas. First, by the number of standards or certifications published and the extent to which these instruments are adopted, trusted, and reduce fragmentation in practice. The first measure of progress will be the degree of recognition that core standards such as ISO/IEC 42001 or IEEE P2863 achieve across jurisdictions. Formal adoption by regulators, coupled with mutual recognition agreements, would indicate that the coherence infrastructure is working as intended. Similarly, certification uptake provides another essential signal. Over time, an increasing share of AI systems deployed across borders should carry accredited certification, whether through joint management-system and product schemes or through recognition by international accreditation bodies. Parallel to this, the growth of assurance ecosystems can be tracked by the expansion of accredited auditors, certifiers, and trainers able to verify compliance against established standards. Second, representation is also key and will be reflected in broader participation by SMEs, Indigenous communities, and the Global South within mirror committees and technical working groups. Data on membership and leadership roles in these processes can serve as indicators of whether inclusion is being meaningfully achieved, or whether standards continue to be shaped primarily by a narrow set of actors. Third, success can also be measured in the coherence tools themselves. The publication of regulatory crosswalks, reference architectures, and mapping instruments – and their uptake by both regulators and industry – would provide tangible evidence that the roadmap is helping align otherwise divergent frameworks. Market impact is another lens: reductions in compliance costs reported by SMEs, or evidence that procurement contracts increasingly require certified AI systems, would show that standards and certification are not only being produced but are delivering practical benefits. Fourth, and lastly, innovation speed should be monitored. The ability of SSOs to produce technical reports or other pre-normative outputs within 18 months of a proposal (or ISO/IEC's *Publicly Available Specification* submissions[57]) can serve as a proxy for whether the standards community is keeping pace with the technologies it seeks to govern. Together, these measures – adoption, certification uptake, assurance ecosystem growth, inclusion, coherence tools, market impact, and innovation speed – offer a composite picture of whether standards-setting and international organizations are succeeding in anchoring credibility and interoperability in the global AI ecosystem.

---

[57] Publicly Available Specification (PAS) is a procedure used by ISO/IEC JTC 1 to quickly turn specifications developed by external organizations into international standards.

### (b) NGO and Civil Society Roadmap for AI Regulatory and Technical Interoperability

As demonstrated in the selected cases, while the voices of NGOs have been given places to be heard – like the *Internet Governance Forum* and *Internet Corporation for Assigned Names and Numbers* – it is largely commercial and governmental interests that have dominated these discussions. This challenge persists despite the participation of NGOs and civil society organisations in various regulatory and technical interoperability-setting exercises across the Internet and then the Web's history. Nevertheless, NGOs and civil society organisations have an important and multifaceted role to play. As neutral and transnational actors, NGOs can facilitate multi-stakeholder collaboration among governments, industry, academia, and civil society, enabling the alignment of technical standards and governance frameworks across jurisdictions. Furthermore, NGOs serve as critical advocates for evidence-based policy development, conveners of diverse stakeholder perspectives, independent monitors of implementation, builders of technical and governance capacity, and guardians of human rights and ethical principles in AI development. As we have seen, the dominance of commercial and governmental entities amounts to an ability to set agendas, capitalize on favourable political environments, limit outside participation, utilize their uniquely deep resources, and obviate objective assessments of assurance that their work product is fit for purpose. An exploration of these abilities and the critical role of NGOs within each of these elements is described below, followed by a suggested way forward.

#### (i) <u>Element 1: Agenda Setting</u>

While the early days of 'AI regulation' (circa late 2022) found NGOs and civil society well-positioned to stake their claims to influence, the larger picture tells a different story. The initial push for regulation of frontier AI models quickly coalesced around defining and defending so-called "guardrails" to prevent the riskiest actors taking the riskiest actions. From the EU *AI Act*, to the (2023) *Bletchley Accords*, to the (2023) *Biden White House AI Executive Order*, that guardrail's theme was consistent. Both the AI platform companies and NGOs agreed on that basic point, with disagreement centered on the types of risks to proscribe, and the safeguards that should be applied. NGOs and civil society organizations have played an essential advocacy role in articulating the values and principles – including human rights, democratic accountability, and equitable access – that should ground agenda-setting discussions. Through policy advocacy and research, these organizations have worked to ensure that discussions extend beyond technical guardrails to encompass broader societal impacts, worker protections, environmental considerations, and implications for marginalized communities. On its own terms, of course, that focus left largely unchecked those commercial activities taking place within the guardrails. And yet, as the large Web and AI companies began to challenge NGO advocates in those venues, the governmental authorities one by one began to give way. In California, for example, intense lobbying by the Web/AI companies led a Democratic Governor to veto the one supposedly comprehensive AI regulation bill that had been placed on his desk. In Colorado, the only US State to pass a bill that mirrored the "risky guardrails" framework of the EU *AI Act*, a second round of provisions intended to clarify and perhaps extend the Act's reach did not survive the 2025 legislative session. The capacity of NGOs to harmonize diverse policy perspectives and articulate coherent frameworks that reflect broad-based stakeholder input has been significantly constrained by the retreat of commercial and governmental voices from collaborative agenda-setting processes, leaving civil society advocates isolated and their advocacy efforts substantially weakened.

### (ii) Element 2: Political Environment

Of course, the change in presidential administration in the US quickly tilted the balance away from NGOs and civil society, and towards a more nationalistic bend. Any hint of what was termed 'woke AI' was met with approbation, and outright bans via executive order. The States were not spared this pushback, with a proposed 10-year moratorium morphing into the federal Government wielding the power of the purse to discourage any further State-sponsored regulatory actions (Kang, 2025). NGOs and civil society organizations have found themselves in an increasingly difficult position as political environments have shifted, their watchdog and monitoring functions rendered less effective when governmental actors themselves retreat from regulatory commitment. The ability of civil society to serve as independent monitors of governmental AI policy implementation – a critical accountability mechanism in democratic governance – has been undermined by the politicization of AI regulation and the delegitimization of civil society perspectives. With the rise of AI agents, there appears to be no determined effort in Europe or the US to create a new regulatory framework. In some cases, existing consumer protection laws are touted as adequate. Again, any NGO or civil society voices raising concerns do not appear to be gaining traction. Nevertheless, NGOs retain a critical capacity to advocate for the continued prioritization of human rights protections, ethical principles, and public interest considerations within regulatory frameworks, even when political environments grow hostile. Through convening multi-stakeholder dialogues and fostering international cooperation, civil society can work to build consensus around fundamental principles that transcend political cycles and national boundaries, ensuring that democratic values and human dignity remain central to AI governance discourse.

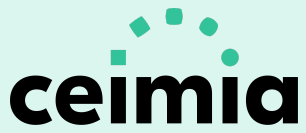

**(iii) Element 3: Participation**

On the technical interoperability side, the situation is starker. As noted above, no standard-setting body to date has produced an AI-to-AI interoperability standard. Instead, it was Google's A2A standard that came to market in April 2025. While Google's A2A standard was an open-source protocol and pro-user, some experts began to raise pointed questions about security risks and limitations and the need for more robust threat modeling frameworks (Huang, 2025; Louck et al., 2025; paloalto network, 2025). Others observed a disturbing lack of basic privacy safeguards (Roosa, 2025) and pointed to everything wrong with the A2A protocol (Kekulawala, 2025). NGOs and civil society organizations, particularly those with technical expertise and independent standing, are positioned to serve a vital monitoring and watchdog function – conducting rigorous assessments of technical standards against established benchmarks for privacy, security, and ethical alignment. Beyond monitoring, civil society actors can contribute capacity building and knowledge-sharing initiatives that strengthen the technical literacy of governments, policymakers, and affected communities regarding AI interoperability standards. Despite these concerns, the industry itself has begun widespread adoption of the A2A protocol with some 150 organizations adopting it (A2A Protocol, 2025; Surapaneni & Stephens, 2025). It seems the market has spoken, quickly and directly despite these unresolved concerns. NGOs and civil society organizations, particularly those in Canada, have a substantial and well-listened to voice that can be used to address A2A's concerns, harmonize stakeholder positions on technical standards, advocate for human rights protections embedded within interoperability frameworks, and aid organizations in making informed A2A adoption decisions. Through convening technical experts, policymakers, and representatives of affected communities, civil society can facilitate dialogue that ensures technical standards reflect both security requirements and the public interest, preventing the consolidation of private technical decisions without adequate public accountability mechanisms.

**(iv) Element 4: Resources**

Further, without a steady and independent stream of money and other resources, NGOs and civil society cannot maintain their position. In 2025, many of the corporate funders have cut back, or even cut ties. The large philanthropic foundations also find a far more challenging environment, particularly with US NGOs and civil society organizations as their actions are sometimes being challenged by the Trump Administration, even as that same administration curtails funding that otherwise would support a vibrant NGO and civil society ecosystem. The resource constraints facing civil society organizations directly impair their capacity to fulfill multiple critical functions: conducting rigorous policy advocacy and research; monitoring regulatory implementation and technical standard adoption; building governance and technical capacity among weaker stakeholders; harmonizing stakeholder positions through convening and dialogue; and safeguarding human rights and ethical principles in AI governance. Without adequate, independent funding – whether from governments, philanthropic institutions, or multilateral organizations committed to civil society support – NGOs and civil society cannot sustain the expertise, staff, and infrastructure necessary to participate meaningfully in multi-stakeholder processes or to serve as effective watchdogs on behalf of the public interest. The erosion of civil society funding represents not merely a resource challenge but a fundamental threat to democratic accountability in AI governance.

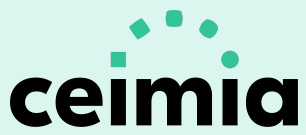

### (v) Element 5: Assurance

Of course, one observation is that many, if not all these purported issues with the new protocol could have been addressed had they been vetted through the Web-normal processes of open standards bodies. Interestingly, most of these raised concerns came from within the corporate AI ecosystem rather than those from NGOs or civil society. NGOs and civil society organizations bring distinctive value to assurance processes by serving as independent auditors of technical and policy standards, evaluating whether systems and frameworks adequately protect human rights, ensure ethical alignment, and meet commitments to transparency and accountability. These organizations possess the legitimacy and credibility to convene disparate stakeholders around shared assurance benchmarks and to conduct monitoring that holds both industry and government accountable to established principles. These internal concerns raise another disturbing point that perhaps civil society has been "hollowed out" by recent funding cuts and empty seats at decision-making bodies, such that they no longer bring to bear the necessary expertise or public forum to express any of their concerns. Restoring civil society capacity for independent assurance requires not only adequate resources but also meaningful participation in standards-setting and governance bodies from the outset, ensuring that NGOs and civil society organizations are positioned as recognized partners in vetting, monitoring, and validating that policies and technical standards meet appropriate benchmarks for human rights protection, privacy, security, democratic values, and ethical alignment.

### (vi) A Way Forward

NGOs and civil society organizations must continue to be brought into these conversations and efforts, be sufficiently funded by governments, firms, international institutions, and philanthropic organizations, and continue to be listened to. NGOs and civil society must be empowered to actively shape AI governance trajectories through robust policy advocacy grounded in evidence and values-driven analysis. To better accomplish this ambitious agenda, existing and incumbent players can undertake the following actions. Existing and incumbent players must proactively include NGOs and civil society in agenda-setting processes from their inception, recognizing these actors' capacity to advocate for evidence-based policy frameworks that harmonize commercial, governmental, and public interests while ensuring human rights considerations are foregrounded. Rather than allowing agendas to be set unilaterally by powerful commercial and governmental actors, multi-stakeholder bodies should be structured to give civil society a meaningful voice in defining scope, priorities, and the fundamental values – like equity, transparency, and accountability – that should guide regulatory and technical discussions. NGOs should be resourced to conduct independent policy research and analysis that challenges narrow commercial interests and articulates the public interest perspective. Furthermore, civil society organizations should be empowered to convene stakeholders around alternative policy framings and to synthesize diverse viewpoints into coherent, values-aligned governance proposals.

NGOs and civil society organizations must have the independence, resources, and protection to fulfill their watchdog and monitoring functions even when political environments become hostile to regulatory oversight or civil society engagement. Governments and international institutions should commit to structural protections for civil society participation in AI governance, including guarantees of meaningful consultation, access to decision-making forums, and protection from political retribution. Civil society actors can work to build resilience through convening networks of international NGOs and civil society organizations that collectively advocate for human rights protections and ethical principles in AI governance, transcending individual political jurisdictions and electoral cycles. Investment in civil society's capacity to monitor governmental compliance with AI governance commitments – and to publicly report on implementation gaps – is essential to maintaining democratic accountability even as political environments shift.

Incumbent technical standard-setting bodies and emerging platforms must genuinely include NGOs and civil society organizations – particularly those with technical expertise, human rights focus, and public accountability – in technical interoperability discussions and decision-making. Civil society should be resourced to conduct independent technical analysis and monitoring of standards such as the A2A protocol, publishing rigorous assessments that evaluate these standards against benchmarks for privacy, security, human rights protection, and ethical alignment. NGOs can convene multi-stakeholder technical working groups that bring together industry experts, academic researchers, government representatives, and civil society actors to harmonize approaches to technical interoperability while ensuring that human rights and public interest considerations are embedded in technical design. Furthermore, civil society organizations should develop and deliver capacity-building initiatives that strengthen the technical literacy of governments, affected communities, and smaller organizations regarding AI interoperability standards, enabling broader participation in technical governance discussions.

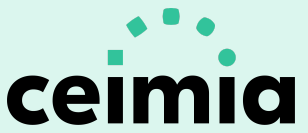

Governments, philanthropic foundations, international institutions, and responsible corporate actors must commit to sustained, independent funding for NGOs and civil society organizations engaged in AI governance work. This funding must be structured to protect civil society autonomy and independence, ensuring that organizations can fulfill their watchdog, advocacy, convening, and capacity-building functions without constraint from funders' political or commercial interests. Funding mechanisms should be designed to support both established organizations and emerging civil society actors, particularly those representing marginalized communities, developing country perspectives, and underrepresented expertise. The protection and expansion of civil society funding is not merely a philanthropic gesture but an investment in democratic governance and the protection of human rights in AI systems.

Governments, standards-setting bodies, and industry actors must recognize NGOs and civil society organizations as essential partners in assurance processes, providing these actors with the authority, access, and resources to conduct independent monitoring and evaluation of whether policies, standards, and implementations meet established benchmarks for human rights protection, privacy, security, democratic accountability, and ethical alignment. Civil society should be positioned from the design phase to evaluate proposed frameworks, monitoring implementation over time and publicly reporting on gaps or failures. NGO and civil society organizations should be resourced to develop and maintain assurance frameworks – including metrics, audit mechanisms, and reporting structures – that can be applied across jurisdictions and technical domains, creating harmonized approaches to accountability in AI governance. These organizations should convene regular multi-stakeholder reviews of policy and technical standards, creating forums for transparent dialogue about whether existing frameworks remain fit for purpose and where adaptation is required to better protect human rights, ensure ethical alignment, and serve the broader public interest.

### (c) Public Sector Roadmap for AI Regulatory and Technical Interoperability

The public sector occupies a critical position in advancing AI regulatory and technical interoperability, serving as regulator, procurer, and AI system user. Drawing from the lessons learned in Part 2's case studies and addressing the governance gaps identified in Part 3, this roadmap provides a structured pathway for governments to build coherent AI governance frameworks that promote domestic innovation and international cooperation. This roadmap recognizes that successful interoperability requires moving beyond the current fragmented landscape of proliferating standards, principles, and regulations toward coordinated frameworks that try to balance standardization and the flexibility required for rapidly evolving AI systems. This roadmap is divided into four phases that represent paths the public sector can take and how to utilize their critical position while doing so to aim toward AI interoperability.

#### (i) Phase 1: Foundation Building and Early Standardization (Months 1-18)

Governments must begin by establishing early definition and measurement standards, as demonstrated by the critical importance of this lesson from the NanoDefine project. The EC's (2011) *Recommendation on the Definition of Nanomaterial* created immediate regulatory implementation problems precisely because standardized measurement methods were absent, requiring the costly NanoDefine project to retrofit solutions that could have been more efficiently implemented during the technology's early development phase. For AI governance, this translates to an urgent need for governments to establish common definitional frameworks for AI systems, risk categories, and measurement protocols before incompatible approaches become institutionalized. During this foundation phase, public sector actors should prioritize the incorporation of global AI policy frameworks and standards into domestic regulatory development. For example, incorporating frameworks like the OECD's (2019) *AI Principles*, G7's (2023b) *Hiroshima Process Guidelines*, and emerging technical standards from bodies like the ISO, IEC, and IEEE. The NanoDefine project's success stemmed from building upon existing international frameworks while creating practical implementation tools, demonstrating how domestic policy can achieve global coherence without sacrificing national priorities. Governments should establish dedicated interoperability coordination offices within their AI governance structures, like the collaborative approach that enabled the International Telegraph Union's early success. These offices would serve as focal points for mapping existing regulatory requirements against international frameworks, identifying gaps in current approaches, and developing cross-government coordination mechanisms. The telecommunications case study illustrates how early institutional frameworks, even when voluntary, can create lasting foundations for international cooperation when supported by clear economic incentives and mutual benefits.

#### (ii) **Phase 2: Adaptive Framework Development and Pilot Implementation (Months 12-36)**

Adding to the foundation building phase, governments should design adaptive governance frameworks that can evolve alongside rapid AI system developments. The EU's *INSPIRE Directive* provides a cautionary example of how rigid 2007-era specifications designed for static datasets clashed with emerging technologies, like real-time sensor networks and AI-driven analytics. To avoid similar pitfalls, public sector AI governance frameworks ought to incorporate built-in mechanisms for updating risk assessments as AI capabilities advance and adapt compliance requirements to emerging use cases. This phase should include the development of regulatory sandboxes and pilot programs that enable controlled testing of regulatory and technical interoperability frameworks across different use cases and jurisdictions. The UK and Singapore's sandbox efforts provide models for how governments can create safe regulatory environments for testing innovative approaches. For AI interoperability, these sandboxes should focus on sectors where cross-border coordination is most critical, such as healthcare, finance, transportation, and defense, while establishing partnerships with regional bodies like APEC, ASEAN, or through bilateral agreements. Governments should begin implementing 'sovereignty-compatible' approaches that allow for regulatory coherence without requiring identical rules across jurisdictions. The Internet architecture case study demonstrates how the TCP/IP protocol's success stemmed from enabling a narrow IP layer that permitted considerable innovation. Similarly, AI governance frameworks should establish core interoperability protocols for risk assessment and accountability tracking and jurisdictional adaptations for transparency requirements, oversight mechanisms, and enforcement approaches.

#### (iii) **Phase 3: Trust Infrastructure and Verification Systems (Months 24-48)**

The telecommunications case study's evolution from trust-based systems to post-Snowden zero-trust architectures provides useful insights for AI interoperability efforts. The original International Telegraph Union framework thrived for decades based on mutual trust, but WWI exposed the fundamental vulnerability of trust-dependent systems when Britain strategically severed Germany's undersea cables. For AI systems, which will operate in potentially adversarial environments from the outset, governments must build verification mechanisms rather than relying solely on cooperative goodwill. During this phase, public sector actors should develop comprehensive audit and verification frameworks that enable independent assessment of AI system compliance and performance. The NanoDefine project's success in building trust across diverse stakeholders stemmed from creating transparent inter-laboratory studies and decision support frameworks that allowed stakeholders to verify compliance independently. AI governance requires analogous mechanisms, such as standardized technical approaches to AI explainability, bias assessment, and risk evaluation that function as verification tools rather than mere guidelines. Governments should establish mutual recognition agreements for AI auditing and certification processes, building on the collaborative validation model demonstrated by NanoDefine. These agreements would enable AI systems certified under one jurisdiction's framework to be recognized in partner jurisdictions, reducing compliance burdens while maintaining safety and accountability standards. The telecommunications sector's shift to zero-trust architectures shows how verification-based systems can maintain functionality even when geopolitical trust breaks down, a crucial consideration given the current tensions in AI development and governance between major powers.

**(iv) <u>Phase 4: Cross-Border Integration and Scaling (Months 36-60)</u>**

The final implementation phase focuses on scaling successful pilot programs and establishing permanent mechanisms for cross-border AI interoperability coordination. Governments should leverage the economic and trade benefits of interoperability to build broader coalitions, as demonstrated by the telegraph's early success in facilitating international commerce. The telecommunications case study illustrated how shared economic interests can sustain cooperative frameworks even during periods of political tension, suggesting that emphasizing AI interoperability's trade facilitation benefits can build durable support for coordination efforts. Public sector procurement policies should actively incentivize AI interoperability by requiring compliance with established international standards and frameworks. Government purchasing power can serve as a significant market signal, like how US government adoption of TCP/IP helped establish it as the dominant internet protocol. By requiring AI vendors to demonstrate interoperability with established frameworks, governments can accelerate market adoption of standardized approaches while avoiding vendor lock-in effects. This phase should include the establishment of permanent institutional mechanisms for ongoing coordination and adaptation. The Internet's success stemmed partly from creating multistakeholder governance bodies like the *Internet Engineering Task Force* that could continuously evolve protocols based on practical implementation experience. AI governance requires similar adaptive institutions that can incorporate new technical developments, address emerging risks, and facilitate ongoing coordination between different jurisdictions and stakeholder communities.

#### (v) Implementation Timeline and Success Metrics

This roadmap spans approximately five years, with overlapping phases that allow for iterative development and continuous refinement. Success metrics should include the percentage of government AI systems that comply with established interoperability frameworks; the number of mutual recognition agreements established with partner jurisdictions; the reduction in compliance costs for organizations operating across multiple jurisdictions; and the degree of technical standardization achieved across different AI applications and sectors. Governments implementing this roadmap must recognize that regulatory or technical interoperability is not an endpoint but an ongoing negotiation between competing interests, as demonstrated across all four case studies. This roadmap's success depends on maintaining continuous dialogue between policymakers, technologists, and end-users to understand their challenges and if those challenges are subsiding while ensuring that frameworks evolve alongside both technological and geopolitical realities. The window for implementing coordinated AI governance frameworks may be rapidly closing as technologies mature and regulatory approaches become institutionalized, making early action on this roadmap essential for avoiding the costly fragmentation patterns that have characterized other technological domains. The public sector's unique position as both regulator and user of AI systems provides governments with multiple intervention points for promoting interoperability. By leveraging their regulatory authority, procurement power, and convening capacity in a coordinated manner, governments can address the current AI governance fragmentation while building foundations for sustained international cooperation in this critical technological domain.

### (d) Private Sector Roadmap for AI Regulatory and Technical Interoperability

Drawing from the lessons learned in the telecommunications, environment, nanotechnology, and internet architecture cases, the private sector faces a critical window of opportunity to proactively implement AI interoperability efforts before regulatory fragmentation and technical lock-in create exponentially higher costs. The current AI governance landscape, characterized by proliferating standards and fragmented and competing regulatory frameworks, mirrors the early stages observed in these case studies, where early interventions proved more effective than retroactive harmonization efforts. The below five-year roadmap provides a five-phase structured approach for private sector organizations to aim toward AI interoperability.

### (i) **Phase 1: Strategic Positioning and Organizational Readiness (Months 1-6)**

The first phase emphasizes establishing adaptive governance frameworks, informed by the *INSPIRE Directive's* lesson that rigid specifications designed for static environments quickly become obsolete when technology evolves rapidly. Private sector organizations must begin by conducting interoperability assessments that evaluate their current and planned AI systems against emerging standards, like Google's (2025) *Agent-to-Agent* protocol and Anthropic's (2024) *Model Context Protocol*. These assessments should identify existing technical debt, integration complexity, and potential lock-in risks like those observed in the early internet architecture case, where organizations that built on open standards rather than proprietary systems achieved greater long-term flexibility.

Where possible, and budget permitting, organizations should establish cross-functional interoperability teams comprising technical, legal, compliance, and business stakeholders, mirroring the collaborative approach that made the NanoDefine project successful across 11 European countries. These teams can develop organizational AI principles that prioritize interoperability from the outset, recognizing that retrofitting interoperability into existing systems requires exponentially more resources than designing for it initially. The foundation phase must also include establishing partnerships with standards organizations like ISO, IEEE, and industry associations or consortiums to stay informed about emerging interoperability frameworks – following the telecommunications industry's model of early participation in standard-setting bodies. During this phase, organizations should begin implementing basic technical and regulatory interoperability measures, like standardized APIs, common data formats, and modular system architectures that can accommodate future interoperability requirements. The key output is an organizational interoperability strategy document that defines clear objectives, identifies priority areas for interoperability implementation, and establishes governance structures for ongoing coordination across business units and external partners.

### (ii) **Phase 2: Building Interoperable Systems Architecture (Months 4-12)**

Building on Phase 1, organizations should develop technical infrastructure that enables internal integration with emerging interoperability standards while maintaining security and compliance requirements. The telecommunications case study's evolution from trust-based systems to zero-trust architectures provides crucial guidance here, demonstrating that interoperability frameworks must incorporate robust verification mechanisms from inception rather than assuming cooperative conditions will persist. Organizations can implement standardized protocols for AI system communication, drawing lessons from the internet architecture case where the success of TCP/IP stemmed from its focus on connecting disparate networks through open protocols. This phase requires significant investment in data governance infrastructure that enables interoperability while protecting sensitive information. Organizations must implement data classification systems, establish clear data sharing protocols, and develop privacy-preserving techniques that allow AI systems to collaborate without exposing proprietary or personal information. The NanoDefine project's emphasis on validated measurement protocols and transparent decision support frameworks provides a model for creating technical standards that build trust among diverse stakeholders while maintaining scientific rigor. Technical infrastructure development must also address the regulatory fragmentation challenge by building systems that can accommodate different compliance requirements across jurisdictions (at least for firms operating in more than one jurisdiction). This infrastructure development involves implementing configurable compliance frameworks that can adapt to varying regulatory requirements without requiring complete system redesigns, like how the early telegraph system succeeded by establishing voluntary standards that nations could adopt without threatening their sovereignty.

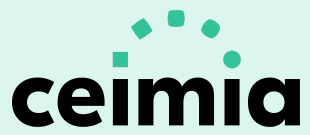

#### (iii) Phase 3: Pilots, Controlled Deployment, and Validation (Months 8-18)

Phase 3 focuses on deploying interoperability capabilities in controlled environments. These deployments are used to validate technical approaches and identify implementation challenges. Organizations should select pilot projects that are of value to their goals and that can be tested across different AI systems, data sources, and business processes. Following the NanoDefine project's approach of systematic validation and transparent reporting, organizations should establish comprehensive testing protocols during their pilots that evaluate technical performance and business outcomes. These protocols should include interoperability stress testing, security validation, regulatory compliance verification, and user experience assessment, among other firm-specific factors. These pilots can be used to develop internal expertise in interoperability implementation and management through training on standards, developing internal best practices for interoperable AI system design, and creating organizational maintenance processes. The pilot's implementation should also include efforts to establish partnerships with other organizations to test cross-enterprise interoperability scenarios. Likewise, the pilot phase should address the security and trust challenges seen in the telecommunications case study by implementing comprehensive monitoring and verification systems. Organizations should develop capabilities to continuously validate the integrity and performance of interoperable systems, establish incident response procedures for interoperability failures, and create transparency mechanisms that build stakeholder confidence in interoperable AI systems.

#### (iv) <u>Phase 4: Enterprise-Wide Implementation and External Partnerships (Months 12-36)</u>

Phase 4 involves scaling successful pilot implementations and establishing interoperability partnerships. Organizations should develop standardized deployment procedures for consistent interoperability implementation across different business units and geographic locations. This standardization should balance consistency with flexibility, allowing local adaptations while maintaining core interoperability principles, like how the *INSPIRE Directive's* phased implementation acknowledged diverse national contexts while pursuing common objectives. External partnership development becomes critical during this phase, as true interoperability requires coordination across organizational boundaries. Organizations should establish formal partnerships with suppliers, customers, and industry peers to implement cross-enterprise interoperability capabilities. These partnerships can include technical integration agreements, data sharing protocols, and joint governance structures that enable effective coordination while protecting each organization's interests. Organizations should also engage regulatory authorities to ensure their interoperability implementations align with evolving regulatory requirements. This engagement should include participating in regulatory consultations, providing input on emerging standards and guidelines, and sharing experiences and best practices with regulators to support their policy development(s).

#### (v) <u>Phase 5: Optimization and Continuous Evolution (Months 24-60 and Ongoing)</u>

The final phase aims to establish ongoing processes for interoperability optimization and adaptation to technical and regulatory changes. Organizations should implement continuous monitoring systems that track interoperability performance, identify emerging challenges, and detect opportunities for improvement or expansion. This monitoring should include technical performance metrics, business outcome measurements, regulatory compliance assessments, and stakeholder satisfaction indicators. Similarly, organizations should establish formal processes for evaluating and adopting new interoperability standards and technologies, such as participating in standards development activities, piloting new protocols and frameworks, and being aware of regulatory developments. Organizations must also use this phase to contribute to broader industry interoperability efforts through knowledge sharing, collaborative research, and participation in industry initiatives.

### (vi) Implementation and Success Metrics

The roadmap timeline allows for overlapping phases to accommodate the shifting nature of interoperability implementation and the varying complexity of different organizational contexts. Organizations with existing strong data governance and technical infrastructure can accelerate through early phases, while those with complex legacy systems or regulatory requirements may require extended timelines for certain phases. Success in this roadmap is characterized by the following activities across each phase. For phase 1, completing an organizational assessment, establishing governance structures, developing interoperability strategy, and beginning basic technical preparations. For phase 2, is it critical to implement core technical infrastructure, adopt initial interoperability standards, and establish data governance frameworks. For phase 3, organizations should deploy pilot implementations, validate technical approaches, and develop operational expertise. For phase 4, it is important to scale implementations enterprise-wide, establish external partnerships, and engage with regulatory processes. And for phase 5, organizations ought to optimize implementations, adapt to evolving requirements, and establish strategic positioning.

Success in implementing this roadmap should be measured through quantitative metrics including reduction in system integration costs, improvement in cross-system data accuracy and consistency, reduction in compliance costs across multiple jurisdictions, and increases in operational efficiency from interoperable AI systems. Qualitative indicators include organizational capability to rapidly adopt new interoperability standards, stakeholder confidence in AI system transparency and accountability, and industry recognition of the organization's interoperability leadership. The roadmap's success will be demonstrated by the organization's ability to operate effectively in an increasingly complex and fragmented AI governance landscape while maintaining competitive advantages through superior interoperability capabilities. Organizations that successfully implement this roadmap will be positioned to capitalize on the economic and strategic benefits of AI interoperability while avoiding the costly retrofitting and fragmentation challenges that have characterized other technological domains.

### Table 2: Summary of Sectoral Roadmaps

| Sector | Phase 1 | Phase 2 | Phase 3 | Phase 4 | Phase 5 | Success Metrics/Ways Forward |
|---|---|---|---|---|---|---|
| Standard-Setting and IOs | Develop shared crosswalks between regimes (EU AI Act, OECD AI Principles, ISO/IEC 42001); establish global reference architecture for AI assurance and governance | Shorten standards development cycles; fast-track work on AI agents and cross-agent protocols; ensure transparency, iteration, and validation in accelerated outputs | Promote joint certification models integrating management-system and product-level assessments; extend international accreditation networks (IAF, ILAC) to AI; leverage government procurement signals | Identify baseline standards serving as 'common backbone' for interoperability while allowing local adaptation; establish sovereignty-compatible certification schemes with global baselines | Expand participation of SMEs, Indigenous communities, and Global South countries; implement funding mechanisms and diversity mandates; ensure standards reflect diverse deployment contexts | Number of standards/certifications published and degree of adoption/recognition across jurisdictions; representation growth in mirror committees and working groups; publication and uptake of regulatory crosswalks and reference architectures; innovation speed - ability to produce outputs within 18 months. Track certification uptake, assurance ecosystem growth, compliance cost reductions, and procurement contract requirements |
| NGO and Civil Society | Stake claims to influence AI regulation and technical standards; define scope of debate and decision-making; maintain | Navigate challenging political landscapes; maintain advocacy effectiveness despite changing administrations; | Ensure meaningful participation in technical interoperability discussions; provide expertise on protocols like A2A; address | Secure steady and independent funding streams; address funding cuts from corporate funders and philanthropic foundations; | Ensure policies and standards meet appropriate benchmarks for privacy, security, and competition; maintain | Set agenda for scope of debate; score points within a constrained political environment; limit external influence while focusing on reaching conclusions; rely on unique position and expertise; ensure grounded assurance that policies and standards |

| | relevance despite commercial and governmental interests dominating discussions | address pushback against regulatory frameworks; raise concerns about emerging AI agent frameworks | concerns about security risks, privacy safeguards, and protocol limitations; maintain voice in SSB | maintain organizational capacity and expertise; ensure continued presence at decision-making bodies | expertise to vet protocols through open standards processes; prevent 'hollowing out' of civil society capacity | meet appropriate benchmarks. Must continue to be brought into conversations, sufficiently funded by governments and firms, and actively listened to, particularly in Canada |
|---|---|---|---|---|---|---|
| Public | Months 1-18: Establish early definition and measurement standards; incorporate global AI policy frameworks and standards into domestic regulatory development; establish dedicated interoperability coordination offices | Months 12-36: Design adaptive governance frameworks that can evolve alongside rapid AI developments; develop regulatory sandboxes and pilot programs; implement 'sovereignty-compatible' approaches | Months 24-48: Develop comprehensive audit and verification frameworks; establish mutual recognition agreements for AI auditing and certification processes | Months 36-60: Scale successful pilot programs; leverage economic and trade benefits to build broader coalitions; establish permanent institutional mechanisms for ongoing coordination | N/A (4 phases only) | Percentage of government AI systems complying with interoperability frameworks; number of mutual recognition agreements; reduction in compliance costs across jurisdictions; degree of technical standardization achieved. Timeline: ~5 years with overlapping phases |
| Private | Months 1-6: Conduct interoperability assessments; establish | Months 4-12: Develop technical infrastructure enabling integration with | Months 8-18: Deploy interoperability capabilities in controlled | Months 12-36: Scale successful pilot implementations; develop | Months 24-60 and Ongoing: Establish ongoing processes for | Reduction in system integration costs; improvement in cross-system data accuracy; reduction in compliance costs across |

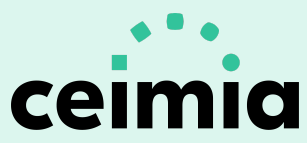

| | | | | | | |
|---|---|---|---|---|---|---|
| | cross-functional interoperability teams; develop organizational AI principles prioritizing interoperability; establish partnerships with standards organizations | emerging standards; implement standardized protocols for AI system communication; implement data governance infrastructure; build configurable compliance frameworks | environments; establish comprehensive testing protocols; develop internal expertise; establish partnerships for cross-enterprise scenarios | standardized deployment procedures; establish formal partnerships with suppliers, customers, and industry peers; engage regulatory authorities | interoperability optimization; implement continuous monitoring systems; establish formal processes for evaluating new standards; contribute to broader industry efforts | jurisdictions; increases in operational efficiency. Qualitative: capability to adopt new standards; stakeholder confidence; industry recognition. Organizations with strong infrastructure can accelerate; those with legacy systems may need extended timelines |

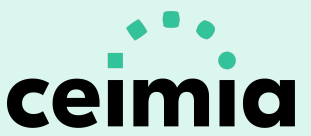

## 4. Conclusion

The current landscape of AI governance reflects both unprecedented opportunity and mounting risk. While significant initiatives like the G7's (2023a) *Hiroshima AI Process* and emerging technical standards such as Google's (2025) *A2A Protocol* represent meaningful progress, the five critical gaps identified – geopolitical disruptions, regulatory fragmentation, governance shortfalls, security and trust deficits, and framework incompatibility – threaten to fragment the global AI ecosystem in ways that could prove irreversible. The window for establishing coherent, interoperable approaches to AI governance is rapidly narrowing, making coordinated action across all sectors not merely advisable but essential for maintaining the potential benefits of AI while mitigating its risks. The sectoral roadmaps developed in this Part of the Report acknowledge that achieving meaningful AI interoperability requires differentiated but coordinated action across public, private, standards, and civil society sectors.

The economic implications of these roadmaps extend beyond compliance costs to encompass fundamental questions about market structure and competitive dynamics in the AI economy. Organizations that successfully implement interoperability frameworks will benefit from reduced integration costs, expanded market access, and decreased regulatory compliance burdens across multiple jurisdictions. Conversely, those that delay implementation or pursue purely proprietary approaches risk being locked-out of emerging AI ecosystems that increasingly depend on standardized and verified compliance mechanisms. The network effects inherent in interoperable systems mean that early adoption advantages can compound rapidly, creating winner-take-all dynamics that reward coordinated action while penalizing fragmentation.

The geopolitical stakes of AI interoperability extend beyond economic considerations to encompass national security, technological sovereignty, and global governance questions. The current trajectory toward fragmented AI ecosystems risks creating technological blocs, particularly as AI sovereignty discussions ramp up, that mirror geopolitical divisions, potentially undermining the collaborative approaches needed to address technical and regulatory interoperability. The sovereignty-compatible approaches emphasized in the standards organization roadmap offer a potential pathway for maintaining cooperative frameworks while respecting national priorities, but their success requires sustained commitment from major powers during a period of increasing technological nationalism.

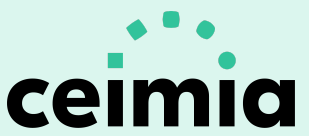

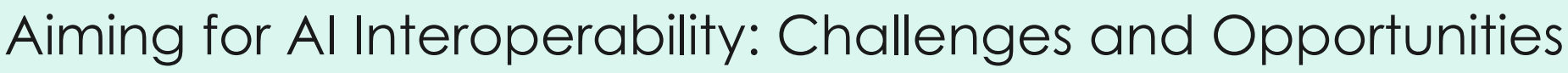

The timeline pressures identified across all roadmaps reflect a reality in which the current phase of AI development may represent the last opportunity to establish interoperable foundations before path dependencies and network effects lock-in fragmented approaches. The rapid pace of AI capability advancement, combined with increasing regulatory attention and geopolitical competition, creates a compressed window for coordinated action. Historical precedents suggest that interoperability frameworks become considerably more difficult to implement once incompatible systems are widely deployed and institutional interests become entrenched around proprietary approaches. Moving forward, success in implementing these roadmaps requires recognition that AI interoperability is not a technical problem with technical solutions, but a complex sociotechnical challenge that demands sustained coordination across multiple dimensions. Technical standards must be developed in parallel with regulatory frameworks, governance mechanisms must accommodate diverse national approaches while maintaining core commonalities, and verification systems must build trust across organizational and jurisdictional boundaries. The roadmaps provide frameworks for this coordination, but their implementation will require unprecedented levels of cooperation among actors who traditionally compete or work in isolation.

The stakes of this coordination effort cannot be overstated. Failure to achieve meaningful AI interoperability will result in a fragmented global ecosystem where AI systems cannot effectively communicate, regulatory compliance becomes prohibitively complex, and the benefits of AI innovation are unevenly distributed. Success, conversely, offers the prospect of AI systems that can collaborate across organizational and national boundaries while maintaining appropriate oversight and accountability mechanisms. The choice between these futures will be determined by actions taken over the next five years, making the implementation of these sectoral roadmaps not merely a policy preference but an urgent imperative for realizing AI's potential while managing its risks effectively.

# References

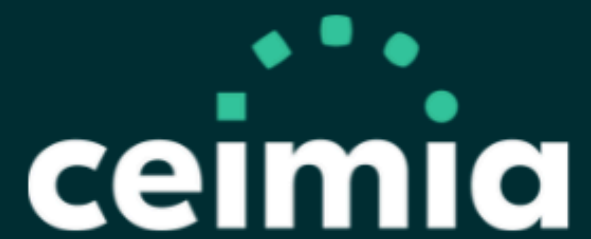

# References

A2A Protocol. (2025). *Partners*. A2A Protocol Documentation. https://a2a-protocol.org/latest/partners/

AI for Good. (2025). AI Standards Exchange Database. *AI for Good*. https://aiforgood.itu.int/ai-standards-exchange/

AI Gov. (2025). *AI Action Plan*. AI Gov. https://www.ai.gov/action-plan

AI Standards Hub. (2025). *AI Standards Hub—The New Home of the AI Standards Community*. AI Standards Hub. https://aistandardshub.org/

Allan, J., Belz, S., Hoeveler, A., Hugas, M., Okuda, H., Patri, A., Rauscher, H., Silva, P., Slikker, W., Sokull-Kluettgen, B., Tong, W., & Anklam, E. (2021). Regulatory landscape of nanotechnology and nanoplastics from a global perspective. *Regulatory Toxicology and Pharmacology*, *122*, 104885. https://doi.org/10.1016/j.yrtph.2021.104885

Anex-Ries, Q. (2025, January 10). Regulating Public Sector AI: Emerging Trends in State Legislation. *Center for Democracy and Technology*. https://cdt.org/insights/regulating-public-sector-ai-emerging-trends-in-state-legislation/

Anthropic. (2024). *Introducing the Model Context Protocol*. Anthropic. https://www.anthropic.com/news/model-context-protocol

APEC. (2025). *APEC Economies Advance AI Standards for Trust and Innovation*. APEC. https://www.apec.org/press/news-releases/2025/apec-economies-advance-ai-standards-for-trust-and-innovation

Arai, M. (2023). *Discerning signal from noise: New report explores the state of global AI standardization initiatives*. Schwartz Reisman Institute for Technology and Society. https://srinstitute.utoronto.ca/news/discerning-signal-from-noise-new-sri-report-explores-the-state-of-global-ai-standardization

Arnold, G. W., Wollman, D. A., FitzPatrick, G., Prochaska, D., Holmberg, D., Su, D. H., Jr, A. R. H., Golmie, N. T., Brewer, T. L., Bello, M., & Boynton, P. A. (2010). NIST Framework and Roadmap for Smart Grid Interoperability Standards, Release 1.0. *NIST*. https://doi.org/10.6028/NIST.sp.1108

ASEAN Secretariat. (2023, August 19). Digital Economy Framework Agreement (DEFA): ASEAN to leap forward its digital economy and unlock US$2 Tn by 2030. *ASEAN*. https://asean.org/asean-defa-study-projects-digital-economy-leap-to-us2tn-by-2030/

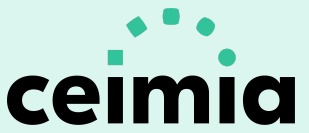

Ashdown, N. (2021, August 5). CS Alert – Offensive Cyber in 1914. *Offensive Cyber Working Group*. https://offensivecyber.org/2021/08/05/cs-alert-offensive-cyber-in-1914/

Asia-Pacific Economic Cooperation. (2023). *Economic Impact of Adopting Digital Trade Rules: Evidence from APEC Member Economies* (APEC Committee on Trade and Investment, pp. 1–69). Asia-Pacific Economic Cooperation. https://www.apec.org/publications/2023/04/economic-impact-of-adopting-digital-trade-rules-evidence-from-apec-member-economies

Auld, G., Casovan, A., Clarke, A., & Faveri, B. (2022). Governing AI through ethical standards: Learning from the experiences of other private governance initiatives. *Journal of European Public Policy*, *29*(11), 1822–1844. https://doi.org/10.1080/13501763.2022.2099449

Basu, K. (2025, February 18). *Companies Look to Jumble of AI Rules in Absence of US Direction*. Bloomberg Law. https://news.bloomberglaw.com/artificial-intelligence/companies-look-to-jumble-of-ai-rules-in-absence-of-us-direction

Bayda, S., Adeel, M., Tuccinardi, T., Cordani, M., & Rizzolio, F. (2019). The History of Nanoscience and Nanotechnology: From Chemical–Physical Applications to Nanomedicine. *Molecules*, *25*(1), 112. https://doi.org/10.3390/molecules25010112

Beckman, L. (2023). Three Conceptions of Law in Democratic Theory. *Canadian Journal of Law & Jurisprudence*, *36*(1), 65–82. https://doi.org/10.1017/cjlj.2022.22

Boghardt, T. (2012). *The Zimmermann Telegram: Intelligence, Diplomacy, and America's Entry into World War I* (Illustrated edition). Naval Institute Press.

Bremer, E. S. (2013). Incorporation by Reference in an Open-Government Age. *Harvard Journal of Law & Public Policy*, *36*(1), 131–210.

Brinsmead, S. (2021). *Essential Interoperability Standards: Interfacing Intellectual Property and Competition in International Economic Law*. Cambridge University Press. https://doi.org/10.1017/9781108913706

Brüngel, R., Rückert, J., Wohlleben, W., Babick, F., Ghanem, A., Gaillard, C., Mech, A., Rauscher, H., Hodoroaba, V.-D., Weigel, S., & Friedrich, C. M. (2019). NanoDefiner e-Tool: An Implemented Decision Support Framework for Nanomaterial Identification. *Materials*, *12*(19), Article 19. https://doi.org/10.3390/ma12193247

CAIDP. (2024). *Artificial Intelligence and Democratic Values Index—2023* (p. 1634). Center for AI and Digital Policy. https://www.caidp.org/reports/aidv-2023/

Carpenter, J., & Watts, P. (2013). *Assessing the value of OS OpenData to the economy of Great Britain—Synopsis* (pp. 1–32). Ordnance Survey.

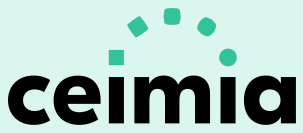

https://www.gov.uk/government/publications/ordnance-survey-open-data-economic-value-study#:~:text=OS%20OpenData%20is%20a%20portfolio,commercially%2C%20from%201%20April%202010.

Casovan, A., Lo, A., Mueller, H., Rivero-Huguet, M., & Shankar, V. (2025). *Canada and the UK compare AI governance efforts and reflect on regulating AI.* https://oecd.ai/en/wonk/canada-and-the-uk-compare-ai-governance-efforts-and-reflect-on-regulating-ai

CEIMIA. (2024). *A Comparative Framework for AI Regulatory Policy: Phase 2*. CEIMIA. https://doi.org/10.5281/zenodo.12584393

Cerf, V. G., & Kahn, R. E. (1974). A Protocol for Packet Network Intercommunication. *IEEE Transactions on Communications*, *22*(5).

Cetl, V., Nunes de Lima, V., Tomas, R., Lutz, M., D'Eugenio, J., Nagy, A., & Robbrecht, J. (2017). *Summary report on status of implementation of the INSPIRE directive in EU* (No. EUR 28930; JRC Technical Reports, pp. 1–32). European Commission. https://data.europa.eu/doi/10.2760/143502

Cho, G., & and Crompvoets, J. (2019). The INSPIRE directive: Some observations on the legal framework and implementation. *Survey Review*, *51*(367), 310–317. https://doi.org/10.1080/00396265.2018.1454686

Cihon, P. (2019). *Standards for AI Governance: International Standards to Enable Global Coordination in AI Research & Development* (pp. 1–41) [Technical Report]. Center for the Governance of AI. https://www.fhi.ox.ac.uk/wp-content/uploads/Standards_-FHI-Technical-Report.pdf

Cihon, P., Maas, M. M., & Kemp, L. (2020). Fragmentation and the Future: Investigating Architectures for International AI Governance. *Global Policy*, *11*(5), 545–556. https://doi.org/10.1111/1758-5899.12890

Copernicus. (2019). *Coordination of Information on the Environment (CORINE) Land Cover*. Ipbes. https://www.ipbes.net/node/29763

Craglia, M., & Annoni, A. (2007). *INSPIRE: An Innovative Approach to the Development of Spatial Data Infrastructures in Europe Concepts*. JRC Publications Repository. https://publications.jrc.ec.europa.eu/repository/handle/JRC37929

Data Spaces Support Centre. (2025). *Cross-data space interoperability considerations in data space design and operation*. Data Spaces Support Centre. https://dssc.eu/space/BVE2/1071252241/Cross-data+space+interoperability+considerations+in+data+space+design+and+operation

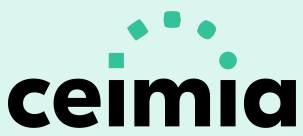

Dawidowicz, A., Kulawiak, M., Zysk, E., & Kocur-Bera, K. (2020). System architecture of an INSPIRE-compliant green cadastre system for the EU Member State of Poland. *Remote Sensing Applications: Society and Environment*, *20*, 100362. https://doi.org/10.1016/j.rsase.2020.100362

Dennis, C., Clare, S., Hawkins, R., Simpson, M., Behrens, E., Diebold, G., Kara, Z., Wang, R., Trager, R., Maas, M., Kolt, N., Anderljung, M., Pilz, K., Reuel, A., Murray, M., Heim, L., & Ziosi, M. (2024). *What Should Be Internationalised in AI Governance?* (pp. 1–75) [Whitepaper]. Oxford Martin AI Governance Initiative. https://www.oxfordmartin.ox.ac.uk/publications/what-should-be-internationalised-in-ai-governance

DIACC. (2020). *Pan-Canadian Trust Framework Model* (No. Final Recommendation V1.0). Digital ID & Authentication Council of Canada. https://diacc.ca/wp-content/uploads/2020/09/PCTF-Model-Final-Recommendation_V1.0.pdf

Dwyer, M. (2025, January 10). State Government Use of AI: The Opportunities of Executive Action in 2025. *Center for Democracy and Technology*. https://cdt.org/insights/state-government-use-of-ai-the-opportunities-of-executive-action-in-2025/

EAIDB. (2024). *The Responsible AI Ecosystem Market Map*. EAIDB. https://www.eaidb.org/map

Elsaid, K., Olabi, A. G., Abdel-Wahab, A., Elkamel, A., Alami, A. H., Inayat, A., Chae, K.-J., & Abdelkareem, M. A. (2023). Membrane processes for environmental remediation of nanomaterials: Potentials and challenges. *Science of The Total Environment*, *879*, 162569. https://doi.org/10.1016/j.scitotenv.2023.162569

Élysée. (2025, February 11). *Statement on Inclusive and Sustainable Artificial Intelligence for People and the Planet*. Elysee.Fr. https://www.elysee.fr/en/emmanuel-macron/2025/02/11/statement-on-inclusive-and-sustainable-artificial-intelligence-for-people-and-the-planet

European Commission. (2010). *INSPIRE WS legal issues—QA v 1.2*. Workshop on Legal Issues: Questions and answers on the implementation of the INSPIRE Directive 2007/2/EC, Brussels, Belgium.

European Commission. (2011). *Recommendation on the Definition of Nanomaterial*. Official Journal of the European Union. https://eur-lex.europa.eu/legal-content/EN/TXT/HTML/?uri=CELEX:32011H0696

European Commission. (2014). *eIDAS Regulation*. European Commission. https://digital-strategy.ec.europa.eu/en/policies/eidas-regulation

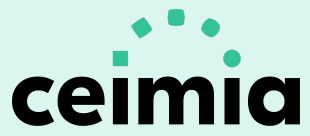

European Commission. (2017). *NANODEFINE (Development of an integrated approach based on validated and standardized methods to support the implementation of the EC recommendation for a definition of nanomaterial)*. European Commission. https://cordis.europa.eu/project/id/604347/reporting

European Commission. (2022a). *COMMISSION STAFF WORKING DOCUMENT EVALUATION of DIRECTIVE 2007/2/EC establishing an Infrastructure for Spatial Information in the European Community (INSPIRE)*. European Commission. https://eur-lex.europa.eu/legal-content/EN/TXT/?uri=CELEX%3A52022SC0195

European Commission. (2022b). *Recommendation on the Definition of Nanomaterial*. Official Journal of the European Union. https://eur-lex.europa.eu/legal-content/EN/TXT/?uri=oj:JOC_2022_229_R_0001

European Commission. (2024). *Harmonised Standards for the European AI Act—European Commission*. European Commission. https://ai-watch.ec.europa.eu/news/harmonised-standards-european-ai-act-2024-10-25_en

European Commission. (2025a). *Danish Basic Data Program*. European Commission. https://ec.europa.eu/digital-building-blocks/sites/digital-building-blocks/sites/pages/viewpage.action?pageId=533365971

European Commission. (2025b). *INSPIRE in your Country*. INSPIRE Knowledge Base. https://knowledge-base.inspire.ec.europa.eu/tools/inspire-your-country_en

European Commission. (2025c). *The General-Purpose AI Code of Practice*. European Commission. https://digital-strategy.ec.europa.eu/en/policies/contents-code-gpai

European Parliament. (2016). *Regulatory Fitness and Performance Programme*. European Parliament. https://www.europarl.europa.eu/doceo/document/TA-8-2016-0104_EN.html

European Trade Union Institute. (2011). *The work of CEN and CENELEC in the context of the New approach*. Etui. https://www.etui.org/topics/health-safety-working-conditions/safety-of-machinery-directives-and-standards/standardisation/the-work-of-cen-and-cenelec-in-the-context-of-the-new-approach

European Union. (1992). *Directive—92/43—Habitats Directive*. European Union. https://eur-lex.europa.eu/eli/dir/1992/43/oj/eng

European Union. (2000). *Directive—2000/60—Water Framework Directive*. European Union. https://eur-lex.europa.eu/eli/dir/2000/60/oj/eng

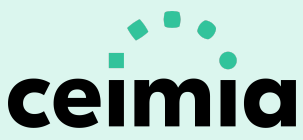

European Union. (2006). *Registration, Evaluation, Authorisation and Restriction of Chemicals*. European Union. http://data.europa.eu/eli/reg/2006/1907/2022-12-17/eng

European Union. (2007). *Directive—2007/2—INSPIRE Directive*. European Union. https://eur-lex.europa.eu/eli/dir/2007/2/oj/eng

European Union. (2008). *Directive—2008/56—Marine Strategy Framework Directive*. European Union. https://eur-lex.europa.eu/eli/dir/2008/56/oj/eng

European Union. (2012). *Biocidal Products Regulation*. European Union. http://data.europa.eu/eli/reg/2012/528/oj/eng

European Union. (2017, April 5). *Medical Devices Regulation*. European Union. http://data.europa.eu/eli/reg/2017/745/oj/eng

Executive Office of the President. (2023, November 1). *Safe, Secure, and Trustworthy Development and Use of Artificial Intelligence*. Federak Register. https://www.federalregister.gov/documents/2023/11/01/2023-24283/safe-secure-and-trustworthy-development-and-use-of-artificial-intelligence

FATF. (2025). *The FATF Recommendations*. FATF. https://www.fatf-gafi.org/en/publications/Fatfrecommendations/Fatf-recommendations.html

Faveri, B. (2025). *Exploring Influencing Factors to Incorporation by Reference Use: A Comparative Analysis of Canada and the United States*. SSRN. https://doi.org/10.2139/ssrn.5217207

Faveri, B., & Auld, G. (2025). Assurance Actors as Intermediaries in AI Risk Governance. *Journal of Comparative Policy Analysis*, 1–24. https://doi.org/10.1080/13876988.2025.2578357

Faveri, B., & Auld, G. (2023). *Informing Possible Futures for the use of Third-Party Audits in AI Regulations* (pp. 1–41). Carleton University. http://doi.org/10.22215/sppa-rgi-nov2023

Faveri, B., & Auld, G. (2024). *Lessons for Responsible AI Standardization from Transnational Governance* (SSRN Scholarly Paper No. 4911923). https://papers.ssrn.com/abstract=4911923

Faveri, B., Johnson-León, M., & Sylvester, P. (2025). *Towards A Global AI Auditing Framework: Assessment and Recommendations* (Scientific Panel on Global Standards for AI Audits, pp. 1–79). https://ipie.info/10.61452/ZWED1485

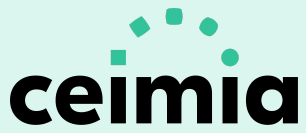

Faveri, B., & Marchant, G. (2024a). *Artificial Intelligence (AI) and AI-Related Legislative Efforts at the United States' Federal and State Levels* [Dataset]. Harvard Dataverse. https://doi.org/10.7910/DVN/I5RO4Y

Faveri, B., & Marchant, G. (2024b). *International Artificial Intelligence (AI) and AI-Related Standards Landscape: ISO, IEEE, ETSI, and ITU* [Dataset]. Harvard Dataverse. https://doi.org/10.7910/DVN/LBIXU6

Faveri, B., Shank, C., Whitt, R., & Dawson, P. (2025a, April 16). *The Need for and Pathways to AI Regulatory and Technical Interoperability*. Tech Policy Press. https://techpolicy.press/the-need-for-and-pathways-to-ai-regulatory-and-technical-interoperability

Faveri, B., Shank, C., Whitt, R., & Dawson, P. (2025b, July 10). *Learning from Past Successes and Failures to Guide AI Interoperability*. Tech Policy Press. https://techpolicy.press/learning-from-past-successes-and-failures-to-guide-ai-interoperability

Faveri, B., Shank, C., Whitt, R., & Dawson, P. (2025, October 15). *Closing the Gaps in AI Interoperability*. Tech Policy Press. https://techpolicy.press/closing-the-gaps-in-ai-interoperability

Fazlioglu, M. (2023). *The Snowden disclosures, 10 years on*. IAPP. https://iapp.org/news/a/the-snowden-disclosures-10-years-on

Federal Register. (2025, January 31). *Removing Barriers to American Leadership in Artificial Intelligence*. Federal Register. https://www.federalregister.gov/documents/2025/01/31/2025-02172/removing-barriers-to-american-leadership-in-artificial-intelligence

Femenia-Ribera, C., Mora-Navarro, G., & Pérez, L. J. S. (2022). Evaluating the use of old cadastral maps. *Land Use Policy*, *114*, 105984. https://doi.org/10.1016/j.landusepol.2022.105984

Ferguson, C. (1999). *High Stakes, No Prisoners: A Winner's Tale of Greed and Glory in the Internet Wars*. Crown Business.

G7. (2023). *G7 Hiroshima AI Process*. Hiroshima AI Process. https://www.soumu.go.jp/hiroshimaaiprocess/en/index.html

G7. (2023b). *Hiroshima Process International Guiding Principles for Organizations Developing Advanced AI System*. G7. https://www.mofa.go.jp/files/100573471.pdf

G7. (2024). *Friends Group*. Hiroshima AI Process. https://www.soumu.go.jp/hiroshimaaiprocess/en/meeting.html

Gabele, F. (2008). *The implementation and the transposition of the INSPIRE Directive within the Belgian Federal context*. FIG Open Symposium, Verona, Fiere. https://www.fig.net/resources/proceedings/2008/verona_am_2008_comm7/ppt/14_sept/5_1_gabele.pdf

Gaillard, C., Mech, A., Wohlleben, W., Babick, F., Hodoroaba, V.-D., Ghanem, A., Weigel, S., & Rauscher, H. (2019). A technique-driven materials categorisation scheme to support regulatory identification of nanomaterials. *Nanoscale Advances*, *1*(2), 781–791. https://doi.org/10.1039/C8NA00175H

Gil, H. M., Price, T. W., Chelani, K., Bouillard, J.-S. G., Calaminus, S. D. J., & Stasiuk, G. J. (2021). NIR-quantum dots in biomedical imaging and their future. *iScience*, *24*(3), 102189. https://doi.org/10.1016/j.isci.2021.102189

Google. (2024). *Recommendations for Regulating AI* (pp. 1–20). Google. https://ai.google/static/documents/recommendations-for-regulating-ai.pdf

Gottardo, S., Mech, A., Drbohlavová, J., Małyska, A., Bøwadt, S., Riego Sintes, J., & Rauscher, H. (2021). Towards safe and sustainable innovation in nanotechnology: State-of-play for smart nanomaterials. *NanoImpact*, *21*, 100297. https://doi.org/10.1016/j.impact.2021.100297

Government of Canada. (2023, July 31). *Public Key Infrastructure Configuration Requirements*. Government of Canada. https://www.canada.ca/en/government/system/digital-government/policies-standards/enterprise-it-service-common-configurations/public-key-infrastructure-configuration-requirements.html

GOV.UK. (2025). *The Bletchley Declaration by Countries Attending the AI Safety Summit, 1-2 November 2023*. GOV.UK. https://www.gov.uk/government/publications/ai-safety-summit-2023-the-bletchley-declaration/the-bletchley-declaration-by-countries-attending-the-ai-safety-summit-1-2-november-2023

Hadfield, G., Arai, M., Casovan, A., Shankar, V., Narayan, J., Bodkin, R., Cobey, C., Zhu, Y., Dawson, P., Dymond, J., Green-Noble, L., Clavell, G., Khlaaf, H., Mittelstadt, B., Scott, A., Payette, D., Rubtsov, A., Stoyanovich, J., & Shank, C. (2024). *Artificial Intelligence Certification: Unlocking the power of AI through innovation and trust* (Certification Working Group). Schwartz Reisman Institute for Technology and Society. https://srinstitute.utoronto.ca/news/ai-certification-ecosystem

Harrell, P. (2025). *Managing the Risks of China's Access to U.S. Data and Control of Software and Connected Technology*. Carnegie Endowment for International Peace. https://carnegieendowment.org/research/2025/01/managing-the-

risks-of-chinas-access-to-us-data-and-control-of-software-and-connected-technology?lang=en

Higgins, P., & Eckersley, P. (2011, December 15). *An Open Letter From Internet Engineers to the U.S. Congress*. Electronic Frontier Foundation. https://www.eff.org/deeplinks/2011/12/internet-inventors-warn-against-sopa-and-pipa

High-Level Advisory Body on AI. (2024). *Governing AI for Humanity* (pp. 1–101). United Nations.

Huang, K. (2025). Threat Modeling Google's A2A Protocol [Cloud Security Alliance]. *Industry Insights*. https://cloudsecurityalliance.org/blog/2025/04/30/threat-modeling-google-s-a2a-protocol-with-the-maestro-framework

IAPP. (2025a). *Global AI Governance Law and Policy: Canada*. IAPP. https://iapp.org/resources/article/global-ai-governance-canada/

IAPP. (2025b). *US State AI Governance Legislation Tracker*. IAPP. https://iapp.org/resources/article/us-state-ai-governance-legislation-tracker/

Information Commissioner's Office. (2025). *Our work on Artificial Intelligence*. Information Commissioner's Office; ICO. https://ico.org.uk/about-the-ico/what-we-do/our-work-on-artificial-intelligence/

Innovation Science and Economic Development Canada. (2024, November 21). *The International Network of AI Safety Institutes: Mission statement*. Government of Canada; Innovation, Science and Economic Development Canada. https://ised-isde.canada.ca/site/ised/en/international-network-ai-safety-institutes-mission-statement

INSPIRE Knowledge Base. (2025). *Implement*. European Commission. https://knowledge-base.inspire.ec.europa.eu/overview/implement_en

International Network of AI Safety Institutes. (2024). *International Network of AI Safety Institutes: Mission Statement*. https://www.nist.gov/document/international-network-ai-safety-institutes-mission-statement

Internet Governance Forum. (2024). *The AI Governance We Want, Call to Action: Liability, Interoperability, Sustainability & Labour* (No. PNAI Policy Brief 2024; Policy Network on AI, pp. 1–111). Internet Governance Forum. https://www.dataeconomypolicyhub.org/_files/ugd/c77f62_6f7fd3e26733408388e3920a348cead6.pdf

ISO. (1984). *ISO 7498:1984 Information processing systems—Open Systems Interconnection—Basic Reference Model*. ISO. https://www.iso.org/standard/14252.html

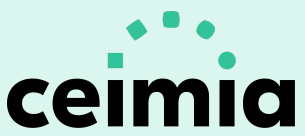

ISO. (2023). *ISO/IEC 22123-3:2023 Information technology—Cloud computing—Part 3: Reference architecture*. ISO. https://www.iso.org/standard/82759.html

ITU. (1865). *International Telegraph Convention*. ITU. https://search.itu.int/history/HistoryDigitalCollectionDocLibrary/4.1.43.fr.201.pdf

ITU. (2020). *Plenipotentiary Conferences*. ITU. https://www.itu.int:443/en/history/Pages/PlenipotentiaryConferences.aspx?conf=4.1

ITU. (2024). *Overview of ITU's History*. ITU. https://www.itu.int:443/en/history/Pages/ITUsHistory.aspx

ITU. (2025). *The 1865 International Telegraph Conference*. ITU. https://www.itu.int:443/en/history/Pages/ITUBorn1865.aspx

Jobin, A., Ienca, M., & Vayena, E. (2019). The global landscape of AI ethics guidelines. *Nature Machine Intelligence*, *1*(9), Article 9. https://doi.org/10.1038/s42256-019-0088-2

Kang, C. (2025, July 1). Defeat of a 10-Year Ban on State A.I. Laws Is a Blow to Tech Industry. *The New York Times*. https://www.nytimes.com/2025/07/01/us/politics/state-ai-laws.html

Keil, M. (2017). *CORINE Land Cover products for Germany, created by DLR-DFD on behalf of the Federal Environment Agency (UBA) – an Overview* (pp. 1–10). German Aerospace Center. https://www.dlr.de/en/eoc/research-transfer/projects-missions/corine-land-cover

Kekulawala, C. (2025, May 22). Everything wrong with Agent2Agent (A2A) Protocol. *Medium*. https://medium.com/@ckekula/everything-wrong-with-agent2agent-a2a-protocol-7e5ae8d4ab2b

Kerry, C., Meltzer, J., Renda, A., & Wyckoff, A. (2025). Network architecture for global AI policy. *Brookings*. https://www.brookings.edu/articles/network-architecture-for-global-ai-policy/

Kjellson, B. (2017). *Country Report of Sweden: Swedish Spatial Data Infrastructure and the National Geodata Strategy* (pp. 1–16). United Nations Committee of Experts on Global Geospatial Information Management. https://ggim.un.org/country-reports/documents/Sweden-2017-country-report.pdf

Kore, J., & Dhawan, E. (2025). The Global Landscape of AI Safety Institutes. *All Tech Is Human*. https://alltechishuman.org/all-tech-is-human-blog/the-global-landscape-of-ai-safety-institutes

Kumar, N. (2024, June 16). G7 Summit, 2024 Italia: G7 Leaders' Communiqué, The Official Statement of G7 leaders. *The International Parliament Journal (IPJ)*. https://parliamentjournal.com/2024/06/16/g7-summit-2024-italia-g7-leaders-communique-the-official-statement-of-g7-leaders/

Kumar, P., & Raj, A. (2025). Nanoencapsulation Techniques for Controlled Release of Agrochemicals. In *Sustainable Era of Nanomaterials* (pp. 147–184). Springer, Singapore. https://doi.org/10.1007/978-981-96-4471-1_8

Land Monitoring Service. (2021). *CORINE Land Cover*. Copernicus. https://land.copernicus.eu/en/products/corine-land-cover

Laszio, M. (2025). *Colorado's AI law delayed until June 2026: What the latest setback means for businesses*. Clark Hill PLC. https://www.clarkhill.com/news-events/news/colorados-ai-law-delayed-until-june-2026-what-the-latest-setback-means-for-businesses/

Liu, F., Tong, J., Mao, J., Bohn, R. B., Messina, J. V., Badger, M. L., & Leaf, D. M. (2011). NIST Cloud Computing Reference Architecture. *NIST*. https://doi.org/10.6028/NIST.SP.500-292

Louck, Y., Stulman, A., & Dvir, A. (2025). *Improving Google A2A Protocol: Protecting Sensitive Data and Mitigating Unintended Harms in Multi-Agent Systems* (No. arXiv:2505.12490; Version 3). arXiv. https://doi.org/10.48550/arXiv.2505.12490

Lyon, D. (2014). Surveillance, Snowden, and Big Data: Capacities, consequences, critique. *Big Data & Society*, *1*(2). https://doi.org/10.1177/2053951714541861

Malik, S., Muhammad, K., & Waheed, Y. (2023). Nanotechnology: A Revolution in Modern Industry. *Molecules (Basel, Switzerland)*, *28*(2), 661. https://doi.org/10.3390/molecules28020661

Marchant, G. E., & Gutierrez, C. I. (2023). Soft Law 2.0: An Agile and Effective Governance Approach for Artificial Intelligence. *Minnesota Journal of Law, Science and Technology*, *24*(2), 375–424.

Marsh, A. (2018). Morse Code's Vanquished Competitor: The Dial Telegraph. *IEEE Spectrum*. https://spectrum.ieee.org/morse-codes-vanquished-competitor-the-dial-telegraph

Mech, A., Rauscher, H., Babick, F., Hodoroaba, V.-D., Ghanem, A., Wohlleben, W., Marvin, H., Weigel, S., Brüngel, R., Friedrich, C. M., Rasmussen, K., Löschner, K., & Gilliland, D. (2020). *The NanoDefine methods manual*. Publications Office of the European Union. https://data.europa.eu/doi/10.2760/79490

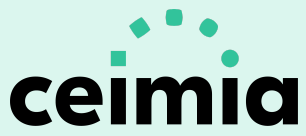

Mech, A., Wohlleben, W., Ghanem, A., Hodoroaba, V.-D., Weigel, S., Babick, F., Brüngel, R., Friedrich, C. M., Rasmussen, K., & Rauscher, H. (2020). Nano or Not Nano? A Structured Approach for Identifying Nanomaterials According to the European Commission's Definition. *Small*, *16*(36), 2002228. https://doi.org/10.1002/smll.202002228

Menon, N. (2025, July 24). The Rise of the Global South in Standard-Setting: Opportunity or Overdue? *The BSB EDGE Blog*. https://blog.bsbedge.com/index.php/2025/07/24/the-rise-of-the-global-south-in-standard-setting-opportunity-or-overdue/

Metz, C. (2012). Meet the Man Who Invented the Instructions for the Internet. *Wired*. https://www.wired.com/2012/05/steve-crocker/

NANDA. (2025). *NANDA - The Internet of AI Agents*. MIT NANDA. https://nanda.media.mit.edu/

NIST. (2023). *AI Risk Management Framework*. NIST. https://www.nist.gov/itl/ai-risk-management-framework

OECD. (2019). *The OECD AI Principles*. OECD.AI Policy Observative. https://oecd.ai/en/ai-principles

OECD.AI. (2025). *OECD's live repository of AI strategies & policies*. OECD.AI Policy Observatory. https://oecd.ai/en/dashboards

Office of the United States Trade Representative. (2020). *United States-Mexico-Canada Agreement*. United States Trade Representative. https://ustr.gov/trade-agreements/free-trade-agreements/united-states-mexico-canada-agreement/agreement-between

Office of the United States Trade Representative. (2021). *U.S.-E.U. Trade and Technology Council (TTC)*. United States Trade Representative. https://ustr.gov/useuttc

Ogryzek, M., Tarantino, E., & Rząsa, K. (2020). Infrastructure of the Spatial Information in the European Community (INSPIRE) Based on Examples of Italy and Poland. *ISPRS International Journal of Geo-Information*, *9*(12), Article 12. https://doi.org/10.3390/ijgi9120755

Onikepe, A. (2024, July 17). *Interoperability in AI governance: A Work in Progress*. Tech Policy Press. https://techpolicy.press/interoperability-in-ai-governance-a-work-in-progress

Open Voice Interoperability Initiative. (2020). *About Interoperable Conversational Assistants*. Open Voice Interoperability. https://voiceinteroperability.ai/about-interoperability/

OpenAI. (2025). *Recognition of International and Federal AI Safety Frameworks for State Law Compliance*. https://cdn.openai.com/pdf/oai_ca-safety-letter_8-11-25.pdf

Ostrom, E. (1990). *Governing the Commons: The Evolution of Institutions for Collective Action*. Cambridge University Press. https://doi.org/10.1017/CBO9780511807763

OWASP. (2025). Agent Name Service (ANS) for Secure AI Agent Discovery v1.0. *OWASP Gen AI Security Project*. https://genai.owasp.org/resource/agent-name-service-ans-for-secure-al-agent-discovery-v1-0/

paloalto network. (2025, August 14). *Safeguarding AI Agents: An In-Depth Look at A2A Protocol Risks and Mitigations*. Paloalto. https://live.paloaltonetworks.com/t5/community-blogs/safeguarding-ai-agents-an-in-depth-look-at-a2a-protocol-risks/ba-p/1235996

Racicot, R., & Simpson, K. (2025). *China's AI Governance Initiative and Its Geopolitical Ambitions*. Centre for International Governance Innovation. https://www.cigionline.org/articles/chinas-ai-governance-initiative-and-its-geopolitical-ambitions/

Rannestig, E., & Nilsson, M. (2008). *INSPIRE in Sweden—An Important Part of the National Geodata Strategy*. https://www.fig.net/resources/proceedings/fig_proceedings/fig2008/papers/ts01a/ts01a_03_%20rannestig_nilsson_2658.pdf

Rasmussen, K., González, M., Kearns, P., Sintes, J. R., Rossi, F., & Sayre, P. (2016). Review of achievements of the OECD Working Party on Manufactured Nanomaterials' Testing and Assessment Programme. From exploratory testing to test guidelines. *Regulatory Toxicology and Pharmacology*, *74*, 147–160. https://doi.org/10.1016/j.yrtph.2015.11.004

Rogers, M., & Eden, G. (2017). Digital Citizenship and Surveillance | The Snowden Disclosures, Technical Standards, and the Making of Surveillance Infrastructures. *International Journal of Communication*, *11*(0).

Roosa, S. (2025, June 30). AI Armageddon Series. *Data Protection Report*. https://www.dataprotectionreport.com/2025/06/ai-armageddon-series/

Rose, S., Borchert, O., Mitchell, S., & Connelly, S. (2020). *Zero Trust Architecture* (No. NIST Special Publication (SP) 800-207). National Institute of Standards and Technology. https://doi.org/10.6028/NIST.SP.800-207

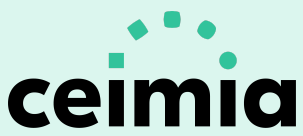

Rossi, A. (2018). How the Snowden Revelations Saved the EU General Data Protection Regulation. *The International Spectator*, *53*(4), 95–111. https://doi.org/10.1080/03932729.2018.1532705

Russell, A. L. (2014). *Open Standards and the Digital Age: History, Ideology, And Networks*. Cambridge University Press.

Science Museum. (2018). *Standardising time: Railways and the electric telegraph*. Science Museum. https://www.sciencemuseum.org.uk/objects-and-stories/standardising-time-railways-and-electric-telegraph

Shank, C. E. (2021). Credibility of Soft Law for Artificial Intelligence—Planning and Stakeholder Considerations. *IEEE Technology and Society Magazine*, *40*(4), 25–36. https://doi.org/10.1109/MTS.2021.3123737

Siemens. (2025). *Telegraphy and Telex*. Siemens. https://www.siemens.com/global/en/company/about/history/technology/information-and-communications-technology/telegraphy-and-telex.html

Soler, G. J., De, N. S., Bassani, E., Sanchez, I., Evas, T., André, A.-A., & Boulangé, T. (2024). *Harmonised Standards for the European AI Act*. European Commission. https://publications.jrc.ec.europa.eu/repository/handle/JRC139430

Standage, T. (2014). *The Victorian Internet: The Remarkable Story of the Telegraph and the Nineteenth Century's On-line Pioneers* (Second Edition, Revised). Bloomsbury USA.

Surapaneni, R., Jha, M., Vakoc, M., & Segal, T. (2025). *Announcing the Agent2Agent Protocol (A2A)*. Google for Developers. https://developers.googleblog.com/en/a2a-a-new-era-of-agent-interoperability/

Surapaneni, R., & Stephens, P. (2025). Agent2Agent protocol (A2A) is getting an upgrade. *Google Cloud Blog*. https://cloud.google.com/blog/products/ai-machine-learning/agent2agent-protocol-is-getting-an-upgrade

Sylvester, D. J., Abbott, K. W., & Marchant, G. E. (2009). Not again! Public perception, regulation, and nanotechnology. *Regulation & Governance*, *3*(2), 165–185. https://doi.org/10.1111/j.1748-5991.2009.01049.x

Tech for Good Institute. (2024, November 15). The ASEAN Digital Economy Framework Agreement: A Vision for a Future-Ready Region. *Tech For Good Institute*. https://techforgoodinstitute.org/blog/expert-opinion/the-asean-digital-economy-framework-agreement-a-vision-for-a-future-ready-region/

Thumfart, J. (2025). Digital Sovereignty in China, Russia, and India: From NWICO to SCO and BRICS. In L. Belli & M. Jiang (Eds.), *Digital Sovereignty in the BRICS*

*Countries: How the Global South and Emerging Power Alliances Are Reshaping Digital Governance* (pp. 41–62). Cambridge University Press. https://doi.org/10.1017/9781009531085.004

Tsiavos, P. (2010). *INSPIREd by Openness: The case of the implementation of Directive 2007/2/EC in Greece as a general model for open data regulation within the context of Public Sector Information* (Topic Report No. 16; pp. 1–18). European Public Sector Information Platform.

UNECE. (1998). *Aarhus Convention*. https://unece.org/DAM/env/pp/documents/cep43e.pdf

United Nations. (2023). *High-Level Advisory Body on Artificial Intelligence*. UN Office for Digital and Emerging Technologies. https://www.un.org/digital-emerging-technologies/ai-advisory-body

Van, N. P. (2011). *Impact analysis of the Joint Research Centre and its direct actions under the EU Research Framework Programmes* (No. JRC66774; pp. 1–74). European Commission. https://doi.org/10.2788/6724

van Rijn, J., Afantitis, A., Culha, M., Dusinska, M., Exner, T. E., Jeliazkova, N., Longhin, E. M., Lynch, I., Melagraki, G., Nymark, P., Papadiamantis, A. G., Winkler, D. A., Yilmaz, H., & Willighagen, E. (2022). European Registry of Materials: Global, unique identifiers for (undisclosed) nanomaterials. *Journal of Cheminformatics*, *14*(1), Article 1. https://doi.org/10.1186/s13321-022-00614-7

Vanderhaegen, M., & Muro, E. (2005). Contribution of a European spatial data infrastructure to the effectiveness of EIA and SEA studies. *Environmental Impact Assessment Review*, *25*(2), 123–142. https://doi.org/10.1016/j.eiar.2004.06.011

Villasenor, J. (2025). How overly aggressive bans on AI chip exports to China can backfire. *Brookings*. https://www.brookings.edu/articles/how-overly-aggressive-bans-on-ai-chip-exports-to-china-can-backfire/

Webster, G., Creemers, R., Kania, E., & Triolo, P. (2017). Full Translation: China's "New Generation Artificial Intelligence Development Plan" (2017). *Stanford University: DigiChina*. https://digichina.stanford.edu/work/full-translation-chinas-new-generation-artificial-intelligence-development-plan-2017/

Wenzlhuemer, R. (2007). The Development of Telegraphy, 1870–1900: A European Perspective on a World History Challenge. *History Compass*, *5*(5), 1720–1742. https://doi.org/10.1111/j.1478-0542.2007.00461.x

Werbach, K. (2007). Only Connect. *Berkeley Technology Law Journal*, *22*, 1233–1289.

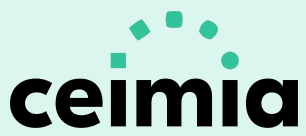

White House. (2023, October 30). Executive Order on the Safe, Secure, and Trustworthy Development and Use of Artificial Intelligence. *White House*. https://bidenwhitehouse.archives.gov/briefing-room/presidential-actions/2023/10/30/executive-order-on-the-safe-secure-and-trustworthy-development-and-use-of-artificial-intelligence/

White House Office of Science and Technology Policy. (2022). *Blueprint for an AI Bill of Rights* (pp. 1–73). The White House. https://www.whitehouse.gov/wp-content/uploads/2022/10/Blueprint-for-an-AI-Bill-of-Rights.pdf

White House, T. W. (2025, January 14). *Executive Order on Advancing United States Leadership in Artificial Intelligence Infrastructure*. The White House. https://bidenwhitehouse.archives.gov/briefing-room/presidential-actions/2025/01/14/executive-order-on-advancing-united-states-leadership-in-artificial-intelligence-infrastructure/

Whitt, R. (with Burgess, N., & Sperry, M.). (2024). *Reweaving the Web: How together we can create a human-centered Internet of trust* (A. Green, Ed.).

Whitt, R. S. (2013). *A Deference to Protocol: Fashioning a Three-Dimensional Public Policy Framework for the Internet Age* (SSRN Scholarly Paper No. 2031186). Social Science Research Network. https://doi.org/10.2139/ssrn.2031186

Whitt, R. S. (2019). Hiding in the Open: How Tech Network Policies Can Inform Openness by Design (and Vice Versa). *Georgetown Law Technology Review*, *3*(1).

Wiggers, K. (2025a, March 26). OpenAI adopts rival Anthropic's standard for connecting AI models to data. *TechCrunch*. https://techcrunch.com/2025/03/26/openai-adopts-rival-anthropics-standard-for-connecting-ai-models-to-data/

Wiggers, K. (2025b, April 9). Google to embrace Anthropic's standard for connecting AI models to data. *TechCrunch*. https://techcrunch.com/2025/04/09/google-says-itll-embrace-anthropics-standard-for-connecting-ai-models-to-data/

Woolf, A. (2010). Powered by standards – new data tools for the climate sciences. *International Journal of Digital Earth*, *3*(1), 85–102. https://doi.org/10.1080/17538941003672268

World Bank. (2024). *Global Trends in AI Governance: Evolving Country Approaches* (pp. 1–105). World Bank. https://hdl.handle.net/10986/42500

World Trade Organization. (1995). *Agreement on Technical Barriers to Trade*. World Trade Organization. https://www.wto.org/english/docs_e/legal_e/tbt_e.htm

Zajko, M. (2023). *Canada is failing to regulate AI amid fear and hype*. Policy Options. https://policyoptions.irpp.org/magazines/june-2023/canada-failing-ai-regulation-fear-hype/

Zanatta, R., & Rielli, M. (2024, December 10). The Artificial Intelligence Legislation in Brazil: Technical Analysis of the Text to Be Voted on in the Federal Senate Plenary. *Data Privacy Brasil Research*. https://www.dataprivacybr.org/en/the-artificial-intelligence-legislation-in-brazil-technical-analysis-of-the-text-to-be-voted-on-in-the-federal-senate-plenary/

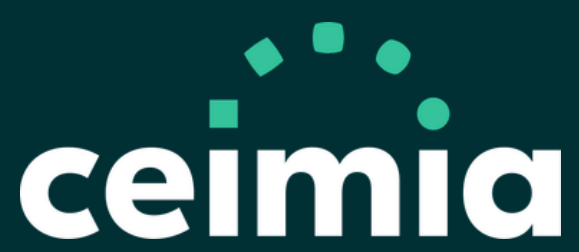

7260 Rue Saint-Urbain, Suite 602, Montréal,
QC H2R 2Y6, Canada
info@ceimia.org
ceimia.org

Follow us on

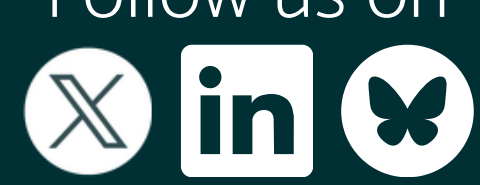